\documentclass{aa}

\usepackage{graphicx}
\usepackage{txfonts}
\usepackage{tikz,tikz-3dplot}\usetikzlibrary{calc}
\usepackage{mathtools}
\usepackage{caption}
\usepackage{subcaption}
\usepackage{xcolor}
\usepackage{hyperref}
\hypersetup{colorlinks,breaklinks,
  linkcolor=[rgb]{0.368417, 0.506779, 0.709798},
  citecolor=[rgb]{0.368417, 0.506779, 0.709798},
  urlcolor=[rgb]{0.368417, 0.506779, 0.709798}}
\usepackage{siunitx}
\usepackage{pifont}

\usepackage{amstext}  
\usepackage{comment}

\newcommand{\mbf}{\mathbf}
\newcommand{\mrm}{\mathrm} 

\newcommand{\be}{\begin{equation}}                                 
\newcommand{\ee}{\end{equation}}                                   
\newcommand{\bea}{\begin{eqnarray}}                                
\newcommand{\eea}{\end{eqnarray}}

\usepackage{amsmath}
\usepackage{upgreek}
\usepackage{booktabs}

\begin{document}

\title{The effects of the spin and quadrupole moment of SgrA* on the orbits of S stars}

   \author{K. Abd El Dayem
          \inst{1}
          \and
          F. H. Vincent\inst{1}
          \and
          G. Heissel\inst{1,2}          
          \and
          T. Paumard\inst{1}
          \and
          G. Perrin\inst{1}
          }

   \institute{LIRA, Observatoire de Paris, Université PSL, CNRS, Sorbonne Université, Université Paris Cité, 5 place Jules Janssen, 92195 Meudon, France\\
              \email{karim.abdeldayem@obspm.fr}
            \and
            Advanced Concepts Team, European Space Agency, TEC-SF, ES-TEC, Keplerlaan 1, NL-2201, AZ Noordwijk, The Netherlands
             }


 
  \abstract
   {Measuring the astrometric and spectroscopic data of stars orbiting the central black hole in our galaxy (Sgr A*) offers a promising way to measure relativistic effects. In principle, the “no-hair” theorem can be tested at the Galactic Center by monitoring the orbital precession of S-stars due to the angular momentum (or spin) and quadrupole moment of Sgr A*. For this, closer-in stars that are more affected by the rotation of the black hole might be required. It is possible that future observations of GRAVITY+ could detect such inner stars that might have been too faint to be detected by GRAVITY.
   }
   {We want to analytically and numerically characterize the orbital reorientations induced by spin-related effects of Sgr A* up to second post-Newtonian (2PN) order. 
   }
   {We use the two-timescale method to derive the 2PN analytical expressions of the secular evolution of the orbital parameters that are related to the observer. In order to study the interaction between the orientation of an orbit and that of the black hole, we introduce quantities that are straightforwardly derived from the orbital parameters, while being observer-independent, and giving us a theoretical insight into the impact of the Kerr geometry. We used a post-Newtonian code (OOGRE) to simulate putative stars that orbit closer to Sgr A*, thus being much more affected by the spin and quadrupole moment effects. This allows us to test the code against analytical expressions that we derived.} 
   {We exhibit three orbital-timescale precession rates that encode the in-plane pericenter shift and the out-of-plane redirection of the osculating ellipse. We provide the 2PN expressions of these precession rates and express the orbit-integrated associated angular shifts of the pericenter and of the ellipse axes. We relate these orbital-timescale precession rates to the secular-timescale precession of the orbital angular momentum around the black hole spin axis. We consider that the theoretical insight we provide in this article will be useful in constraining the spin effect of Sgr A* with GRAVITY+ observations.
   }
   {}


   \maketitle

\section{Introduction}

After years of monitoring S stars in the central parsec of the Milky Way studies have demonstrated the presence of a supermassive compact object called Sgr A* at the center of the Galaxy \citep{Eckart+96,Ghez+98,Ghez+03,Ghez+08,Schoedel+02,GillessenEtAl2009,Gillessen+17,GRAVITY+22_mass_distribution}, which is very likely a supermassive black hole (SMBH). According to the "no-hair" theorem, a black hole can be completely characterized by only three externally observable classical parameters: the mass, the angular momentum (hereafter referred to as the spin), and the electric charge. The latter is constrained for Sgr A* by recent General Relativistic Particle-In-Cell (GRPIC) simulations to a fraction of Wald's charge (B. Cerutti, private communication), that is, the charge acquired by a black hole immersed in a vertical magnetic field \citep{Wald_1974}. The magnetic field needed for this charge to become dynamically important for bodies orbiting Sgr A* is many orders of magnitude higher than the typical magnetic field in the strong-field region close to Sgr A*. We can thus safely neglect the BH charge and consider that it is fully characterized by only two parameters, its mass and spin, and described within general relativity by the Kerr metric. 
This theorem thus implies that if we consider a higher moment of the gravity field, like the mass quadrupole, it must be linked to the mass and spin, meaning that this theorem can be tested by independently measuring these 3 quantities and verifying the relation between them. Dozens of S star orbits are currently known \citep{Gillessen+17}, including the highly elliptical one of the star S2 with a 16-year period, reaching $R \approx 1400R_S \approx 120$AU from Sgr A* at its pericenter, where $R_S = 2GM_{\bullet}/c^2$ with $G$ and $M_{\bullet}$ being the gravitational constant and the black hole's mass respectively. By combining the infrared light collected by the 4 Unit Telescopes of the Very Large Telescope (VLT) at Paranal, the interferometric instrument GRAVITY was able to estimate the mass of Sgr A* at $M_{\bullet}=(4.2996\pm 0.0118) \times10^6 M_{\odot}$ \citep{GRAVITY+24}. Even though the spin and quadrupole moment of the black hole are still unknown, the future monitoring of S stars could, in principle, provide constraints on these parameters \citep{Will2008,Angelil+14} which in turn would allow us to test the “no-hair” theorem. 

As stated in \citet{Yu+16}, understanding the spin direction of a supermassive black hole is also important because it can reveal clues about its growth history. For example, if the black hole spin is aligned with the young stellar disk in the Galactic Center, it may suggest that the black hole grew mainly from a gas disk that once matched the stellar disk. However, if the spin direction is very different from the stellar disk, it could mean that the black hole's growth came from multiple chaotic accretion events, rather than one major episode.

The different relativistic effects that can be observed in the vicinity of a strong gravitational field, like the one surrounding Sgr A*, include:
\begin{list}{$\circ$}{}
  \item the Schwarzschild precession;
  \item the Shapiro time delay;
  \item the relativistic redshifts (transverse Doppler shift in special relativity as well as the gravitational redshift appearing in general relativity);
  \item the gravitational lensing effect; 
  \item the relativistic aberration. 
\end{list}
In the case of a rotating black hole, we need to add onto the previous effects the Kerr ones that arise from the rotation of the black hole: 
\begin{list}{$\circ$}{}
  \item the Lense-Thirring (LT) effect, which occurs because the rotation of the body (a spinning black hole for example) distorts spacetime, causing nearby inertial frames to be "dragged" along with the rotation (also known as the frame-dragging effect). It impacts both the star orbit and photon trajectory and varies depending on the norm and the direction of the spin relative to the star's orbit and affects both the astrometry and the spectroscopy;
  \item the quadrupole moment effect, which is due to the oblateness of the black hole arising from its rotation. Similarly to the Lense-Thirring effect, this one also impacts both the star orbit and photon trajectory, varies depending on the norm and the direction of the spin relative to the star's orbit and affects both the astrometry and the spectroscopy.  
\end{list}

From a precession point of view, the Schwarzschild precession is the most dominant relativistic effect, leading to a pericenter advance. If the black hole is rotating, then the Lense-Thirring and the quadrupole moment effects will generate both apsidal (in-plane) precession and nodal (out-of-plane) precession as opposed to the Schwarzschild case which only has an in-plane effect. The detection of the Schwarzschild precession was made possible with S2 and other stars \citep{GRAVITY+20_Schwarzschild_prec, GRAVITY+24}, but the higher order Kerr effects will prove to be more challenging, especially the quadrupole moment. By measuring the evolution of the orbital orientations of multiple closer-in stars \citep{Will2008}, it is possible to constrain the value and orientation of the spin, along with the magnitude of the quadrupole moment. However, considering how quickly these higher order effects fall with distance from Sgr A*, we would likely need to work with stars that have shorter periods and/or higher eccentricities if we want to test the no-hair theorem in a relatively short timescale \citep{Grould+17,Waisberg+18}. Therefore, we will consider putative stars multiple times closer to Sgr A*, and thus much more affected by Kerr effects. Such stars may exist if they are too faint to have been already detected by GRAVITY. 
Furthermore, GRAVITY+, the upgrade of GRAVITY and of the VLTI infrastructure, is now close to completion and will very soon have the potential to track fainter stars closer to the black hole than S2 \citep{GRAVITY+22_mass_distribution}. Thus, it might become possible to not only constrain the spin parameters, but also the quadrupole moment.

We know that the different effects that have the most potential to interfere with the detection the Lense-Thirring or quadrupole moment effects, that we do not consider in this work, include \citep{Alush+22}: the mass-precession \citep{MerrittEtAl2010}, the star's spin \citep{Dixon1970}, the vector resonant relaxation \citep{KocsisEtAl2015, MerrittEtAl2010}, and the tidal disruption \citep{PsaltisEtAl2013, FabryckyEtAl2007}.

This paper is organized as follows: we present in section \ref{sec2} an overview of the framework that is used in our study. The derivation of the relativistic orbital precessions in the post-Newtonian approximation is laid out in section \ref{sec4}. Then, in section \ref{sec5}, we use the OOGRE code to study how the spin and quadrupole moment affect the evolution of orbits. Finally, we summarize and discussing the applicability of our results to upcoming observations in section \ref{sec_summary}.


\section{Reference frames}\label{sec2}

Let us start this work by properly laying out the framework and present the four reference frames\footnote{All reference frames in this paper are right-handed and centered on Sgr A*.} that will be needed.

\subsection{Fundamental, orbital, and black hole frames}\label{}

Following the convention of \citet{PoissonWill2014} and \citet{Merritt2013}, we introduce a so-called fundamental frame $(\mbf X, \mbf Y, \mbf Z)$ that we use as a reference for describing the problem. The choice of this frame is arbitrary and will be made in subsection \ref{choice_fund_frame}. The problem we want to study is about a star in a bound orbit around a rotating black hole, meaning that we need to describe two vectors in the fundamental frame; the angular momentum of the orbit and the angular momentum of the black hole, which will naturally define two other frames, that of the orbit and that of the black hole.

\begin{figure}[htp]
    \centering
    \includegraphics[width=9cm]{ 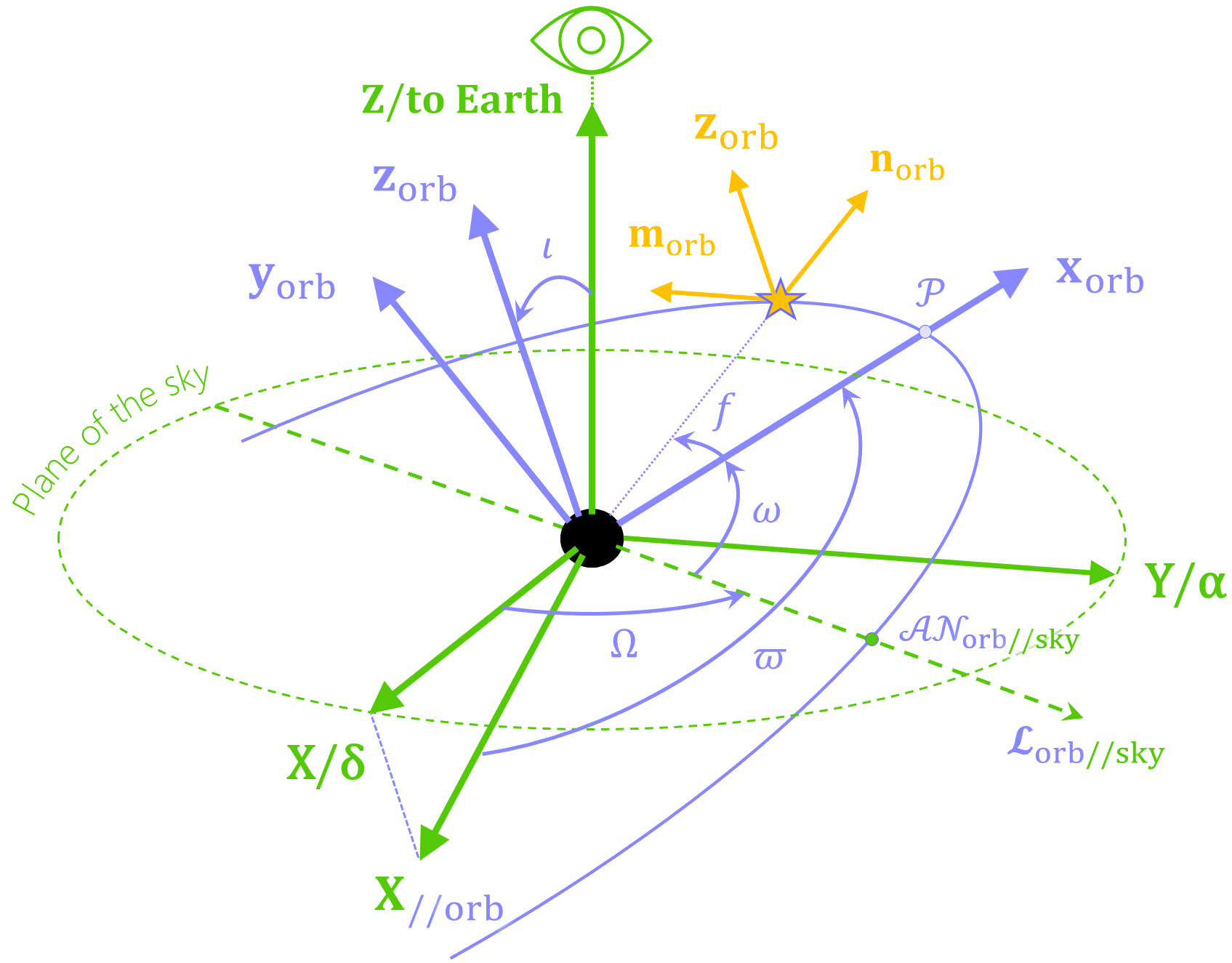}
    \caption{\label{frame_orb} The orbit frame $(\mbf x_{\mathrm{orb}} , \mbf y_{\mathrm{orb}} , \mbf z_{\mathrm{orb}} )$ and orbital elements are represented in purple; and the fundamental frame $(\mbf X, \mbf Y, \mbf Z)$ in green. The pericenter, represented by the point $\mathcal{P}$, is also added. Finally, $\mrm{\mbf{\mathcal{L}_\mbf{orb//sky}}}$ denotes the line of nodes of the orbit frame and sky plane. We also represent the Gaussian frame of the star $(\mbf n_{\mrm{orb}},\mbf m_{\mrm{orb}},\mbf z_{\mrm{orb}})$, the projection of the $\mbf X$ axis on the orbit plane, and the angle $\varpi$ which will be used in section \ref{sec5}. Here we illustrate the fundamental frame when it matches that of a distant observer (see subsection \ref{choice_fund_frame}). Also, it is useful to note the the azimuthal angle $\varphi$ is such that $\varphi=\omega+f$.}
\end{figure}

The orbit frame $(\mbf x_{\mathrm{orb}} , \mbf y_{\mathrm{orb}} , \mbf z_{\mathrm{orb}} )$ is illustrated by figure \ref{frame_orb}. The axis $\mbf z_{\mathrm{orb}}$ is along the angular momentum of the orbit, such that the plane $(\mbf x_{\mathrm{orb}} , \mbf y_{\mathrm{orb}})$ matches the instantaneous plane\footnote{The plane containing the black hole, the star and the three-velocity.} of the orbit. Also, $\mbf x_{\mathrm{orb}}$ always points to the instantaneous pericenter of the orbit (we remind that the pericenter of a relativistic orbit is not fixed, see the notion of osculating orbital parameters reminded in appendix \ref{sec3}). Further details about the line of nodes $\mrm{\mbf{\mathcal{L}_\mbf{orb//sky}}}$ and the ascending node of the orbit $\mathcal{AN}_{\mrm{orb//sky}}$ are provided in appendix \ref{app_frames}. The orbital parameters that are included in figure \ref{frame_orb} will be discussed further in appendix \ref{sec3.1}. For the time being, it is useful to note that the orientation of an orbit is characterized by three angles: the two angles $\iota$ and $\Omega$ define the orientation of the orbital plane relative to the fundamental frame, and the third angle $\omega$ defines the position of the pericenter inside the orbital plane. In addition, it is useful to introduce the Gaussian frame\footnote{Also denoted as $(\vec n,\vec m, \vec k)$ in Eq. (4.55) of \citet{Merritt2013}.} $(\mbf n_{\mrm{orb}},\mbf m_{\mrm{orb}},\mbf z_{\mrm{orb}})$, which follows the movement of the star as illustrated in figure \ref{frame_orb}; $\mbf n_{\mrm{orb}}$ is the radial vector and $\mbf m_{\mrm{orb}}$ completes the orthonormal basis.

\begin{figure}[htp]
    \centering
    \includegraphics[width=9cm]{ 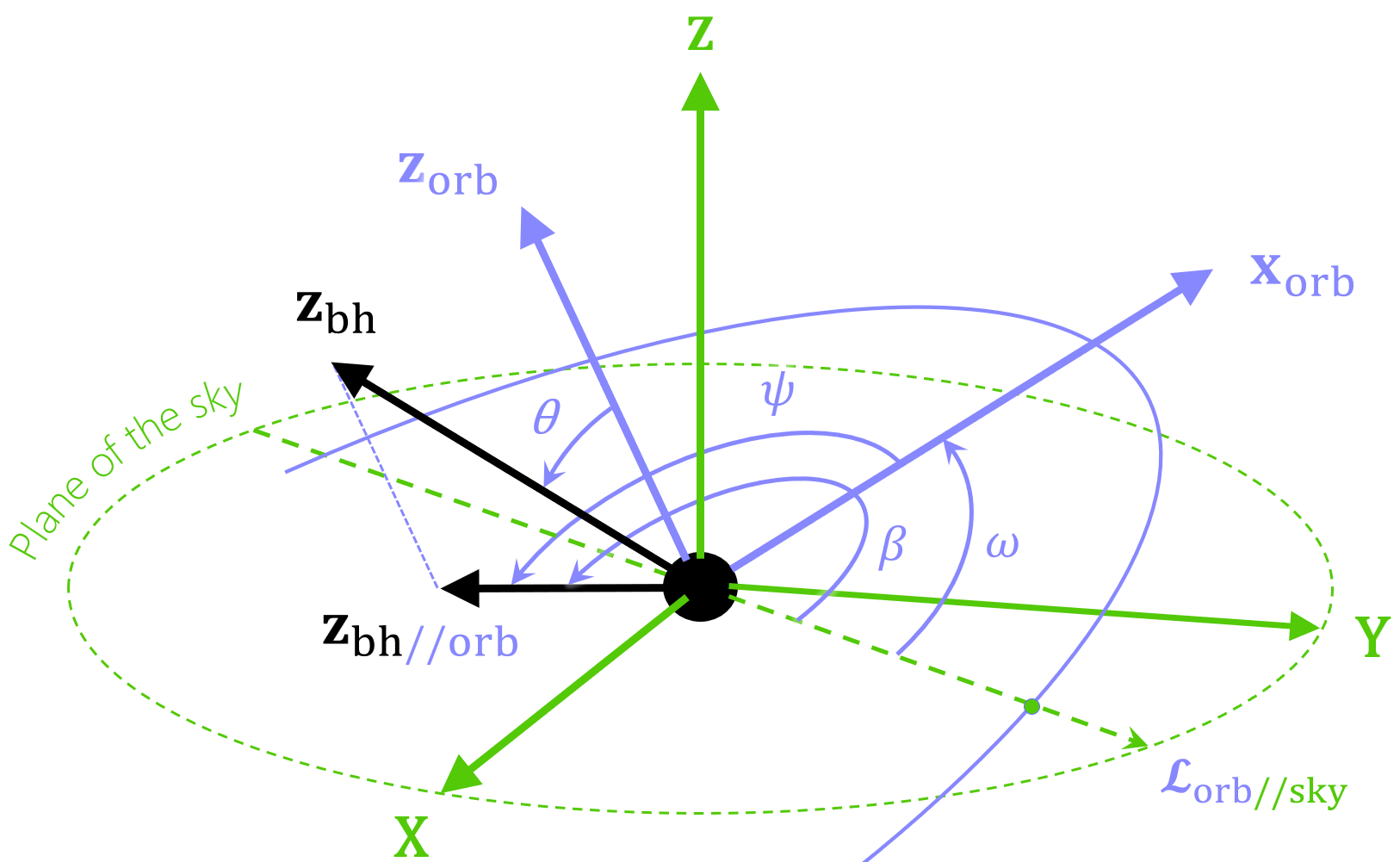}
    \caption{\label{frame_bh_orb} The angles $(\theta,\beta)$ in purple characterize the orientation of the spin vector relative to the orbit frame. We also add the angle $\psi\equiv\beta-\omega$, so that $(\theta,\psi)$ are the spherical coordinates of the spin axis in the orbit frame.}
\end{figure}

The black hole frame $(\mbf x_{\mathrm{bh}} , \mbf y_{\mathrm{bh}} , \mbf z_{\mathrm{bh}} )$, labeled in cartesian coordinates, is illustrated in figure \ref{frame_bh_orb}. It is defined such that the angular momentum, or spin vector of the black hole is along $\mbf z_{\mathrm{bh}}$. The orientation of $\mbf x_{\mathrm{bh}}$ and $\mbf y_{\mathrm{bh}}$ is arbitrary. We introduce the angles $(\theta,\beta)$ that allow to orient $\mbf z_{\mathrm{bh}}$ in the orbit frame as defined in figure \ref{frame_bh_orb}:
\begin{list}{$\circ$}{}  
\item $\theta \in [0;\pi]$ is the rotation angle from $\mbf z_{\mathrm{orb}}$ to $\mbf z_{\mathrm{bh}}$;
\item $\beta \in [0;2\pi]$ is the rotation angle about $\mbf z_{\mathrm{orb}}$, from $\mrm{\mbf{\mathcal{L}_\mbf{orb//sky}}}$ to $\mbf z_{\mathrm{bh//orb}}$.
\end{list}
We also introduce an alternative definition of angles $(i',\Omega')$ in appendix \ref{app_frames} that are useful for performing multi-star fits. These angles will not be considered further in this article.

\subsection{Our choice of the fundamental frame: observer's frame}\label{choice_fund_frame}

At this point, we need to make a choice of a fundamental frame. Considering that we eventually want to model the observable in the sky plane, it is natural to choose the observer's frame as the fundamental one. As illustrated in figures \ref{frame_orb}, to do so, we take $\mbf Z$ as pointing towards the observer and $(\mbf X, \mbf Y) = (\vec \delta, \vec \alpha) = (\mbf{DEC}, \mbf{RA})$ corresponding to the observer's screen with $\vec \alpha$ or RA, and $\vec \delta$ or DEC referring to the right ascension and declination, respectively. In other words, we define the fundamental plane as the sky plane and center it on the apparent position of Sgr A*.
This choice has a crucial impact on the definition of orbital parameters considering that the angular parameters are defined with respect to the fundamental frame. In literature, we find different definitions of the observer's frame and fundamental frame. This can often lead to different results and thus cause apparent contradictions, as will be mentioned in the following sections and appendices.

\section{Post-Newtonian expressions of the secular shifts}\label{sec4}

\subsection{Post-Newtonian acceleration}\label{sec4.1}

Here we want to focus on the post-Newtonian (PN) formalism which allows us to separate the Lense-Thirring effect from the quadrupole moment effect. For the post-Newtonian formalism, we will be using harmonic coordinates which are commonly used for PN works \citep{PoissonWill2014}. In the PN approximation, the equation of motion of a star of negligible mass $m_{\mrm star}$ compared to the mass of the black hole $M_{\bullet}$ writes \citep{PoissonWill2014}: 
\begin{align}\label{acc}
    \ddot{\mbf r} = -\frac{Gm}{r^2}\mbf n_{\mrm{orb}} + \mbf a_{\mrm{PN}},
\end{align}
where $m = M_{\bullet}+m_{\mrm star}\approx$\footnote{This approximation will be referred to as \ding{182} and can be justified by the fact that the small PN parameter $\epsilon=\frac{GM}{c^2p}$ is at least one or two orders of magnitude greater than $\eta=\frac{m_{\mrm star}}{M_{\bullet}}\sim10^{-6}-10^{-5}$ for any hypothetical star relevant for our study ($m_{\mrm star}\sim1-10M_{\odot}$ and with $p \leq  p^\mrm{S2})$.} $ \;M_{\bullet}$, $\mbf r$ is the separation vector from the black hole to the star, and $\mbf n_{\mrm{orb}}=\mbf r/r$ with:
\begin{align}\label{r}
    r=|\mbf r|=\frac{p}{1+e\cos f},
\end{align}
where $p=a_{\mrm{sma}}(1-e^2)$ is the semi-latus rectum\footnote{Sometimes $p$ and $e$ are alternatively defined as constants even in GR using the pericenter and apocenter radii $r_-$ and $r_+$ by: $p=r_-(1+e)=r_+(1-e)$. However, this definition will yield a different definition of the pericenter advance compared to $p=a_{\mrm{sma}}(1-e^2)$,  \citep[see][]{Tucker2019}.}.
A dot in \eqref{acc} denotes derivation with respect to the harmonic-coordinate time $t$ and the PN acceleration $\mbf a_{\mrm{PN}}$ accounts for deviations from Keplerian motion due to GR. Thanks to the use of the osculating orbital elements method, we are able to use the Newtonian relation of Eq. \eqref{r} at each date, while keeping in mind that the involved parameters are time-varying, and that we use the weak-field and small-velocity approximation when working with the PN formalism.\\


If one wants to understand, at leading order, how the Schwarzschild precession and the spin-related effects impact the orbit of S-stars, it is sufficient to look at the dominant order of the analytical expressions of the secular shifts of orbital elements \citep{Will2008}. In practice, however, since the PN formalism appears to be the most suitable for the process of fitting the data of S stars\footnote{Because it allows to disentangle the Lense-Thirring from the quadrupole moment effects.}, one will have to choose one specific PN order for all effects to be consistent in the fitting process. As will be seen in this section, the dominant order of the Lense-Thirring and quadrupole moment effects are at the 1.5PN (corresponding to $(v/c)^3$) and 2PN orders (corresponding to $(v/c)^4$), respectively. Therefore, we need to obtain the analytical expressions up to the 2PN order. 

Considering the second-order PN approximation (2PN) corresponding to $(v/c)^4$, the PN acceleration can be decomposed into 3 parts:
\begin{align}\label{}
    \mbf a_{\mrm{PN}} \approx\footnotemark \;
    \mbf a_{\mrm{2PN}} = \mbf a_{\mrm{Sch}}^{\mrm{2PN}} + \mbf a_\mrm{LT}^{\mrm{1.5PN}} + \mbf a_{Q}^{\mrm{2PN}},
\end{align} \footnotetext{This approximation will be referred to as \ding{183}.}
with $\mbf a_{\mrm{Sch}}^{\mrm{2PN}}$, $\mbf a_\mrm{LT}^{\mrm{1.5PN}}$, $\mbf a_{Q}^{\mrm{2PN}}$ corresponding to the 2PN acceleration of the star due to the Schwarzschild precession, the spin and the quadrupole moment, respectively. The upper index reminds the PN order at which the various effects enter the equation of motion. By considering approximation \ding{182}, and by defining the small PN parameter as:
\be\epsilon=\frac{Gm}{c^2p},\ee
the accelerations can be expressed in terms of $\epsilon$ as \citep{Merritt2013,Will2008,Blanchet2003}:
\begin{align}
    \begin{split}
    &\mbf a_{\mrm{Sch}}^{\mrm{2PN}} = \frac{\epsilon}{p}(1+e\cos f)^2\left(\left(4\frac{Gm}{r } -v^2\right)\mbf n_{\mrm{orb}} +4v_r\vec v \right)\\   
    & \quad\quad\;\; + \frac{\epsilon^2}{p}(1+e\cos f)^3\left( \left(2v_r^2 -9\frac{Gm}{r }\right)\mbf n_{\mrm{orb}}-2v_r\vec v \right),
    \end{split}
\end{align}
\begin{align}
    \begin{split}
    &\mbf a_\mrm{LT}^{\mrm{1.5PN}} = -2\epsilon^{3/2}\sqrt{\frac{G m}{p^3}}(1+e\cos f)^3\chi\\
    &\quad\quad\;\times\left[2\vec v \times \mbf z_{\mrm{bh}}- 3\left(\mbf n_{\mrm{orb}}\cdot\vec v\right)\mbf n_{\mrm{orb}}\times\mbf z_{\mrm{bh}}-3\mbf n_{\mrm{orb}}\left(\mbf n_{\mrm{orb}}\times\vec v\right)\cdot\mbf z_{\mrm{bh}}\right],\label{a_chi_i}
    \end{split}
\end{align}
\begin{align}
    \begin{split}
    & \mbf a_Q^{\mrm{2PN}} = \frac{3}{2}\epsilon^2\frac{Gm}{p^2}(1+e\cos f)^4\chi^2\\
    &\quad\quad\times\Big(5 \mathbf n_{\mrm{orb}} \big( \mathbf n_{\mrm{orb}}\cdot\mbf z_{\mrm{bh}}\big)^2 -2\big( \mathbf n_{\mrm{orb}} \cdot\mbf z_{\mrm{bh}}\big)\mbf z_{\mrm{bh}} - \mathbf n_{\mrm{orb}} \Big) ,
    \end{split}\label{a_Q}
\end{align}
with $v^2 = v_r^2+v_t^2$ and $\chi$ being the spin parameter. The latter is positive and has a maximal value of 1 (for a Kerr black hole). The dimensionless spin vector $\vec\chi=\chi\mbf z_{\mathrm{bh}}$ is linked to the angular momentum $\vec J=J\mbf z_{\mathrm{bh}}$ of a black hole of mass $M_{\bullet}$ by: 
\begin{equation}\label{chi} 
    \vec\chi=\vec J\frac{ c}{GM_{\bullet}^2}.
\end{equation}
As for the quadrupole moment parameter $Q$, we note that it enters Eq. \eqref{a_Q} through:
\begin{equation}\label{Q} 
    Q=-\frac{1}{c^2}\frac{J^2}{M_{\bullet}}=-\frac{G^2M_{\bullet}^3}{c^4}\chi^2,
\end{equation}
given by the no-hair theorem \citep{Poisson1998,Krishnendu2019}. 

We note that the Schwarzschild acceleration is composed of two terms, respectively of 1PN and 2PN orders. The Lense-Thirring and quadrupole moment accelerations are composed of only one term, of order 1.5PN and 2PN, respectively. The leading order of the various effects is thus 1PN, 1.5PN, 2PN, for Schwarzschild, Lense-Thirring, and quadrupole moment effects, respectively. We need to consider the subleading order only for the Schwarzschild term. Lense-Thirring and quadrupole moment subleading terms appear at order above 2PN. In this paper, we consider only the Kerr spacetime, and thus plug the expression of Q given by Eq. \eqref{Q} into our equations. However, if we want to test the no-hair theorem, we should not assume Eq. \eqref{Q} when generating the expression of the secular shifts\footnote{One can thus replace $\chi$ by its expression as a function of $Q$ in the analytical expressions of subsection \ref{sec_evo_orb} to retrieve he analytical expressions as a function of quadrupole moment.}, since $Q$ would need to be measured independently from $\chi$.

\subsection{Lagrange planetary equations}\label{sec4.2}

We can use the orbital perturbation theory \citep{Merritt2013,Will2008} to write the derivative of each, time-varying, osculating orbital parameter with respect to time, as a function of the various $\mathcal{R}$, $\mathcal{S}$ and $\mathcal{W}$ terms, the perturbative accelerations expressed in the Gaussian frame, introduced in appendix \ref{RSW} \citep[see][for details]{PoissonWill2014}. They are called the Lagrange planetary equations\footnote{If one want to study circular orbits where $e=0$, it is advised to use an alternative set of variables to avoid the null denominator in Eq. \eqref{dwdt}. This is done in \citet{WillMaitra2016}. Here we consider realistic orbits where $e\neq0$.} and are expressed as:
\begin{align}
    &\frac{\mrm{d}p}{\mrm{d}t} = 2\sqrt{\frac{p^3}{Gm}} \frac{1}{1+e\cos f} \mathcal{S}; \label{dpdt}\\
    &\frac{\mrm{d}e}{\mrm{d}t} = \sqrt{\frac{p}{Gm}} \left [ \sin f \mathcal{R} +\frac{2\cos f+e(1+\cos^2f)}{1+e\cos f} \mathcal{S} \right ]; \label{dedt}\\
    &\frac{\mrm{d}\iota}{\mrm{d}t} = \sqrt{\frac{p}{Gm}} \frac{\cos(\omega+f)}{1+e\cos f} \mathcal{W}; \label{didt}\\
    &\sin\iota\frac{\mrm{d}\Omega}{\mrm{d}t} = \sqrt{\frac{p}{Gm}} \frac{\sin(\omega+f)}{1+e\cos f} \mathcal{W}; \label{dOmdt}\\
    \begin{split}
    &\frac{\mrm{d}\omega}{\mrm{d}t} = \frac{1}{e}\sqrt{\frac{p}{Gm}} \left [ -\cos f \mathcal{R} +\frac{2+e\cos f}{1+e\cos f}\sin f \mathcal{S}\right.\\
    &\;\;\quad\quad \quad\quad \quad\quad \left.  -e\cot\iota\frac{\sin (\omega+f)}{1+e\cos f} \mathcal{W}\right ]. \label{dwdt}
    \end{split}
\end{align}
No approximations are done to obtain Eqs. \eqref{dpdt} to \eqref{dwdt} from Eqs. \eqref{acc}.

As stated earlier, in the context of a small perturbing force, it is possible to get a good estimation of the orbital dynamics by only looking at the dominant contribution of each effect. To do so, we solve the equations within the framework of perturbation theory and find that, to express the leading order of such an effect for a given orbital parameter, it is appropriate to take the 0PN terms of all other orbital parameters in the right-hand side of Eqs. \eqref{dpdt} to \eqref{dfdt}, before integrating these equations over $t$. Even more convenient would be to use $f$ as an independent variable instead of $t$, considering it has the same fixed integration interval over one period for any star.  It evolves as \citep[see][for details]{PoissonWill2014}:
\begin{align}\label{dfdt}
    \begin{split}
    \frac{\mrm{d}f}{\mrm{d}t} &= \left(\frac{\mrm{d}f}{\mrm{d}t}\right)_{\mrm{Kepler}}-\left(\frac{\mrm{d}\omega}{\mrm{d}t} +\cos\iota \frac{\mrm{d}\Omega}{\mrm{d}t}\right)\\
    &= \sqrt{\frac{Gm}{p^3}}(1+e\cos f)^2 \\
    &\quad+ \frac{1}{e} \sqrt{\frac{p}{Gm}}\left [\cos f \mathcal{R} -\frac{2+e\cos f}{1+e\cos f}\sin f \mathcal{S}\right ],
    \end{split}
\end{align}
which reminds that $f$ is the angle between the \textit{varying} pericenter and the position vector of the star $\mbf r$ (see figure \ref{frame_orb}).
We make use of the fact that the term $\left(\frac{\mrm{d}f}{\mrm{d}t}\right)_{\mrm{Kepler}}=\sqrt{\frac{Gm}{p^3}}(1+e\cos f)^2=\left(\frac{\mrm{d}\varphi}{\mrm{d}t}\right)_{\mrm{Kepler}}$ is Keplerian (0PN) and that the non-Keplerian terms on the right-hand side of Eq. \eqref{dfdt} start at 1PN, 1.5PN and 2PN orders for the Schwarzschild, Lense-Thirring and quadrupole moment effects respectively. This means that we are allowed to express the temporal variation as:
\begin{align}\label{dtdf_0}
    \begin{split}
    \frac{\mrm{d}t}{\mrm{d}f} &= \sqrt{\frac{p^3}{Gm}}\frac{1}{(1+e\cos f)^2}\left[1+\mathcal{O}\left(\epsilon\right)\right],
    \end{split}
\end{align}
before multiplying Eqs. \eqref{dpdt} to \eqref{dwdt} by the zeroth-order term of Eq. \eqref{dtdf_0} to get the dominant order derivative of each orbital parameter with respect to the true anomaly $f$.

\subsection{Analytical integration of the planetary equations}\label{AnaliticalWill2008}

Now that we have the expression of the planetary equation as a function of $f$, we integrate them with $f$ from $f_0$ to $f_0+2\pi$\footnote{This is one way of defining the pericenter advance. Another way to derive the pericenter advance is by directly integrating the timelike geodesic equation which, by means of the conservation of energy and angular momentum, leads to an ordinary differential equation for the radius r in terms of the angle $\phi$ of the Boyer-Lindquist coordinates \citep{BH_eric, Tucker2019}. The angle between successive turning points, or extrema of $r$, can be obtained exactly from a radial integral. The expansion of that result in a post-Newtonian sequence agrees with the osculating element method at lowest order and the differences at higher orders are illusory, because the semilatus rectum and eccentricity have different meanings in the two approaches \citep{Tucker2019}. However, note that the integration of $f$ from $f_0$ to $f_0+2\pi$ is not valid in the case of a "zoom-whirl" behavior \citep{BH_eric}.} to find the dominant terms of the secular evolution of the orbital parameters. By considering the keplerian dominant constant term of $p$, $e$, $\omega$ and $\beta$ in the right hand side of Langrange planetary equations, we find that $p$ and $e$ (and consequently $a_{\mrm{sma}}$ and $P$) do not show any secular variations after one orbit at the dominant orders. As for the other parameters, we find that the precessions per orbit of a star’s argument of periapsis $\Delta\omega$, longitude of the ascending node $\Delta\Omega$, and inclination $\Delta \iota$ verify\footnote{With the different contribution of the Schwarzschild, Lense-Thirring  and quadrupole moment effects denoted with the "Sch", "LT" and "$Q$" subscripts.}:

\begin{align}
    &\Delta\omega_{\mrm{Sch}}|^{1\mrm{PN}} = 6\pi\epsilon, \\
    &\Delta\omega_\mrm{LT}|^{1.5\mrm{PN}} = -4\pi \epsilon^{3/2} (2\cos{\theta}+\cot \iota\sin\theta\sin\beta)\;\chi, \label{Dom_chi} \\
    &\Delta\omega_{Q}|^{2\mrm{PN}} \!\!= -\frac{3\pi}{2} \epsilon^{2}  \left(1\!-\!3\cos^2{\theta}-2\cot \iota\cos\theta\sin\theta\sin\beta \right)\chi^2, \label{Dom_q}
\end{align}
\begin{align}
    &\Delta \iota_{\mrm{Sch}}|^{1\mrm{PN}} = 0,\\
    &\Delta \iota_\mrm{LT}|^{1.5\mrm{PN}} = 4\pi \epsilon^{3/2} \sin\theta\cos\beta\;\;\chi, \label{Di_chi}\\
    &\Delta \iota_{Q}|^{2\mrm{PN}} = -3\pi \epsilon^{2} \cos\theta \sin\theta\cos\beta\;\; \chi^2, \label{Di_q}
\end{align}
\begin{align}
    &\sin \iota\Delta\Omega_{\mrm{Sch}}|^{1\mrm{PN}} = 0,\\
    &\sin \iota\Delta\Omega_\mrm{LT}|^{1.5\mrm{PN}} =4\pi \epsilon^{3/2}  \sin\theta\sin\beta\;\;\chi, \label{DOm_chi}\\
    &\sin \iota\Delta\Omega_{Q}|^{2\mrm{PN}} = -3\pi \epsilon^{2}  \cos\theta\sin\theta\sin\beta\;\; \chi^2 ,\label{DOm_q}
\end{align}
We note that \citet{Will2008} shares the same definition of fundamental frame as in our work, which is why we agree on the variation of the angular orbital parameters. However, our Eqs. \eqref{Di_chi} and \eqref{Di_q} should not be compared with Eqs. (3.23a) of \citet{WillMaitra2016} and (5) of \citet{Tucker2019} since we make a different choice of fundamental frame from theirs, leading to different definitions of $\iota$ and $\Omega$ among other parameters.

These secular trends are valid at 2PN order for the Lense-Thirring and quadrupole moment shifts, for which subleading contributions appear at orders higher than 2PN. However, for the Schwarzschild terms, we need to compute the subleading 2PN contribution in addition to the leading 1PN contribution provided above. To do so, the two-timescale method detailed in appendix \ref{app2timescale} and \citet{WillMaitra2016, Tucker2019} is necessary to include the contribution of the periodic evolution of the orbital parameters. This allows us to obtain the following:
\begin{align}
    &\Delta\omega_{\mrm{Sch}}|^{2\mrm{PN}} = 6\pi\epsilon -\frac{3\pi}{2} \epsilon^2\left(10-e^2\right), \label{w_2PN_sch}\\
    &\Delta \iota_{\mrm{Sch}}|^{2\mrm{PN}} = 0,\\
    &\Delta\Omega_{\mrm{Sch}}|^{2\mrm{PN}} = 0,
\end{align}
which are more precise analytical expressions than the ones given in \citet{Alush+22}. 
In the 1PN term of Eq. \eqref{w_2PN_sch}, $\epsilon=Gm/(c^2\tilde{p})$ with $\tilde{p}$ being the orbit-averaged value of $p$, as detailed in appendix \ref{app2timescale_sch}.
Notice that if $\iota= 0$, $\mrm{\mbf{\mathcal{L}_\mbf{orb//sky}}}$ cannot be defined as it is degenerate, then $\omega$ and $\Omega$, and thus Eqs. \eqref{Dom_chi}, \eqref{Dom_q}, \eqref{DOm_chi} and \eqref{DOm_q} will not be defined either. This issue will be addressed by the introduction of new variables in subsection \ref{sec5}.


\subsection{Numerical integration of the planetary equations}\label{sec_oogre}

In order to numerically integrate the planetary equations to simulate orbits, we use our Python-based code called "Osculating Orbits in General Relativistic Environments" (OOGRE), which was originally developed by \citet{Heissel+22}. It is a perturbed Kepler-orbit model code that is based on the formalism presented in section \ref{sec4.1} and optimized for stars in the Galactic Center. 

The purpose of section \ref{sec5} is to understand the impact of Kerr effects on the geometry of the orbit; the interest will lie in the dynamical interaction between the black hole and one star. Therefore, we do not consider the effect of mass-precession and vector resonant relaxation, and also neglect the tidal disruption and effect of the star's spin. Moreover, as the goal is not to simulate observations here, we turn off the following effects considering that they would get in the way:
\begin{itemize}
  \item the Shapiro time delay, 
  \item the R\o mer effect, 
  \item the gravitational lensing\footnote{Two flavors of gravitational lensing are implemented in OOGRE: a ray-tracing based one and an analytical approximation. Both are turned off in this section.}.
\end{itemize}
Also, the relativistic aberration is not implemented in OOGRE and will not be needed for the following section either. However, we extend this code, which only considered $\mbf a_{\mrm{Sch}}^{\mrm{1PN}}$, to the Kerr metric by adding the spin and quadrupole moment contributions $\mbf a_\mrm{LT}^{\mrm{1.5PN}}$ and $\mbf a_{Q}^{\mrm{2PN}}$ in order to be able to simulate these effects. We also push the Schwarzschild precession to the 2PN order with $\mbf a_{\mrm{Sch}}^{\mrm{2PN}}$ for consistency.

We choose a set of initial conditions as in \citet{Heissel+22}, and fix the osculating time as $t_{\mathrm{osc}}=t_{\mathrm{peri}}-P_\mrm{osc}/2$, an approximate\footnote{It is not the exact value of the time of apocenter passage because here $P_\mrm{osc}= 2\pi \sqrt{a_\mrm{sma, osc}^3/(G M)}$ is the osculating (Keplerian) constant and not the relativistic variable.} value of the apocenter, granting the code more stability. The code integrates Eqs. \eqref{dpdt} to \eqref{dwdt} at the core of OOGRE while only using the approximations \ding{182} and \ding{183}. 
\section{Spin impact on the orientation of the orbit}\label{sec5}


First, we want to understand analytically the effect of spin on the evolution of the orientation of the orbit. This will lead us to introduce several precession movements, either instantaneous or secular. The secular evolution is well known \citep[see e.g][]{KocsisEtAl2011}, it consists of a precession movement of $\mbf z_{\mathrm{orb}}$ about $\mbf z_{\mathrm{bh}}$. Instantaneous evolution is both less known and more useful in the perspective of orbital fits. We will therefore focus on the latter. Second, we will illustrate this analytical understanding by performing simulations for a hypothetical star that we call "S2/10"; a star with orbital parameters identical to those of S2 but with a semi-major axis 10 times smaller.

\subsection{Instantaneous orientation change}\label{inst_formalism}

\subsubsection{Orbital frame orientation change}\label{inst_evo_orb}


Let us characterize the evolution of the orientation of the orbit by expressing the temporal evolution of the orbit frame. Using the Lagrange planetary equations, it is possible to reexpress the temporal variations of these vectors as functions of the temporal variations of the orbital elements. We get \citep[see][]{PoissonWill2014}:
\begin{align}
\begin{split}    
\frac{\mrm{d} \mbf x_{\mrm{orb}}}{\mrm{d} t} &= \frac{\mrm{d} \varpi}{\mrm{d} t} \mbf y_{\mrm{orb}}  + \frac{\mrm{d} \Theta}{\mrm{d} t} \, \mbf z_{\mrm{orb}} \\
&= \frac{\mrm{d} \varpi}{\mrm{d} t} \mbf z_{\mrm{orb}}\times\mbf x_{\mrm{orb}}  - \frac{\mrm{d} \Theta}{\mrm{d} t} \mbf y_{\mrm{orb}}\times\mbf x_{\mrm{orb}},
\label{dx_orb/dt}
\end{split}\\
\begin{split}
\frac{\mrm{d} \mbf y_{\mrm{orb}}}{\mrm{d} t} &=   -  \frac{\mrm{d} \varpi}{\mrm{d} t} \mbf x_{\mrm{orb}} + \frac{\mrm{d} \Xi}{\mrm{d} t} \mbf z_{\mrm{orb}}\\
&= -  \frac{\mrm{d} \varpi}{\mrm{d} t} \mbf y_{\mrm{orb}}\times\mbf z_{\mrm{orb}} +\frac{\mrm{d} \Xi}{\mrm{d} t} \mbf x_{\mrm{orb}}\times\mbf y_{\mrm{orb}},
\label{dy_orb/dt}    
\end{split}\\
\begin{split}    
\frac{\mrm{d} \mbf z_{\mrm{orb}}}{\mrm{d} t} &= - s_\mathcal{W}\sqrt{\left( \frac{\mrm{d} \iota}{\mrm{d} t}\right)^2 + \sin^2 \iota \left( \frac{\mrm{d} \Omega}{\mrm{d} t}\right)^2} \, \mbf m_{\mathrm{orb}}\\
&= - \frac{\mrm{d} \Theta}{\mrm{d} t} \mbf x_{\mrm{orb}} +\frac{\mrm{d} \Xi}{\mrm{d} t} \mbf y_{\mrm{orb}}\\
&= - \frac{\mrm{d} \Theta}{\mrm{d} t} \mbf y_{\mrm{orb}}\times\mbf z_{\mrm{orb}} -\frac{\mrm{d} \Xi}{\mrm{d} t} \mbf x_{\mrm{orb}}\times\mbf z_{\mrm{orb}},
\end{split}\label{dz_orb/dt}
\end{align}
where $s_\mathcal{W}$ is the sign of $\mathcal{W}$ and:
\begin{align}
    &\frac{\mrm{d} \varpi}{\mrm{d} t} \equiv \frac{\mrm{d} \omega}{\mrm{d} t} + \cos \iota \, \frac{\mrm{d} \Omega}{\mrm{d} t};\label{dvarpi/dt}\\
    &\frac{\mrm{d} \Theta}{\mrm{d} t} \equiv \sin \omega \frac{\mrm{d} \iota}{\mrm{d} t} - \sin \iota \cos \omega \frac{\mrm{d} \Omega}{\mrm{d} t};\label{dTheta/dt}\\
    &\frac{\mrm{d} \Xi}{\mrm{d} t} \equiv -\cos \omega \frac{\mrm{d} \iota}{\mrm{d} t} - \sin\iota\sin \omega \frac{\mrm{d} \Omega}{\mrm{d} t}.\label{dXi/dt}
\end{align} 
We add the last lines of Eqs. \eqref{dx_orb/dt} to \eqref{dz_orb/dt} to reveal the precession aspect of the equations and their different components.

We obtain:
\be
\frac{\mrm{d} \mbf x_{\mrm{orb}}}{\mrm{d} t} \cdot \mbf y_{\mrm{orb}} = -\frac{\mrm{d} \mbf y_{\mrm{orb}}}{\mrm{d} t} \cdot \mbf x_{\mrm{orb}} =\frac{\mrm{d} \varpi}{\mrm{d} t}, 
\ee
encoding the change of direction of the major and minor axes in the orbit plane;
\be
\begin{split}
\frac{\mrm{d} \mbf x_{\mrm{orb}}}{\mrm{d} t} \cdot \mbf z_{\mrm{orb}} & = -\frac{\mrm{d} \mbf z_{\mrm{orb}}}{\mrm{d} t} \cdot \mbf x_{\mrm{orb}} =\frac{\mrm{d} \Theta}{\mrm{d} t},
\end{split}
\ee
encoding the change of the major axis orthogonal to the orbit plane, or equivalently the change of the orbital angular momentum direction along the major axis;
\begin{align}\label{I2}
    \frac{\mrm{d} \mbf z_{\mrm{orb}}}{\mrm{d} t} \cdot \mbf y_{\mrm{orb}} & =-\frac{\mrm{d} \mbf y_{\mrm{orb}}}{\mrm{d} t} \cdot \mbf z_{\mrm{orb}}= \frac{\mrm{d} \Xi}{\mrm{d} t} ,
\end{align}
encoding the change of the minor-axis direction orthogonal to the orbit plane, or equivalently the change of the orbital angular momentum along the minor axis. We note that $\mbf x_{\mrm{orb}}\cdot\frac{\mrm{d} \mbf x_{\mrm{orb}}}{\mrm{d} t}=\mbf y_{\mrm{orb}}\cdot\frac{\mrm{d} \mbf y_{\mrm{orb}}}{\mrm{d} t}=\mbf z_{\mrm{orb}}\cdot\frac{\mrm{d} \mbf z_{\mrm{orb}}}{\mrm{d} t}=0$, as it should for unit vectors. 
Therefore, we can decompose the overall precession into two components: the precession of the argument of periapsis within the changing orbital plane, with associated characteristic frequency $\frac{\mrm{d} \varpi}{\mrm{d} t}$, hereafter referred to as "in-plane\footnote{In the case where we have out-of-plane precession, the notion of in-plane becomes more delicate considering that the orbital plane changes all the time. In the following, we use this formulation to refer to instantaneous in-plane precession.} precession"; and the precessions of the orbit plane, with characteristic associated frequencies $\frac{\mrm{d} \Theta}{\mrm{d} t}$ and $\frac{\mrm{d} \Xi}{\mrm{d} t}$, hereafter referred to as "out-of-plane precession". 

These scalar quantities are also presented\footnote{In the second line of equation "(3)" of \citet{Will+23} the unit vector "$\mbf e_X$" should read "$\mbf e_Y$".} in \citet{Will+23}, and are independent from the observer's location, which is obvious from their definition: they depend only on $(\mbf x_{\mrm{orb}},\mbf y_{\mrm{orb}},\mbf z_{\mrm{orb}})$ which depends only on the orbit. As such, $\varpi$, $\Theta$ and $\Xi$ appear to be very useful for expressing observer-independent statements on the variation of the pericenter and angular momentum directions.

\subsubsection{Secular orientation change}\label{sec_evo_orb}

Let us introduce the secular shifts $\Delta\varpi$, $\Delta\Theta$ and $\Delta\Xi$, as the integration over an orbit of the temporal variations of $\varpi$, $\Theta$ and $\Xi$. The secular in-plane shift of the pericenter can be expressed at 2PN order as: 
\begin{align}
    &\Delta\varpi_{\mrm{Sch}}|^{2\mrm{PN}} = \Delta\omega_{\mrm{Sch}}|^{2\mrm{PN}} = 6\pi\epsilon -\frac{3\pi}{2} \epsilon^2\left(10-e^2\right); \label{Dvarpi_sch}\\
    &\Delta\varpi_\mrm{LT}|^{2\mrm{PN}} = -8\pi \epsilon^{3/2} \cos{\theta}\;\chi; \label{Dvarpi_chi}\\
    &\Delta\varpi_{Q}|^{2\mrm{PN}} = -\frac{3\pi}{2} \epsilon^{2} \left(1-3\cos^2{\theta} \right)\;\chi^2.  \label{Dvarpi_Q}
\end{align}

Notice that Eqs. \eqref{Dvarpi_chi} and \eqref{Dvarpi_Q} will still be valid if $\iota=0$, as opposed to Eqs. \eqref{Dom_chi}, \eqref{Dom_q}, \eqref{DOm_chi} and \eqref{DOm_q}. Using $\varpi$ allows us to quantify the precession inside the orbital plane without using the notion of $\mrm{\mbf{\mathcal{L}_\mbf{orb//sky}}}$ which is degenerate in the case of a face-on\footnote{The evolution of the orbital angular momentum direction makes the notion of face-on or edge-on view of the orbit tricky when $\theta \neq 0$, therefore, when talking about a face-on or edge-on view of the orbit in this article, we will be referring to the initial view of the orbit.} orbit. From Eqs. \eqref{Dvarpi_chi} and \eqref{Dvarpi_Q} we see that the Lense-Thirring and quadrupole moment in-plane precession depend on the polar angle $\theta$ but not on the azimuthal angle $\psi$ of the spin vector spherical coordinates defined in figure \ref{frame_bh_orb}. We see that the in-plane precession of the Lense-Thirring effect is maximal for orbits in the equatorial plane of the black hole, i.e for $\theta=0[\pi]$, and vanishes for polar\footnote{Meaning that the orbit plane contains the spin axis of the black hole or equivalently that $\mbf z_{\mathrm{orb}}$ is perpendicular to $\mbf z_{\mathrm{bh}}$. Considering that $\theta$ has a periodic variation due to the Lense-Thirring effect in this case, but not a secular one (see subsection \ref{sec_time_avg}), a polar orbit will only remain polar on a secular scale.} orbits, i.e for $\theta=\pi/2[\pi]$. The quadrupole moment in-plane precession, however, is maximized for both equatorial and polar orbits, and vanishes for $\theta = \arccos\left(\pm1/\sqrt{3}\right)$.

\begin{figure}[htp]
    \centering
    \subfloat[Orbital timescale orientation change]
    {\includegraphics[clip,width=1\columnwidth]{ 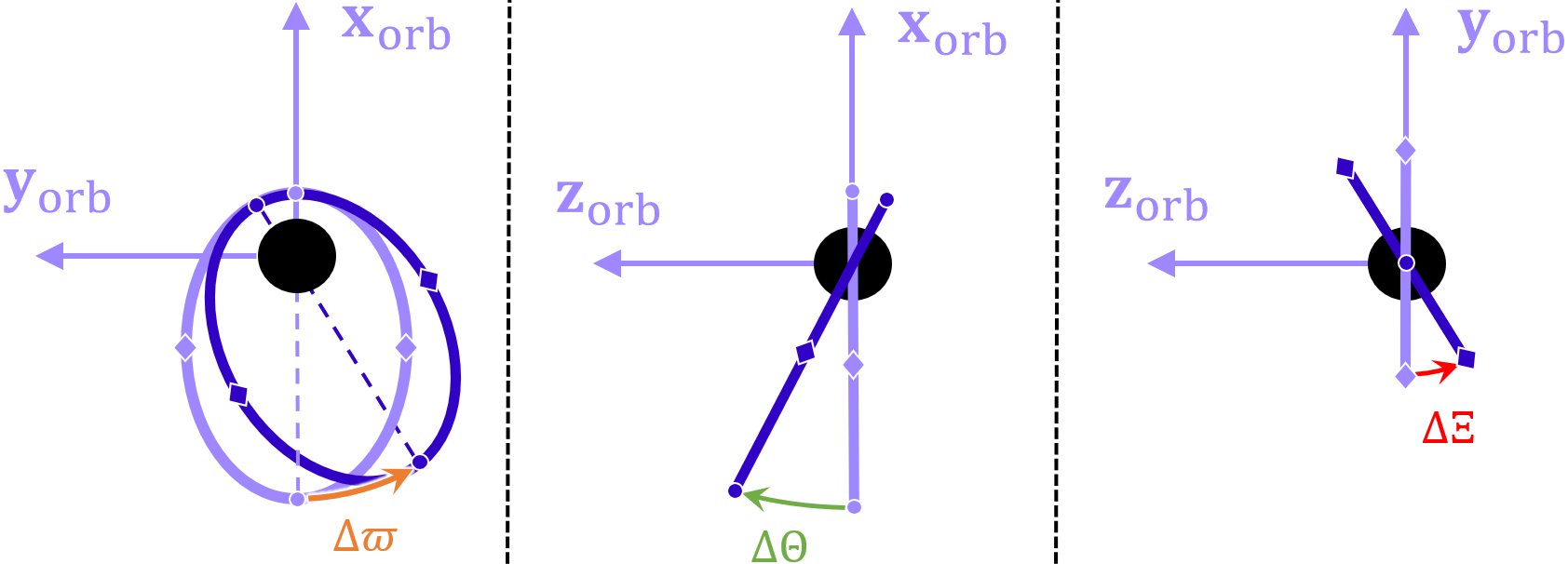}\label{split_out_orb}}\\
    \vspace{3mm}
    \subfloat[Secular timescale orientation change]
    {\includegraphics[trim={0 0 0 2.2cm},clip,width=0.7\columnwidth]{ 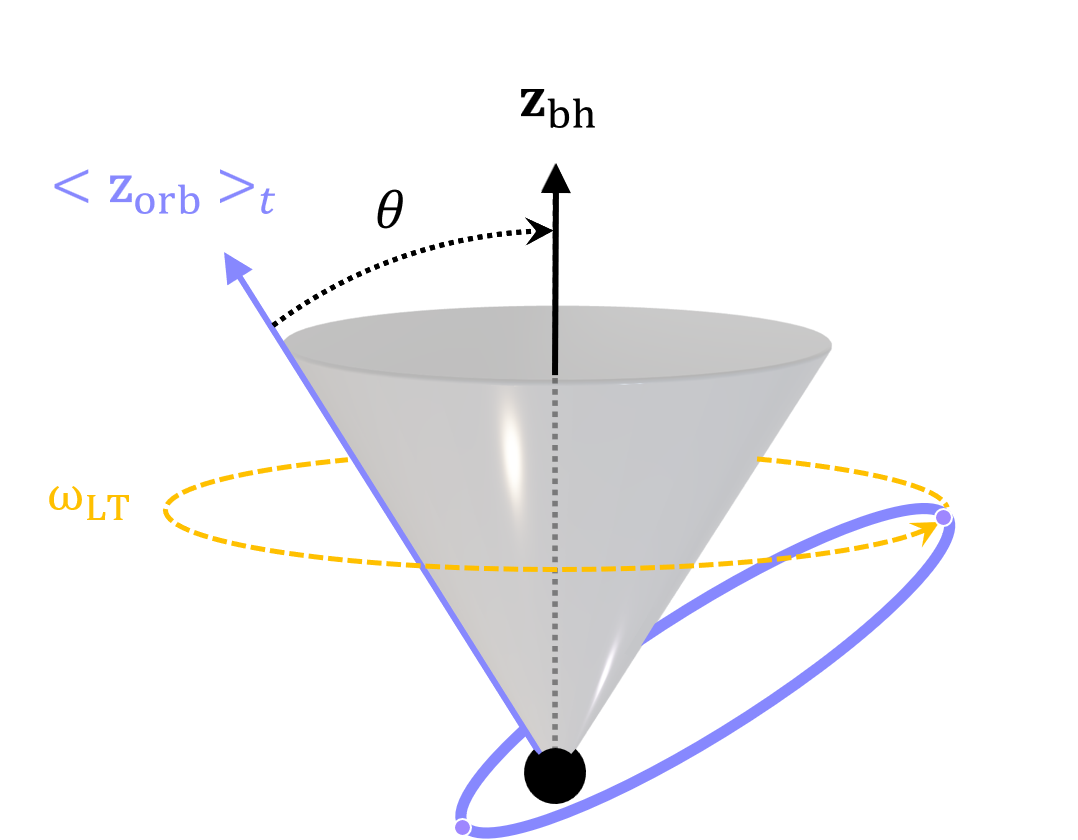}\label{split_out_sec}}\\
    \caption{\label{split_out} Figure \ref{split_out_orb} shows the evolution of the orbit orientation on an orbital period $P$ timescale. Three precessions are experienced: an in-plane precession (left panel) with contributions at 1PN, 1.5PN and 2PN; two out-of-plane precessions, one around the minor axis (middle panel) and another around the major axis (right panel) with contributions at 1.5PN and 2PN. Figure \ref{split_out_sec} shows the evolution of the orbit orientation on a secular timescale $P^\mrm{cone}\gg P$ (see Eq. \eqref{P_cone}). The orbital angular momentum experiences a precession around the black hole spin axis.}
\end{figure}

The secular out-of-plane shift of the major axis can be expressed at 2PN order from Eq. \eqref{dTheta/dt} as:
\begin{align}\label{Delta_Theta}
    &\Delta\Theta = \sin \omega \Delta\iota - \sin \iota \cos \omega \Delta\Omega
\end{align}
to characterize the out-of-plane component of the pericenter precession. When using the secular shifts of the orbital parameters derived in subsection \ref{AnaliticalWill2008} in Eq. \eqref{Delta_Theta}, we obtain the following expressions:
\begin{align}
    &\Delta \Theta_\mrm{LT}|^{2\mrm{PN}} = -4\pi \epsilon^{3/2} \sin\theta\sin\psi\;\;\chi; \label{DTh_chi}\\
    &\Delta \Theta_{Q}|^{2\mrm{PN}} = 3\pi \epsilon^{2} \cos\theta\sin\theta\sin\psi\;\; \chi^2, \label{DTh_q}
\end{align}
corresponding to the out-of-plane movement of the major axis precession (also referred to as out-of-plane apocenter shift) due to the Lense-Thirring and quadrupole moment effects, as illustrated in figure \ref{split_out}. The Schwarzschild precession does not contribute to the out-of-plane shifts at any PN order.

Finally, the secular out-of-plane shift of the minor axis can be expressed at 2PN order from Eq. \eqref{dXi/dt} as:
\begin{align}\label{Delta_Xi}
    &\Delta\Xi = -\cos\omega \Delta\iota -\sin\omega \sin\iota \Delta\Omega.
\end{align}
We obtain the following expressions:
\begin{align}
    &\Delta \Xi_\mrm{LT}|^{2\mrm{PN}} = -4\pi \epsilon^{3/2} \sin\theta\cos\psi\;\;\chi; \label{DXi_chi}\\
    &\Delta \Xi_{Q}|^{2\mrm{PN}} = 3\pi \epsilon^{2} \cos\theta\sin\theta\cos\psi\;\; \chi^2. \label{DXi_q}
\end{align}

The expressions of $\Delta\Theta$ and $\Delta\Xi$ above show that these quantities depend on both $\theta$ and $\psi$. If we take $\psi=\pi/2[\pi]$, i.e when the projection of the spin axis of the black hole on the orbit is along the minor axis, then $\Delta\Theta$ would be maximized, while $\Delta\Xi$ would be null, i.e the major axis precesses in a plane orthogonal to the minor axis. Conversely, with $\psi=0[\pi]$, i.e when the projection of the spin axis on the orbit is along the major axis, $\Delta\Theta$ would be null, while $\Delta\Xi$ would be maximized, i.e the minor axis precesses in a plane orthogonal to the major axis.
Note that due to the in-plane precession, this can be true only instantaneously.

The Lense-Thirring effect generates a maximal out-of-plane precession for polar orbits and a vanishing one for equatorial orbits. We also see that the configuration that allows for the most apocenter displacement due to the Lense-Thirring effect (the maximum of $\sqrt{\Delta\varpi_\mrm{LT}^2+\Delta\Theta_\mrm{LT}^2}$), would be an equatorial orbit.

As for the quadrupole moment effect, it does not induce any out-of-plane precessions for polar orbits. In addition, we observe that the out-of-plane precession is maximized for $\theta=\pi/4[\pi]$. 
This behavior is similar to the Newtonian case of a particle orbiting an oblate mass ; if the particle is mid-inclined with respect to the equatorial plane of the central mass\footnote{which is flattened around the poles.}, the particle would experience the highest level of asymmetries along the orbit leading to the highest out-of-plane shifts among all possible relative orientations \citep{Boain2004}. 
In terms of overall apocenter angular shift, we also see that the configuration that maximizes $\sqrt{\Delta\varpi_Q^2+\Delta\Theta_Q^2}$ would be an equatorial orbit. 

Even though $\omega$ and $\beta$ are observer dependent, $\psi=\beta-\omega$, the angle between $\mbf z_{\mathrm{bh//orb}}$ and $\mbf x_{\mathrm{orb}}$, rotating around $\mbf z_{\mathrm{orb}}$ in figure \ref{split_out}, is observer independent. Therefore, $\Delta\Theta$ and $\Delta\Xi$ are as well.

\subsection{Secular shifts for S2 and "S2/10"}\label{S2&S2/10}

Let us consider the star S2, as well as a hypothetical star that can potentially be detected with GRAVITY+, and that we call "S2/10": a star with an orbit identical to that of S2 but with a semi-major axis 10 times smaller. In table \ref{tab_S2} we compare the various maximal secular shifts of the star S2. In order to compare these shifts using a single set of orbital parameters, we use osculating orbital elements at apocenter. 
\begin{table}[htb]
    \centering
    \caption{Comparing the various maximal secular shifts of the star S2. We use a single set of orbital parameters for all shifts, the osculating one at apocenter. We recall the values od $\theta$ and $\psi$ that maximize each effect. \label{tab_S2} }
    \begin{tabular}{ccccc}
        \multicolumn{1}{c}{$\Delta X^\mrm{2PN}_\mrm{S2}$}
        & \multicolumn{1}{c}{|max($\Delta X$)|}
        & \multicolumn{1}{c}{$\dfrac{|\mrm{max}(\Delta X)|}{\Delta\varpi_\mrm{Sch}}$}
        & \multicolumn{1}{c}{$\theta$}
        & \multicolumn{1}{c}{$\omega-\beta$}\\
        \cline{2-3}
        
        \toprule
        $\Delta\varpi_\mrm{Sch}$ &  $732$ "/rev  & 100\%  & -            & -            \\ \midrule
        $\Delta\varpi_\mrm{LT}$  &  $13.3$ "/rev  & 1.81\% & $0[\pi]$     & -            \\ \midrule
        $\Delta\Theta_\mrm{LT}$  &  $6.63$ "/rev  & 0.91\% & $\pi/2[\pi]$ & $\pi/2[\pi]$ \\ \midrule
        $\Delta\Xi_\mrm{LT}$     &  $6.63$ "/rev  & 0.91\% & $\pi/2[\pi]$ & $0[\pi]$     \\ \midrule
        $\Delta\varpi_Q$         &  $0.0675$ "/rev  & 0.01\% & $0[\pi]$     & -            \\ \midrule
        $\Delta\Theta_Q$         &  $0.0675$ "/rev  & 0.01\% & $\pi/4[\pi]$ & $\pi/2[\pi]$ \\ \midrule
        $\Delta\Xi_Q$            &  $0.0675$ "/rev  & 0.01\% & $\pi/4[\pi]$ & $0[\pi]$     \\ \bottomrule
    \end{tabular}
\end{table}

When computing the secular shifts for "S2/10", we see that 
\begin{align*}
    \Delta\varpi_\mathrm{Sch}^{\mathrm{S2/10}} &= 10\Delta\varpi_\mathrm{Sch}|_\mrm{1PN}^{\mathrm{S2}}+10^2\Delta\varpi_\mathrm{Sch}|_\mrm{2PN}^{\mathrm{S2}}
    \approx 10\, \Delta\varpi_\mathrm{Sch}^{\mathrm{S2}}, \\
    \Delta X_\mathrm{LT}^{\mathrm{S2/10}} &=10^{3/2}\, \Delta X_\mathrm{LT}^{\mathrm{S2}}, \\
    \Delta X_Q^{\mathrm{S2/10}} &= 10^2\, \Delta X_Q^{\mathrm{S2}},
\end{align*}

with $X$ denoting either in-plane ($\varpi$) or out-of-plane ($\Theta$, $\Xi$) precession components. This scaling between the Schwarzschild precession, the Lense-Thirring and the quadrupole moment effects stem from the dependencies in $\epsilon$ (i.e the PN order). 

Not only do all relativistic effects become significantly larger, but having smaller periods allows us to observe more pericenter/apocenter passages, thus accumulating more secular shifts in given amount of time. For instance, the period of the star S2 ($\approx16$yr) corresponds to $10^{3/2}\approx30$ orbits of the star "S2/10". Therefore, when working with "S2/10" instead of S2 over one period of S2, the angular shift due to the Schwarzschild precession is increased by a factor $\approx10^{3/2}\times10\approx300$, the Lense-Thiring precessions by a factor $10^{3/2}\times10^{3/2}=10^{3}$, and the quadrupole moment shifts by a factor $10^{3/2}\times10^{2}=10^{5/2}\approx3000$. These scaling behaviors emphasize the scientific value of observing closer-in stars, providing a powerful means to probe both the spin and higher-order mass moments of Sgr A*.

\subsection{Time-averaged orientation change}\label{sec_time_avg}

After having described the instantaneous orientation change of the orbital frame, we will here discuss its secular evolution over timescales long compared to the orbital period. We know from Eqs. \eqref{dz_orb/dt}, \eqref{didt}, \eqref{dOmdt}, and the corresponding expression of the perturbative acceleration given in \eqref{W_chi} that the Lense-Thirring effect acts on the orbital angular momentum such that:
\be
\frac{\mrm{d} \mbf z_{\mrm{orb}}}{\mrm{d} t}\bigg|_\mrm{LT}\!\!\!\! = -2\frac{G^2m^2}{c^3r^3}\chi\left[\frac{e\sin f}{1+e\cos f}(\mbf z_{\mrm{bh}} \cdot \mbf m_{\mrm{orb}}) +2(\mbf z_{\mrm{bh}} \cdot \mbf n_{\mrm{orb}})\right] \mbf m_{\mrm{orb}}.
\ee
From this we obtain\footnote{by using the following definition for the time-average of a quantity $F$,
$\left\langle F \right\rangle_t = \frac{1}{P} \int_0^{P} F(t) dt= \frac{1}{2\pi a_\mrm{sma}^2\sqrt{1-e^2}} \int_0^{2\pi } r^2F(\varphi) d\varphi$, which is different from the azimuthally-averaged definition of Eq. \eqref{phi_avg} used in the two-timescale analysis.}: 
\be 
\left\langle\frac{\mrm{d} \mbf z_{\mrm{orb}}}{\mrm{d} t}\bigg|_\mrm{LT} \right\rangle_t = \upomega_\mrm{LT}(\mbf z_{\mrm{bh}} \times \mbf z_{\mrm{orb}}), \label{avg_dzorbdt_LT}
\ee
where\footnote{not to be confused with the longitude of the ascending node.}:
\be\label{rate_LT}
\upomega_\mrm{LT} = 2 n \epsilon^{3/2}\chi, 
\ee 
and $n=\sqrt{Gm/a_\mrm{sma}^3}$ is the Newtonian mean motion. 
Moreover, we have the expression (see figure \ref{frame_bh_orb}):
\be \label{zbh_in_orb}
\mbf z_{\mrm{bh}} = \sin\theta\cos\psi \mbf x_{\mrm{orb}} +\sin\theta\sin\psi \mbf y_{\mrm{orb}}+\cos\theta \mbf z_{\mrm{orb}},
\ee
that we insert into Eq. \eqref{avg_dzorbdt_LT}. Then, by identifying to the time average of Eq. \eqref{dz_orb/dt}, we choose to define:
\begin{align}
    &\frac{\mrm{d} \tilde{\Theta}}{\mrm{d} t}\bigg|_\mrm{LT} \equiv -\upomega_\mrm{LT}\sin\theta \sin\psi;\\
    &\frac{\mrm{d} \tilde{\Xi}}{\mrm{d} t}\bigg|_\mrm{LT} \equiv -\upomega_\mrm{LT}\sin\theta \cos\psi,
\end{align} 
and thus get:
\begin{align}
    &\Delta\tilde{\Theta}_\mrm{LT} = P\frac{\mrm{d} \tilde{\Theta}}{\mrm{d} t}\bigg|_\mrm{LT}=\frac{2\pi}{n}\frac{\mrm{d} \tilde{\Theta}}{\mrm{d} t}\bigg|_\mrm{LT}= -4\pi\epsilon^{3/2}\chi\sin\theta \sin\psi;\\
    &\Delta\tilde{\Xi}_\mrm{LT} = P\frac{\mrm{d} \tilde{\Xi}}{\mrm{d} t}\bigg|_\mrm{LT}=\frac{2\pi}{n}\frac{\mrm{d} \tilde{\Xi}}{\mrm{d} t}\bigg|_\mrm{LT}= -4\pi\epsilon^{3/2}\chi\sin\theta \cos\psi.
\end{align} 
This shows that $\frac{\mrm{d} \tilde{\Theta}}{\mrm{d} t}$ and $\frac{\mrm{d} \tilde{\Xi}}{\mrm{d} t}$ are the time-average of $\frac{\mrm{d} \Theta}{\mrm{d} t}$ and $\frac{\mrm{d} \Xi}{\mrm{d} t}$ and that $\Delta\tilde{\Theta}_\mrm{LT}=\Delta\Theta_\mrm{LT}$ and $\Delta\tilde{\Xi}_\mrm{LT}=\Delta\Xi_\mrm{LT}$. In other words, we end up with the same results given by the instantaneous formalism used in subsection \ref{inst_formalism}, which links the two aspects of instantaneous and averaged precession shifts.

If we do the same exercise for the quadrupole moment using Eq. \eqref{dz_orb/dt} and \eqref{W_Q} we get:
\be
\left\langle\frac{\mrm{d} \mbf z_{\mrm{orb}}}{\mrm{d} t}\bigg|_Q \right\rangle_t = -\upomega_Q(\mbf z_{\mrm{bh}} \cdot \mbf z_{\mrm{orb}})(\mbf z_{\mrm{bh}} \times \mbf z_{\mrm{orb}}), \label{avg_dzorbdt_Q}
\ee
where:
\be
\upomega_Q = \frac{3n}{2}\epsilon^2\chi^2 , \label{rate_Q}
\ee
Once again, we see that $\Delta\tilde{\Theta}_Q=\Delta\Theta_Q$ and $\Delta\tilde{\Xi}_Q=\Delta\Xi_Q$ as with the the Lense-Thirring effect above.

We see from Eqs. \eqref{avg_dzorbdt_LT} and \eqref{avg_dzorbdt_Q} that $\mbf z_{\mrm{orb}}$ has an average precession around $\mbf z_{\mrm{bh}}$, meaning that the angle $\theta$ between $\mbf z_{\mrm{orb}}$ and $\mbf z_{\mrm{bh}}$ has no secular evolution. In other words, the secular out-of-plane precession is done at fixed\footnote{$\theta$ can have periodic variations but no secular ones at the 2PN level.} $\theta$, of $\mbf z_{\mathrm{orb}}$ about $\mbf z_{\mathrm{bh}}$ (as illustrated in figure \ref{split_out}), and Eqs. \eqref{rate_LT} and \eqref{rate_Q} are the frequencies at which the out-of-plane precession spans the cone of angle $\theta$ represented in figure \ref{split_out}, due to the Lense-Thirring (at the 1.5PN order) and quadrupole moment (at the 2PN order) effects, respectively. 
We also see that the direction of secular rotation of $\mbf z_{\mrm{orb}}$ due to the Lense-Thirring effect is the same as that of the spin. The direction of secular rotation of $\mbf z_{\mrm{orb}}$ due to the quadrupole moment effect depends on the sign of $-\mbf z_{\mrm{bh}} \cdot \mbf z_{\mrm{orb}}=-\cos\theta$ (see Eq. \eqref{avg_dzorbdt_Q}): in the case of $\theta>\pi/2$ (a retrograde orbit relative to the black hole), the direction of rotation is the same as that of the spin for example, but opposite in the case of prograde orbits relative to the black hole.

In conclusion, we stress that both the in-plane and out-of-plane precessions can be considered differently on two timescales: time-averaged (on the cone defined by Eqs. \eqref{avg_dzorbdt_LT} and \eqref{avg_dzorbdt_Q} for the out-of-plane precession), and instantaneously, which corresponds to a rotation of the axes around each other, given by Eqs. \eqref{dx_orb/dt} and \eqref{dz_orb/dt} (and which can deviate from the cone defined by Eqs. \eqref{avg_dzorbdt_LT} and \eqref{avg_dzorbdt_Q} for the out-of-plane precession).


\subsection{Numerical illustration}\label{sec_simu}

We illustrate the instantaneous variations in the orientation of the orbital frame using simulations tailored to maximize the impact of each relativistic effect (see the analysis of subsection \ref{sec_evo_orb}). To do so, we will align $\mbf z_{\mathrm{bh}}$ on each axis of the orbit frame $\mbf z_{\mathrm{bh}}//\mbf z_{\mathrm{orb}}$ (maximizing the in-plane precession of all effects), $\mbf z_{\mathrm{bh}}//\mbf x_{\mathrm{orb}}$ and $\mbf z_{\mathrm{bh}}//\mbf y_{\mathrm{orb}}$ (maximizing the out-of-plane precession of the Lense-Thirring effect); these three configurations will be referred to as "equatorial orbit", "polar orbit with polar major axis" and "polar orbit with equatorial major axis", respectively. Finally, we defer to appendix \ref{sec_outofplane_mid} the study of "mid-inclined orbits", that maximize the out-of-plane precession of the quadrupole moment, and are such that $\mbf z_{\mathrm{bh}}=-\mbf y^\mrm{pro}_{\mathrm{orb}}+\mbf z^\mrm{pro}_{\mathrm{orb}}=\mbf y^\mrm{ret}_{\mathrm{orb}}-\mbf z^\mrm{ret}_{\mathrm{orb}}$. This will provide intuition on the way the spin of the black hole can impact the orbit, independently from the observer.

The simulated star is taken to have orbital parameters identical to "S2/10" in terms of semi-major axis and eccentricity. We perform a Euclidean projection of the orbit onto a plane—this is for illustrative purposes only and does not aim at reproducing observables. Effects such as the R\o mer delay, gravitational redshift or lensing are deliberately omitted, as they fall outside the scope of this analysis.

We find that the simulated variations in the considered configurations show good agreement with our analytical predictions: the direction and structure of the effects are consistent with post-Newtonian expectations.

We define a prograde and retrograde orbit relative to the black hole, with angular momentum along $\mbf z_{\mathrm{orb}}^\mrm{pro}$ and $\mbf z_{\mathrm{orb}}^\mrm{ret}$, as orbits that verify $\mbf z_{\mathrm{orb}}^\mrm{pro}\cdot\mbf z_{\mathrm{bh}}>0$ and $\mbf z_{\mathrm{orb}}^\mrm{ret}\cdot\mbf z_{\mathrm{bh}}<0$, respectively. 


\subsubsection{In-plane precession: focusing on equatorial orbits}\label{sec_inplane}

In this subsection, we will examine the impact of the in-plane precession due to Kerr effects, independently from the projection on the screen of a distant observer. For this purpose, we simulate obits that are in the equatorial plane of the black hole, thus maximizing the in-plane precession, and eliminating the out-of-plane ones\footnote{We know that the quantities $\mathcal{W}^\mrm{1.5PN}_\mrm{LT}$ and $\mathcal{W}^\mrm{2PN}_Q$, the expressions of which are given in Eqs. \eqref{W_chi} and \eqref{W_Q} yield zero for equatorial orbits, i.e for $\theta = 0[\pi]$. This leads to Eqs. \eqref{didt} and \eqref{dOmdt} also yielding zero, meaning that no out-of-plane movement is induced by the BH's spin. So an initially equatorial orbit will remain so, even in the presence of a spin.}. 

\begin{figure}[htp]
    \centering
    \subfloat[Schematic illustration of the osculating Keplerian orbit]
    {\includegraphics[clip,width=0.7\columnwidth]{ 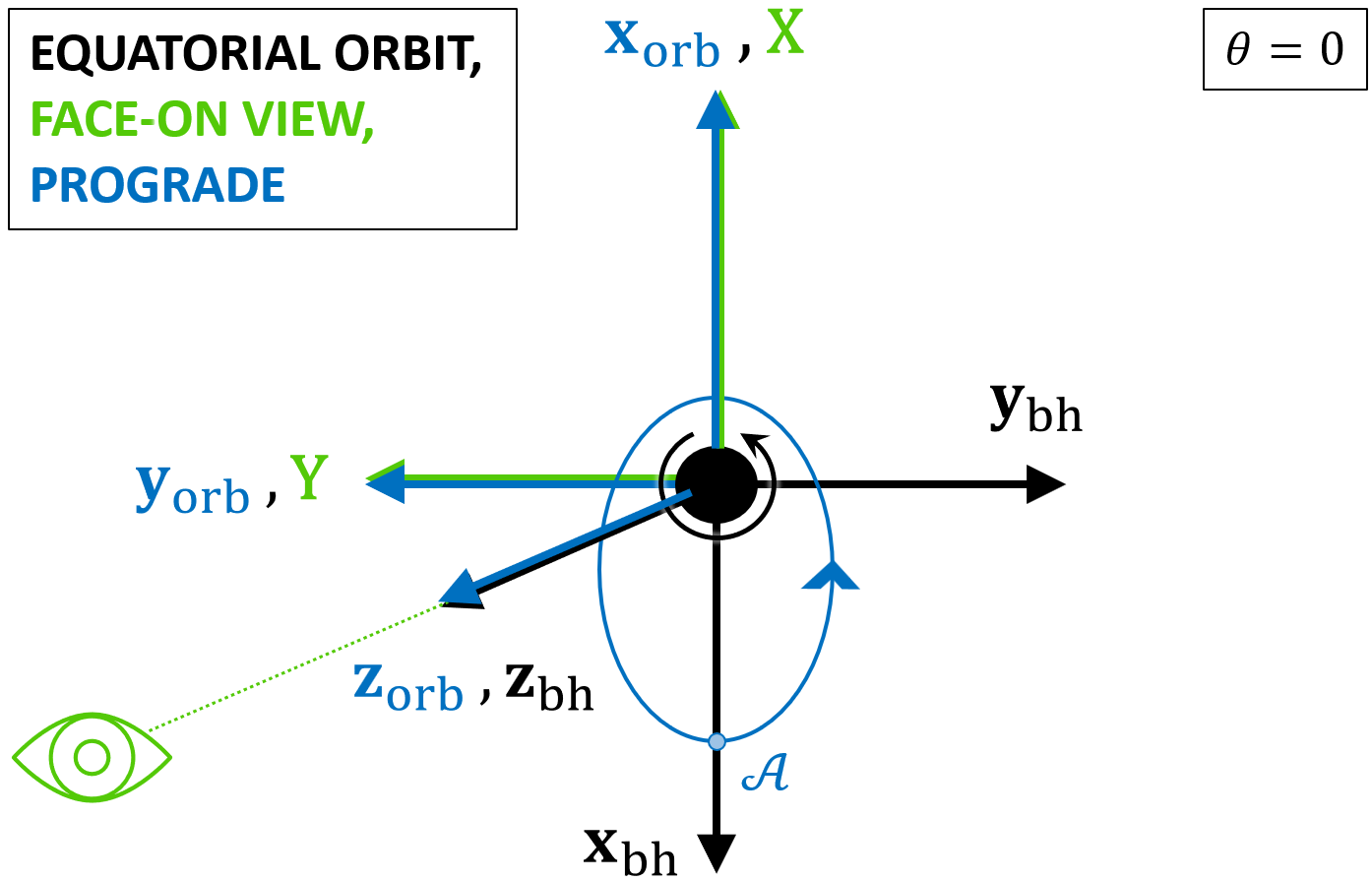}\label{illustr_eq_pro}}\\
    \vspace{3mm}
    \subfloat[Simulation using the 2PN code]
    {\includegraphics[clip,width=0.87\columnwidth]{ 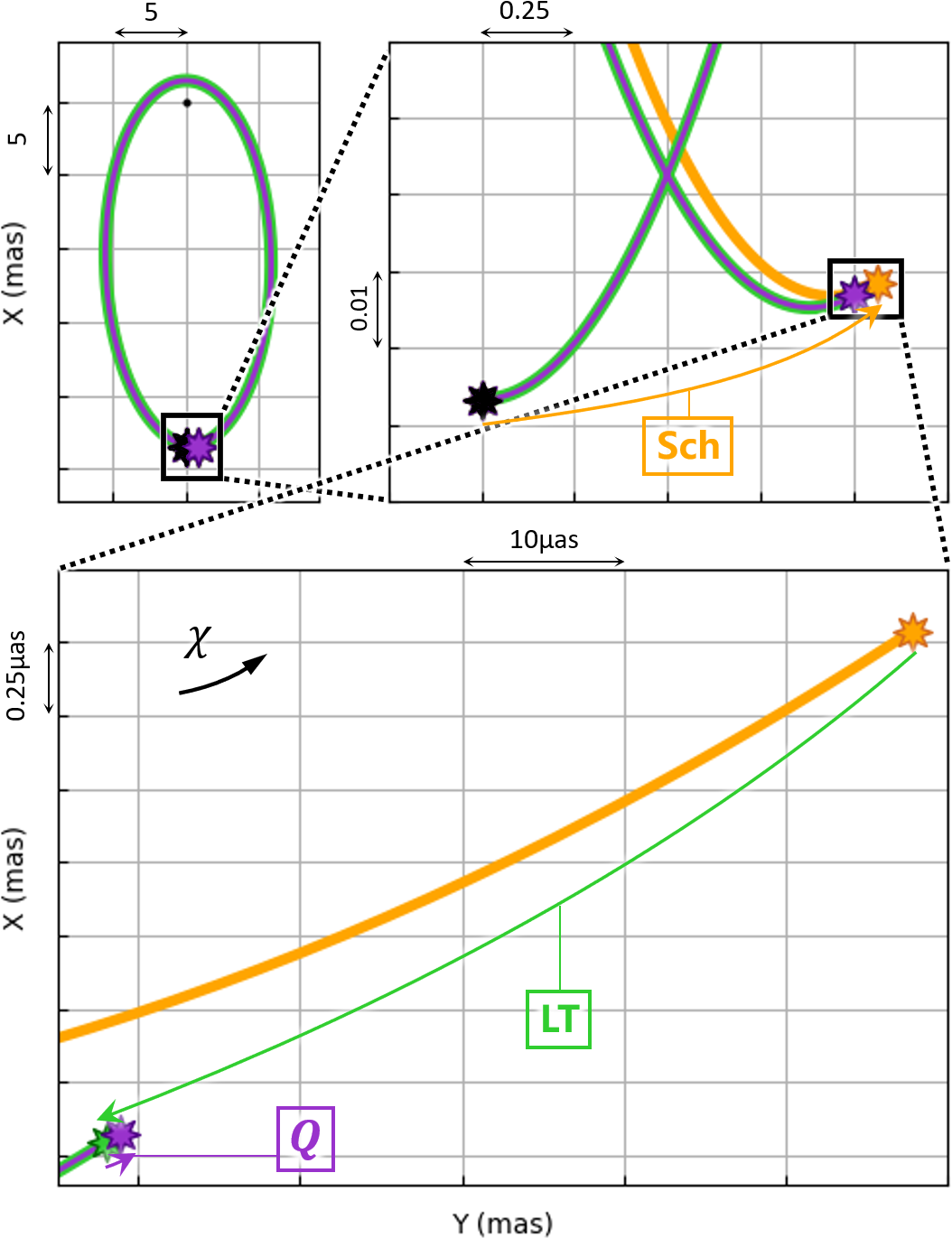}\label{simu_eq_pro}}\\
    \caption{\label{eq_pro} Prograde orbits in the equatorial plane of the black hole with a face-on view. In figure \ref{illustr_eq_pro} the apocenter is represented by the letter $\mathcal{A}$. In figure \ref{simu_eq_pro} we show an orbit of the hypothetical star "S2/10" simulated in  2PN using OOGRE: the "\ding{87}" denotes the apocenters (the first apocenter in black corresponds to the osculating time, which is also the initial date, and is in common for the three orbits). We simulate one orbit for the Schwarzschild precession ("Sch" in orange), one orbit for both the Schwarzschild and Lense-Thirring precessions ("LT" in green), and one orbit for the Schwarzschild, spin and quadrupole moment precessions ("$Q$" in thin purple). We see that the Lense-Thirring and quadrupole moment effects shift the apocenter clockwise and counterclockwise, respectively, as seen by this face-on observer. All numbers are expressed in mas except otherwise noted.}
\end{figure}

\begin{figure}[htp]
    \centering
    \subfloat[Schematic illustration of the osculating Keplerian orbit]
    {\includegraphics[clip,width=0.7\columnwidth]{ 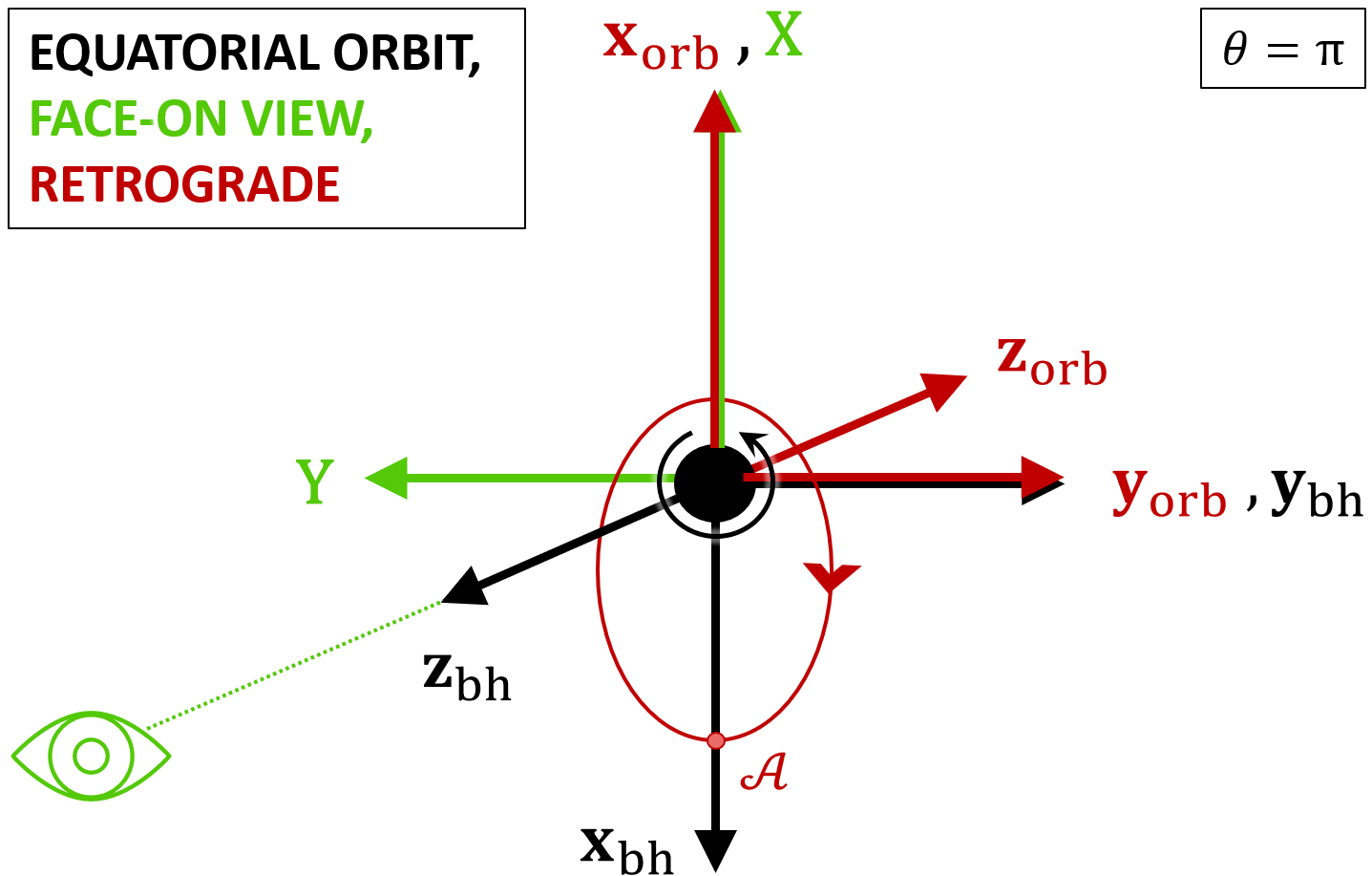}\label{illustr_eq_ret}}\\
    \vspace{3mm}
    \subfloat[Simulation using the 2PN code]
    {\includegraphics[clip,width=0.87\columnwidth]{ 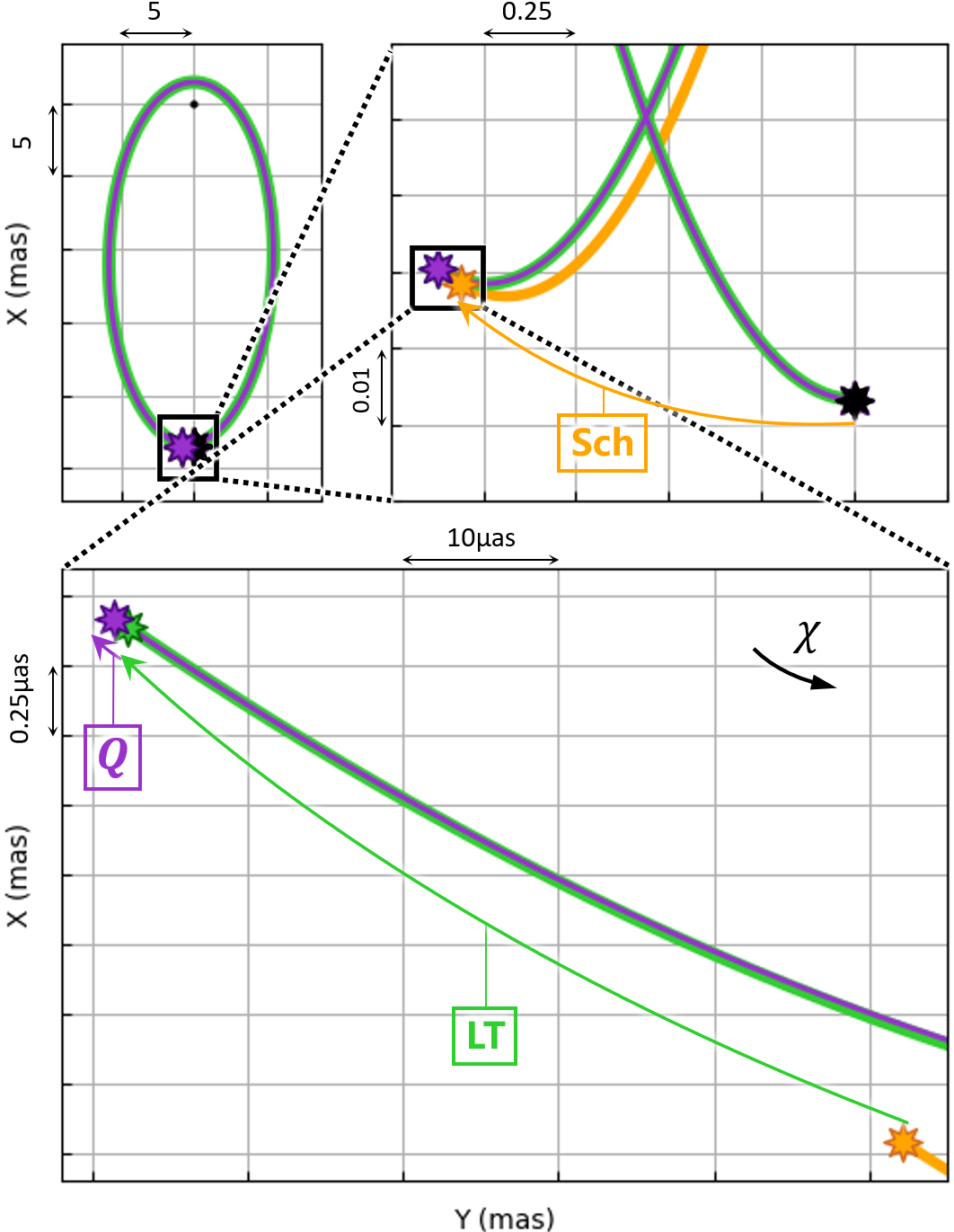}\label{simu_eq_ret}}\\
    \caption{\label{eq_ret} Same as figure \ref{eq_pro} but for retrograde orbits. We see that both the Lense-Thirring and quadrupole moment effects shift the apocenter clockwise as seen by this face-on observer.}
\end{figure}

Let us consider 2 orbits in the equatorial plane of the black hole, one is prograde ($\theta=0$) with $\Delta\varpi_{\mrm{Kerr}}^\mrm{pro } = \Delta\varpi_\mrm{LT}^\mrm{pro } + \Delta\varpi_Q^\mrm{pro }$, and the other is retrograde ($\theta= \pi$) with $\Delta\varpi_{\mrm{Kerr}}^\mrm{ret } = \Delta\varpi_\mrm{LT}^\mrm{ret } + \Delta\varpi_Q^\mrm{ret }$, making $\psi$ degenerate. For understanding the in-plane precession, we perform in figures \ref{eq_pro} and \ref{eq_ret} a Euclidean projection of these orbits onto a face-on plane.


Let us start with the well-known Schwarzschild precession. We know that $\omega$ is measured in the same direction as the movement of the star in its orbit. When looking at the analytical expressions, we see that Eq. \eqref{Dvarpi_sch} is independent from $\theta$, it leads to the same positive $\Delta\varpi_\mrm{sch}$ for prograde and retrograde orbits, regardless of the orientation of the black hole with respect to the orbit. This means that the prograde and retrograde apocenter displacements will go in opposite ways.
We see on figures \ref{simu_eq_pro} and \ref{simu_eq_ret} that the Schwarzschild precession shifts the apocenters of the prograde and retrograde orbits counterclockwise and clockwise, respectively. In simpler words, the simulations agree with Eq. \eqref{Dvarpi_sch} in terms of orientation on the projection plane.

Now, let us examine the Lense-Thirring effect. Figure \ref{simu_eq_pro} indicates an apocenter shift opposite to the direction of the spin, and the same is observed in figure \ref{simu_eq_ret}. Indeed, when looking at Eqs. \eqref{Dvarpi_chi}, we see that for the prograde orbit, $\Delta\varpi_\mrm{LT}^\mrm{pro}<0$ and $\Delta\varpi_\mrm{LT}^\mrm{ret}>0$, meaning that both the prograde and retrograde apocenters will be shifted clockwise, i.e. opposite to the direction of the spin. 

Finally, we see in figures \ref{simu_eq_pro} and \ref{simu_eq_ret} that, similarly to the Schwarzschild precession, the quadrupole moment shifts the apocenters in the same direction as the movement of the star in its orbit. Indeed, when looking at Eq. \eqref{Dvarpi_Q}, we see that the sign of $\Delta\varpi_Q$ depends only on the value of $\cos^2\theta$ meaning that $\Delta\varpi_Q^\mrm{pro}=\Delta\varpi_Q^\mrm{ret}>0$.

Finally, it is important to remember that all these effects do not share the same dominant PN order, and therefore do not act on the same scale. This is very visible in figures \ref{simu_eq_pro} and \ref{simu_eq_ret} where the Schwarzschild precession dominates the Lense-Thirring effect, which in turn dominates the quadrupole moment effect. Also, it is interesting to notice that the total secular shift of orbits in the equatorial plane of a rotating black hole will differ in absolute value in the prograde and retrograde configurations; in the first case the Lense-Thirring and quadrupole moment effects counter each other, whereas in the second case they push the apocenter in the same direction. 

\begin{figure}[htp]
    \centering
    \includegraphics[trim={0 2.2cm 0 2.5cm},clip,width=0.7\columnwidth]{ 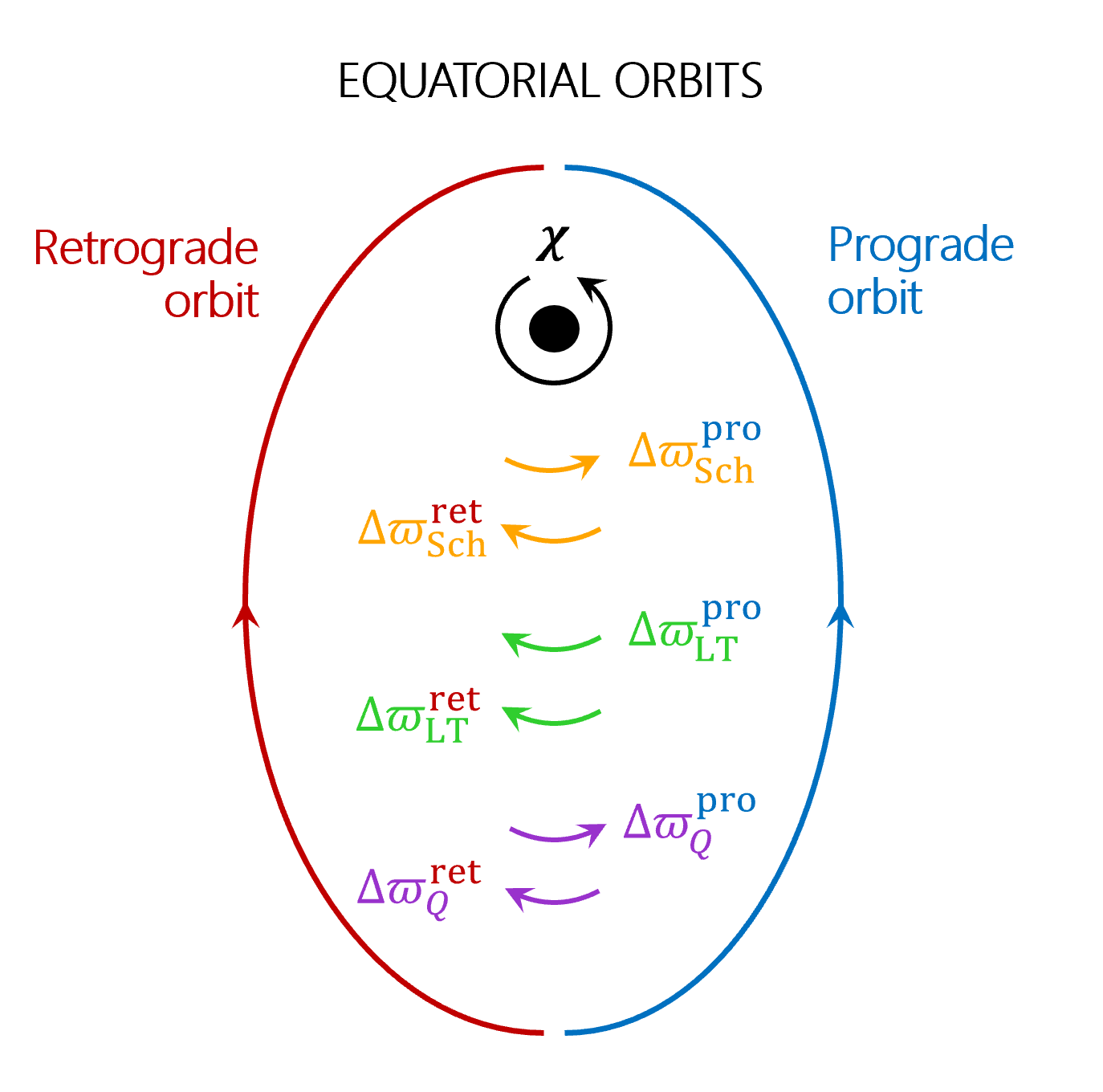}
    \caption{\label{eq_plane_theory} Schematic illustration of the secular precessions for a prograde and a retrograde equatorial orbit. We note that Schwarzschild precession and quadrupole moment  effect push the apocenter in the direction of stellar motion, and the Lense-Thirring precession pushes against the spin.}
\end{figure}

If we summarize all the above, we get the following statements illustrated in figure \ref{eq_plane_theory}:
\begin{enumerate}
      \item The Schwarzschild precession pushes the apocenter in the same direction as the movement of the star in its orbit.
      \item The secular shift due to the Lense-Thirring effect pushes the apocenter of an equatorial orbit in the direction opposite to  the black hole's rotation.
      \item The secular shift due to the quadrupole moment pushes the apocenter of an equatorial orbit in the same direction as the movement of the star in its orbit.
      \item The analytical total Kerr secular shift of equatorial orbits verifies $|\Delta\varpi_{\mrm{Kerr}}^\mrm{pro}| < |\Delta\varpi_{\mrm{Kerr}}^\mrm{ret}|$. 
\end{enumerate}


\subsubsection{Lense-Thirring out-of-plane precession}\label{sec_outofplane}

In this section, we choose to illustrate configurations that maximize the dominant spin-induced effect: the Lense–Thirring precession. As suggested in section \ref{sec_evo_orb}, to better understand the out-of-plane precession of the Lense-Thirring and quadrupole moment effects, it is useful to split it into two subtypes, precession around the major axis of the orbit and precession around the minor axis, encoded by Eqs. \eqref{dXi/dt} and \eqref{dTheta/dt}, respectively. Therefore, we focus on two types of polar orbits, for which the out-of-plane component of this effect is strongest (see subsection \ref{inst_formalism}):
\begin{enumerate}
    \item Polar orbit with polar major axis\footnote{Due to the ever present in-plane precession, the major axis can only instantaneously be polar or equatorial. Here we refer to the osculating state.}: $\mbf z_{\mathrm{bh}}//\mbf x_{\mathrm{orb}}$ i.e $(\theta,\psi)=(\pi/2,0)$;
    \item Polar orbit with equatorial major axis: $\mbf z_{\mathrm{bh}}//\mbf y_{\mathrm{orb}}$ i.e $(\theta,\psi)=(\pi/2,\pi/2)$.
\end{enumerate}

\begin{figure}[htp]
    \centering
    \subfloat[Schematic illustration of the osculating Keplerian orbit]
    {\includegraphics[clip,width=0.7\columnwidth]{ 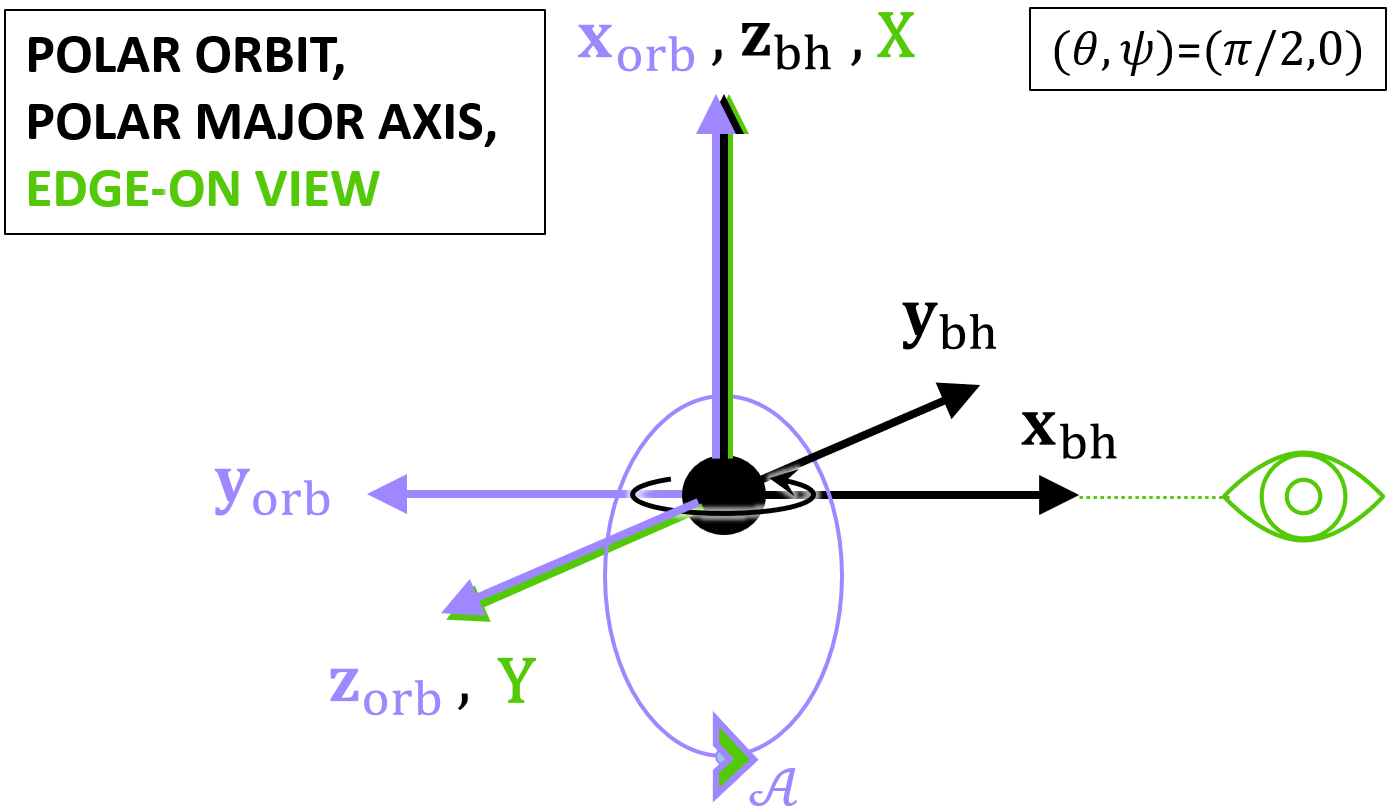}\label{illustr_pol_pol_pro}}\\
    \vspace{3mm}
    \subfloat[Simulation using the 2PN code]
    {\includegraphics[clip,width=0.87\columnwidth]{ 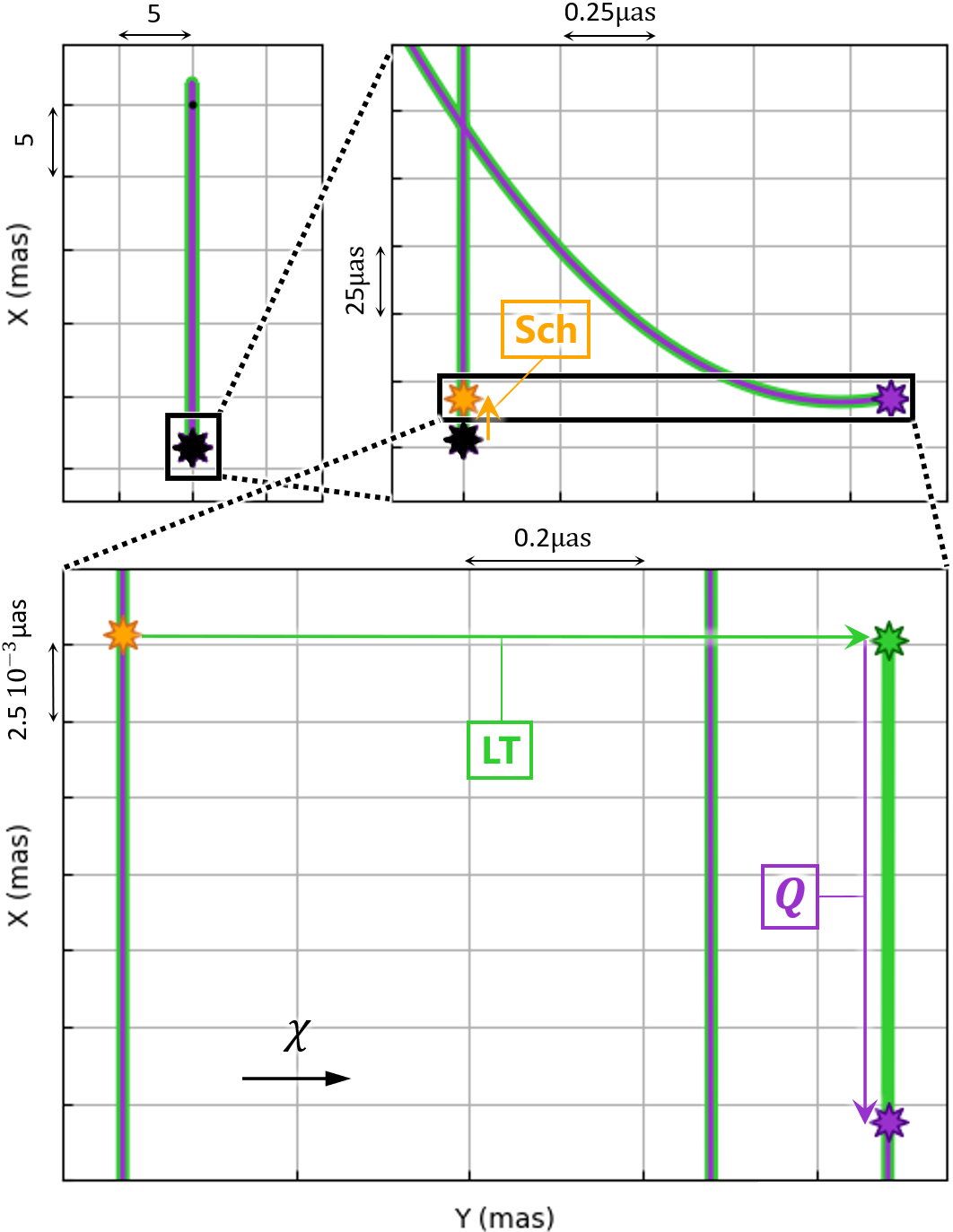}\label{simu_pol_pol_pro}}\\
    \caption{\label{pol_pol_pro} Same as figure \ref{eq_pro} but with orbits in a polar plane and having a major axis on the spin axis of the black hole. We see that the Lense-Thirring effect shifts the minor axis in the same direction as the the spin of the black hole, gradually shifting the orbit plane from edge-on towards face-on. The quadrupole moment effect shifts the apocenter clockwise as seen on face-on projection of figure \ref{illustr_eq_pro}.}
\end{figure}

\begin{figure}[htp]
    \centering
    \subfloat[Schematic illustration of the osculating Keplerian orbit]
    {\includegraphics[clip,width=0.7\columnwidth]{ 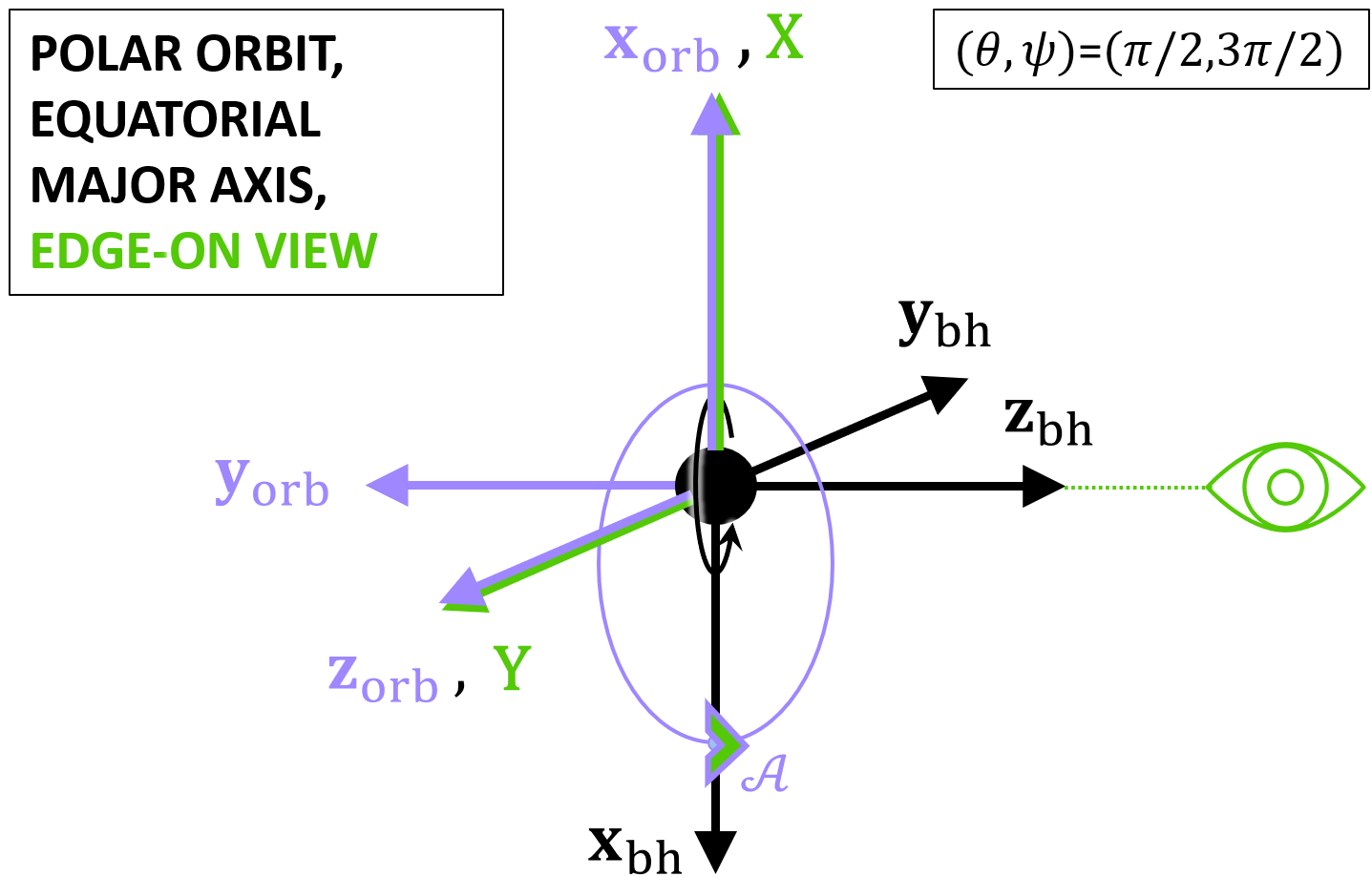}\label{illustr_pol_eq_pro}}\\
    \vspace{3mm}
    \subfloat[Simulation using the 2PN code]
    {\includegraphics[clip,width=0.87\columnwidth]{ 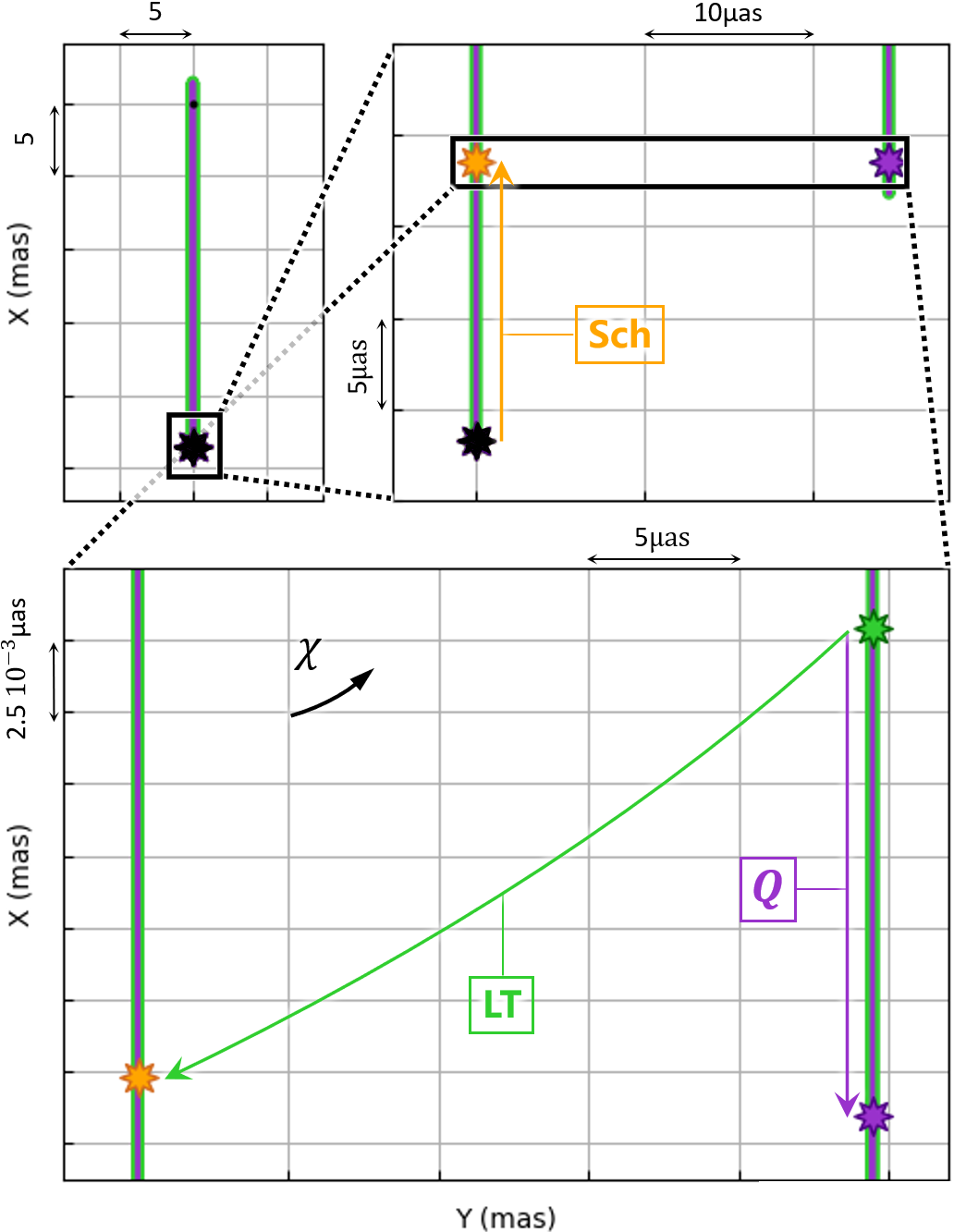}\label{simu_pol_eq_pro}}\\
    \caption{\label{pol_eq_pro} Same as figure \ref{eq_pro} but with orbits in the polar plane and having a major axis on the equatorial plane of the black hole. We see that the Lense-Thirring effect shifts the apocenter in the counterclockwise direction as seen on this projection. The quadrupole moment effect shifts the apocenter clockwise as seen on the face-on projection of figure \ref{illustr_eq_pro}.}
\end{figure}

In order to visualize the out-of-plane components independently from the projection effects, we perform an edge-on Euclidean projection, and consider that the osculating orbit is a vertical line with the apocenter at the bottom of the projection plane (see figures \ref{illustr_pol_pol_pro} and \ref{illustr_pol_eq_pro}).

Let us start with the first subtype, illustrated by figure \ref{pol_pol_pro}, where the orbit lies in the polar plane of the BH and verifies $\mbf z_{\mathrm{bh}}//\mbf x_{\mathrm{orb}}$. As opposed to the equatorial orbits of subsection \ref{sec_inplane}, here the notion of prograde and retrograde orbits is more delicate. Polar orbits cannot be categorized as prograde or retrograde relative to the black hole, and an orbit with an edge-on view cannot be categorized as prograde or retrograde relative to the observer. Therefore, we will instead consider an orbit such that the star is moving towards the observer at apocenter (as illustrated in \ref{illustr_pol_pol_pro}).

We saw in subsection \ref{sec_evo_orb}, that by having $(\theta,\psi)=(\pi/2,0)$ the Lense-Thirring effect maximally contributes to the out-of-plane precession while not contributing to the in-plane one, and that, interestingly enough, the quadrupole moment has the opposite tendency. We remind that having here $\Delta\varpi_\mrm{LT}=0$, $\Delta\Theta_\mrm{LT}=0$ and $\Delta\Xi_\mrm{LT}\neq0$ means that, instantaneously, the apocenter will not experience any secular precession due to the Lense-Thirring effect. However, as soon as the apocenter is shifted out of the spin axis due to other effects, the Lense-Thirring effect will start contributing as well to the apocenter displacement.
Indeed, in figure \ref{simu_pol_pol_pro}, we see that the Lense-Thirring effect consists of an out-of-plane precession that shifts the minor axis in the same direction as the spin of the black hole, gradually turning the orbit plane from being edge-on to face-on, having a counterclockwise rotation in the projection plane. 
Conversely, the quadrupole moment effect is in-plane and shifts the apocenters in the direction opposite to the movement of the star in its orbit. All this is in agreement with Eqs. \eqref{DTh_chi} to \eqref{DXi_q}.

Now, let us consider the second subtype of out-of-plane precession: the one around the minor axis of the orbit. To study this type of precession we take a polar orbit but this time with its minor axis being along the spin axis of the black hole ($\mbf z_{\mathrm{bh}}//\mbf y_{\mathrm{orb}}$), as illustrated in figure \ref{pol_eq_pro}. Since we still have $\theta=\pi/2$ in this configuration, we observe the same tendencies for the evolution of $\Delta \varpi$. However, now we have $\psi=\pi/2$, Eqs. \eqref{DTh_chi} and  \eqref{DTh_q} yield $\Delta\Theta\neq0$ meaning that, as opposed to the previous case, we observe a direct out-of-plane apocenter displacement. Conversely, Eqs. \eqref{DXi_chi} and  \eqref{DXi_q} yield $\Delta\Xi=0$, which means that there is no precession about the major axis, naturally because the spin is along the minor axis.
Indeed, in figure \ref{simu_pol_eq_pro}, we see that the Lense-Thirring effect shifts the major axis in the same direction as the spin of the black hole, while keeping the orbit plane edge-on this time. Similarly to the polar orbit with a polar major axis configuration above, the quadrupole moment effect is in-plane and shifts the apocenters in the direction opposite to the movement of the star in its orbit. Once more, all this is in agreement with Eqs. \eqref{DTh_chi} to \eqref{DXi_q}.

All this shows that a single orbit cannot efficiently reveal all relativistic effects; rather, observing orbits with diverse orientations is key to isolating the different precession components.

\section{Summary and conclusions}\label{sec_summary}



In this article, we have presented the various kinds of precessions undergone by a stellar orbit due to the 1.5- and 2PN-order spin effects. 
We have shown that on an orbital timescale, the orbit frame is subject to an in-plane precession parametrized by the rate $\dot{\varpi}$, and to out-of-plane precessions parametrized by the rates $\dot{\Xi}$ and $\dot{\Theta}$. Integrating over an orbit, we provide the 2PN expressions of $\Delta\varpi$, $\Delta\Xi$, and $\Delta\Theta$ (see Eqs. \eqref{Dvarpi_sch} to \eqref{DXi_q}), which correspond to the tilt angle between two successive pericenters, and to the out-of-plane tilt of the minor and major axes, respectively (see figure \ref{split_out}). We also highlight the contributions due to Schwarzschild, Lense-Thirring, and quadrupole effects in these formulas.
Moreover, we show how these orbital timescale precessions are linked to the well-known secular timescale precession of the orbital angular momentum around the black hole spin axis.
By considering orbital orientations that maximize the various kinds of precessions of the orbit, we illustrate numerically these findings and show the consistency of our PN orbit-integration code OOGRE with our analytical predictions. 
We demonstrate that no single orbital configuration is optimal for detecting all relativistic effects simultaneously; instead, a diversity of orbital orientations is essential for disentangling the different types of precessions.
We also provide estimates of the on-sky astrometric effects associated to these various precessions for the particular case of the S2 star, as well as for "S2/10", a hypothetical star on an orbit ten times smaller than that of S2.
Our results, highlighting the orbital-timescale reorientation of stellar orbits due to the black hole spin, will be very relevant for trying to detect the spin parameter of Sgr A* on S stars at the Galactic center through their astrometric signatures by the GRAVITY instrument.

When comparing the Schwarzschild, Lense-Thirring, and quadrupole moment contributions, their relative magnitudes are governed by their post-Newtonian (PN) order:
\begin{itemize}
    \item The Schwarzschild precession scales as $\epsilon \sim \mathcal{O}(v^2/c^2)$ (1PN),
    \item The Lense-Thirring precession as $\epsilon^{3/2} \sim \mathcal{O}(v^3/c^3)$ (1.5PN),
    \item The quadrupole moment precession as $\epsilon^2 \sim \mathcal{O}(v^4/c^4)$ (2PN),
\end{itemize}
where $\epsilon=GM/c^2p$ is the small PN parameter. To quantify how these effects change for stars on tighter orbits, we considered “S2/10”. In this case, the PN parameter increases by a factor of 10, i.e., $\epsilon^\mathrm{S2/10} \approx 10\epsilon^\mathrm{S2}$. 
Thus, we saw that all relativistic effects grow significantly stronger at smaller orbital radii, and the shorter orbital periods of such stars enable more frequent pericenter and apocenter passages. As a result, secular precessions accumulate more rapidly over a fixed observational timespan. For instance, when working with "S2/10" over one period of S2, the angular shifts are increased by a factor $\approx300$ for the Schwarzschild precession, by a factor $10^{3}$ for the Lense-Thiring precessions, and by a factor $\approx3000$ for the quadrupole moment shifts, compared to the star S2. These scaling properties underscore the high scientific value of monitoring stars on tighter orbits, thus offering a powerful probe of both the spin and multipolar structure of Sgr A*.

Our study highlights the importance of identifying and tracking new S-stars in the inner regions of the Galactic Center, where relativistic effects are strongest. The next-generation GRAVITY+ \citep{GRAVITY+23} and ERIS \citep{ERIS} instruments will significantly enhance the detection and monitoring capabilities, making it possible to refine the orientation of Sgr A* and further test the no-hair theorem.
Even if closer-in stars cannot be detected, an alternative approach is to use multi-star fitting with the currently known S-stars. Since these stars have different orientations and thus exhibit different types of secular shifts, they can provide additional constraints on both the black hole's spin magnitude and its orientation. This method has the potential to reduce the required monitoring time needed to test the no-hair theorem.
Additionally, a deeper understanding of systematic uncertainties---including the role of stellar perturbations and external mass distributions---will be necessary to interpret high-precision orbital data robustly and provide a more comprehensive picture of the astrophysical environment surrounding Sgr A*. 

\begin{acknowledgements}
For fruitful discussions we thank our colleagues Laura Bernard, Éric Gourgoulhon, Aurélien Hees and Alexandre Le Tiec, as well as Clifford M. Will.
\end{acknowledgements}

%
%

\bibliography{BibFile}

\begin{thebibliography}{41}
\expandafter\ifx\csname natexlab\endcsname\relax\def\natexlab#1{#1}\fi

\bibitem[{Alush \& Stone(2022)}]{Alush+22}
Alush, Y. \& Stone, N.~C. 2022, Phys. Rev. D, 106, 123023

\bibitem[{Ang{\'e}lil \& Saha(2014)}]{Angelil+14}
Ang{\'e}lil, R. \& Saha, P. 2014, Monthly Notices of the Royal Astronomical Society, 444, 3780

\bibitem[{{Blanchet} \& {Iyer}(2003)}]{Blanchet2003}
{Blanchet}, L. \& {Iyer}, B. 2003

\bibitem[{{Boain}(2004)}]{Boain2004}
{Boain}, J.~R. 2004

\bibitem[{Brumberg(2017)}]{Brumberg}
Brumberg, V.~A. 2017, Essential relativistic celestial mechanics (CRC Press)

\bibitem[{Collaboration {et~al.}(2022)Collaboration, Akiyama, Alberdi, Alef, Algaba, Anantua, Asada, Azulay, Bach, Baczko, Ball, Balokovi{\'c}, Barrett, Baub{\"o}ck, Benson, Bintley, Blackburn, Blundell, Bouman, Bower, Boyce, Bremer, Brinkerink, Brissenden, Britzen, Broderick, Broguiere, Bronzwaer, Bustamante, Byun, Carlstrom, Ceccobello, Chael, kwan Chan, Chatterjee, Chatterjee, Chen, Chen, Cheng, Cho, Christian, Conroy, Conway, Cordes, Crawford, Crew, Cruz-Osorio, Cui, Davelaar, Laurentis, Deane, Dempsey, Desvignes, Dexter, Dhruv, Doeleman, Dougal, Dzib, Eatough, Emami, Falcke, Farah, Fish, Fomalont, Ford, Fraga-Encinas, Freeman, Friberg, Fromm, Fuentes, Galison, Gammie, Garc{\'\i}a, Gentaz, Georgiev, Goddi, Gold, G{\'o}mez-Ruiz, G{\'o}mez, Gu, Gurwell, Hada, Haggard, Haworth, Hecht, Hesper, Heumann, Ho, Ho, Honma, Huang, Huang, Hughes, Ikeda, Impellizzeri, Inoue, Issaoun, James, Jannuzi, Janssen, Jeter, Jiang, Jim{\'e}nez-Rosales, Johnson, Jorstad, Joshi, Jung, Karami, Karuppusamy, Kawashima, Keating,
  Kettenis, Kim, Kim, Kim, Kim, Kino, Koay, Kocherlakota, Kofuji, Koch, Koyama, Kramer, Kramer, Krichbaum, Kuo, Bella, Lauer, Lee, Lee, Leung, Levis, Li, Lico, Lindahl, Lindqvist, Lisakov, Liu, Liu, Liuzzo, Lo, Lobanov, Loinard, Lonsdale, Lu, Mao, Marchili, Markoff, Marrone, Marscher, Mart{\'\i}-Vidal, Matsushita, Matthews, Medeiros, Menten, Michalik, Mizuno, Mizuno, Moran, Moriyama, Moscibrodzka, M{\"u}ller, Mus, Musoke, Myserlis, Nadolski, Nagai, Nagar, Nakamura, Narayan, Narayanan, Natarajan, Nathanail, Fuentes, Neilsen, Neri, Ni, Noutsos, Nowak, Oh, Okino, Olivares, Ortiz-Le{\'o}n, Oyama, {\"O}zel, Palumbo, Paraschos, Park, Parsons, Patel, Pen, Pesce, Pi{\'e}tu, Plambeck, PopStefanija, Porth, P{\"o}tzl, Prather, Preciado-L{\'o}pez, Psaltis, Pu, Ramakrishnan, Rao, Rawlings, Raymond, Rezzolla, Ricarte, Ripperda, Roelofs, Rogers, Ros, Romero-Ca{\~n}izales, Roshanineshat, Rottmann, Roy, Ruiz, Ruszczyk, Rygl, S{\'a}nchez, S{\'a}nchez-Arg{\"u}elles, S{\'a}nchez-Portal, Sasada, Satapathy, Savolainen, Schloerb,
  Schonfeld, Schuster, Shao, Shen, Small, Sohn, SooHoo, Souccar, Sun, Tazaki, Tetarenko, Tiede, Tilanus, Titus, Torne, Traianou, Trent, Trippe, Turk, van Bemmel, van Langevelde, van Rossum, Vos, Wagner, Ward-Thompson, Wardle, Weintroub, Wex, Wharton, Wielgus, Wiik, Witzel, Wondrak, Wong, Wu, Yamaguchi, Yoon, Young, Young, Younsi, Yuan, Yuan, Zensus, Zhang, Zhao, Zhao, Agurto, Allardi, Amestica, Araneda, Arriagada, Berghuis, Bertarini, Berthold, Blanchard, Brown, C{\'a}rdenas, Cantzler, Caro, Castillo-Dom{\'\i}nguez, Chan, Chang, Chang, Chang, Chang, Chen, Chilson, Chuter, Ciechanowicz, Colin-Beltran, Coulson, Crowley, Degenaar, Dornbusch, Dur{\'a}n, Everett, Faber, Forster, Fuchs, Gale, Geertsema, Gonz{\'a}lez, Graham, Gueth, Halverson, Han, Han, Hasegawa, Hern{\'a}ndez-Rebollar, Herrera, Herrero-Illana, Heyminck, Hirota, Hoge, Schimpf, Howie, Huang, Jiang, Jinchi, John, Kimura, Klein, Kubo, Kuroda, Kwon, Lacasse, Laing, Leitch, Li, Liu, Liu, Lin, Lu, Mac-Auliffe, Martin-Cocher, Matulonis, Maute, Messias,
  Meyer-Zhao, Monta{\~n}a, Montenegro-Montes, Montgomerie, Nolasco, Muders, Nishioka, Norton, Nystrom, Ogawa, Olivares, Oshiro, P{\'e}rez-Beaupuits, Parra, Phillips, Poirier, Pradel, Qiu, Raffin, Rahlin, Ram{\'\i}rez, Ressler, Reynolds, Rodr{\'\i}guez-Montoya, Saez-Madain, Santana, Shaw, Shirkey, Silva, Snow, Sousa, Sridharan, Stahm, Stark, Test, Torstensson, Venegas, Walther, Wei, White, Wieching, Wijnands, Wouterloot, Yu, Yu, \& Zeballos}]{EHT22}
Collaboration, E. H.~T., Akiyama, K., Alberdi, A., {et~al.} 2022, The Astrophysical Journal Letters, 930, L12

\bibitem[{Dixon(1970)}]{Dixon1970}
Dixon, W.~G. 1970, Dynamics of Extended Bodies in General Relativity. I. Momentum and Angular Momentum (Proceedings of the Royal Society of London. Series A, Mathematical and Physical Sciences, Volume 314, Issue 1519, pp. 499-527)

\bibitem[{Eckart \& Genzel(1996)}]{Eckart+96}
Eckart, A. \& Genzel, R. 1996, Nature, 383, 415

\bibitem[{Fabrycky \& Tremaine(2007)}]{FabryckyEtAl2007}
Fabrycky, D. \& Tremaine, S. 2007

\bibitem[{Ghez {et~al.}(2003)Ghez, Duch{\^{e}}ne, Matthews, Hornstein, Tanner, Larkin, Morris, Becklin, Salim, Kremenek, Thompson, Soifer, Neugebauer, \& McLean}]{Ghez+03}
Ghez, A.~M., Duch{\^{e}}ne, G., Matthews, K., {et~al.} 2003, The Astrophysical Journal, 586, L127

\bibitem[{Ghez {et~al.}(1998)Ghez, Klein, Morris, \& Becklin}]{Ghez+98}
Ghez, A.~M., Klein, B.~L., Morris, M., \& Becklin, E.~E. 1998, The Astrophysical Journal, 509, 678

\bibitem[{Ghez {et~al.}(2008)Ghez, Salim, Weinberg, Lu, Do, Dunn, Matthews, Morris, Yelda, Becklin, Kremenek, Milosavljevic, \& Naiman}]{Ghez+08}
Ghez, A.~M., Salim, S., Weinberg, N.~N., {et~al.} 2008, The Astrophysical Journal, 689, 1044

\bibitem[{Gillessen {et~al.}(2009)Gillessen, Eisenhauer, Trippe, Alexander, Genzel, Martins, \& Ott}]{GillessenEtAl2009}
Gillessen, S., Eisenhauer, F., Trippe, S., {et~al.} 2009, The Astrophysical Journal, 692, 1075

\bibitem[{Gillessen {et~al.}(2017)Gillessen, Plewa, Eisenhauer, Sari, Waisberg, Habibi, Pfuhl, George, Dexter, von Fellenberg, Ott, \& Genzel}]{Gillessen+17}
Gillessen, S., Plewa, P.~M., Eisenhauer, F., {et~al.} 2017, The Astrophysical Journal, 837, 30

\bibitem[{Gourgoulhon(Lecture notes)}]{BH_eric}
Gourgoulhon, E. Lecture notes, Geometry and physics of black holes

\bibitem[{{GRAVITY Collaboration} {et~al.}(2024){GRAVITY Collaboration}, {Abd El Dayem, K.}, {Abuter, R.}, {Aimar, N.}, {Amorim, A.}, {Ball, J.}, {Baub\"ock, M.}, {Berger, J. P.}, {Bonnet, H.}, {Bourdarot, G.}, {Brandner, W.}, {Cardoso, V.}, {Cl\'enet, Y.}, {Dallilar, Y.}, {Davies, R.}, {de Zeeuw, P. T.}, {Dexter, J.}, {Drescher, A.}, {Eisenhauer, F.}, {F\"orster Schreiber, N. M.}, {Foschi, A.}, {Garcia, P.}, {Gao, F.}, {Gendron, E.}, {Genzel, R.}, {Gillessen, S.}, {Habibi, M.}, {Haubois, X.}, {Hei\ss{}el, G.}, {Henning, T.}, {Hippler, S.}, {Horrobin, M.}, {Jochum, L.}, {Jocou, L.}, {Kaufer, A.}, {Kervella, P.}, {Lacour, S.}, {Lapeyr\`ere, V.}, {Le Bouquin, J.-B.}, {L\'ena, P.}, {Lutz, D.}, {Ott, T.}, {Paumard, T.}, {Perraut, K.}, {Perrin, G.}, {Pfuhl, O.}, {Rabien, S.}, {Shangguan, J.}, {Shimizu, T.}, {Scheithauer, S.}, {Stadler, J.}, {Stephens, A.W.}, {Straub, O.}, {Straubmeier, C.}, {Sturm, E.}, {Tacconi, L. J.}, {Tristram, K. R. W.}, {Vincent, F.}, {von Fellenberg, S.}, {Widmann, F.}, {Wieprecht, E.},
  {Wiezorrek, E.}, {Woillez, J.}, {Yazici, S.}, \& {Young, A.}}]{GRAVITY+24}
{GRAVITY Collaboration}, {Abd El Dayem, K.}, {Abuter, R.}, {et~al.} 2024, A\&A, 692

\bibitem[{{GRAVITY Collaboration} {et~al.}(2022){GRAVITY Collaboration}, {Abuter, R.}, {Aimar, N.}, {Amorim, A.}, {Ball, J.}, {Baub\"ock, M.}, {Berger, J. P.}, {Bonnet, H.}, {Bourdarot, G.}, {Brandner, W.}, {Cardoso, V.}, {Cl\'enet, Y.}, {Dallilar, Y.}, {Davies, R.}, {de Zeeuw, P. T.}, {Dexter, J.}, {Drescher, A.}, {Eisenhauer, F.}, {F\"orster Schreiber, N. M.}, {Foschi, A.}, {Garcia, P.}, {Gao, F.}, {Gendron, E.}, {Genzel, R.}, {Gillessen, S.}, {Habibi, M.}, {Haubois, X.}, {Hei\ss{}el, G.}, {Henning, T.}, {Hippler, S.}, {Horrobin, M.}, {Jochum, L.}, {Jocou, L.}, {Kaufer, A.}, {Kervella, P.}, {Lacour, S.}, {Lapeyr\`ere, V.}, {Le Bouquin, J.-B.}, {L\'ena, P.}, {Lutz, D.}, {Ott, T.}, {Paumard, T.}, {Perraut, K.}, {Perrin, G.}, {Pfuhl, O.}, {Rabien, S.}, {Shangguan, J.}, {Shimizu, T.}, {Scheithauer, S.}, {Stadler, J.}, {Stephens, A.W.}, {Straub, O.}, {Straubmeier, C.}, {Sturm, E.}, {Tacconi, L. J.}, {Tristram, K. R. W.}, {Vincent, F.}, {von Fellenberg, S.}, {Widmann, F.}, {Wieprecht, E.}, {Wiezorrek, E.},
  {Woillez, J.}, {Yazici, S.}, \& {Young, A.}}]{GRAVITY+22_mass_distribution}
{GRAVITY Collaboration}, {Abuter, R.}, {Aimar, N.}, {et~al.} 2022, A\&A, 657, L12

\bibitem[{{GRAVITY Collaboration} {et~al.}(2023){GRAVITY Collaboration}, {Abuter, R.}, {Alarcon, P.}, {Allouche, F.}and {Amorim, A.}, {Ball, J.}, {Baub\"ock, M.}, {Berger, J. P.}, {Bonnet, H.}, {Bourdarot, G.}, {Brandner, W.}, {Cardoso, V.}, {Cl\'enet, Y.}, {Dallilar, Y.}, {Davies, R.}, {de Zeeuw, P. T.}, {Dexter, J.}, {Drescher, A.}, {Eisenhauer, F.}, {F\"orster Schreiber, N. M.}, {Foschi, A.}, {Garcia, P.}, {Gao, F.}, {Gendron, E.}, {Genzel, R.}, {Gillessen, S.}, {Habibi, M.}, {Haubois, X.}, {Hei\ss{}el, G.}, {Henning, T.}, {Hippler, S.}, {Horrobin, M.}, {Jochum, L.}, {Jocou, L.}, {Kaufer, A.}, {Kervella, P.}, {Lacour, S.}, {Lapeyr\`ere, V.}, {Le Bouquin, J.-B.}, {L\'ena, P.}, {Lutz, D.}, {Ott, T.}, {Paumard, T.}, {Perraut, K.}, {Perrin, G.}, {Pfuhl, O.}, {Rabien, S.}, {Shangguan, J.}, {Shimizu, T.}, {Scheithauer, S.}, {Stadler, J.}, {Stephens, A.W.}, {Straub, O.}, {Straubmeier, C.}, {Sturm, E.}, {Tacconi, L. J.}, {Tristram, K. R. W.}, {Vincent, F.}, {von Fellenberg, S.}, {Widmann, F.}, {Wieprecht, E.},
  {Wiezorrek, E.}, {Woillez, J.}, {Yazici, S.}, \& {Young, A.}}]{GRAVITY+23}
{GRAVITY Collaboration}, {Abuter, R.}, {Alarcon, P.}, {et~al.} 2023, A\&A

\bibitem[{{GRAVITY Collaboration} {et~al.}(2018){GRAVITY Collaboration}, {Abuter, R.}, {Amorim, A.}, {Anugu, N.}, {Baub\"ock, M.}, {Benisty, M.}, {Berger, J. P.}, {Blind, N.}, {Bonnet, H.}, {Brandner, W.}, {Buron, A.}, {Collin, C.}, {Chapron, F.}, {Cl\'enet, Y.}, {dCoud\'e u Foresto, V.}, {de Zeeuw, P. T.}, {Deen, C.}, {Delplancke-Str\"obele, F.}, {Dembet, R.}, {Dexter, J.}, {Duvert, G.}, {Eckart, A.}, {Eisenhauer, F.}, {Finger, G.}, {F\"orster Schreiber, N. M.}, {F\'edou, P.}, {Garcia, P.}, {Garcia Lopez, R.}, {Gao, F.}, {Gendron, E.}, {Genzel, R.}, {Gillessen, S.}, {Gordo, P.}, {Habibi, M.}, {Haubois, X.}, {Haug, M.}, {Hau\ss{}mann, F.}, {Henning, Th.}, {Hippler, S.}, {Horrobin, M.}, {Hubert, Z.}, {Hubin, N.}, {Jimenez Rosales, A.}, {Jochum, L.}, {Jocou, L.}, {Kaufer, A.}, {Kellner, S.}, {Kendrew, S.}, {Kervella, P.}, {Kok, Y.}, {Kulas, M.}, {Lacour, S.}, {Lapeyr\`ere, V.}, {Lazareff, B.}, {Le Bouquin, J.-B.}, {L\'ena, P.}, {Lippa, M.}, {Lenzen, R.}, {M\'erand, A.}, {M\"uler, E.}, {Neumann, U.}, {Ott, T.},
  {Palanca, L.}, {Paumard, T.}, {Pasquini, L.}, {Perraut, K.}, {Perrin, G.}, {Pfuhl, O.}, {Plewa, P. M.}, {Rabien, S.}, {Ram\'{\i}rez, A.}, {Ramos, J.}, {Rau, C.}, {Rodr\'{\i}guez-Coira, G.}, {Rohloff, R.-R.}, {Rousset, G.}, {Sanchez-Bermudez, J.}, {Scheithauer, S.}, {Sch\"oller, M.}, {Schuler, N.}, {Spyromilio, J.}, {Straub, O.}, {Straubmeier, C.}, {Sturm, E.}, {Tacconi, L. J.}, {Tristram, K. R. W.}, {Vincent, F.}, {von Fellenberg, S.}, {Wank, I.}, {Waisberg, I.}, {Widmann, F.}, {Wieprecht, E.}, {Wiest, M.}, {Wiezorrek, E.}, {Woillez, J.}, {Yazici, S.}, {Ziegler, D.}, \& {Zins, G.}}]{GRAVITY+18_redshift}
{GRAVITY Collaboration}, {Abuter, R.}, {Amorim, A.}, {et~al.} 2018, A\&A, 615, L15

\bibitem[{{GRAVITY Collaboration} {et~al.}(2020){GRAVITY Collaboration}, {Abuter, R.}, {Amorim, A.}, {Baub\"ock, M.}, {Berger, J. P.}, {Bonnet, H.}, {Brandner, W.}, {Cardoso, V.}, {Cl\'enet, Y.}, {de Zeeuw, P. T.}, {Dexter, J.}, {Eckart, A.}, {Eisenhauer, F.}, {F\"orster Schreiber, N. M.}, {Garcia, P.}, {Gao, F.}, {Gendron, E.}, {Genzel, R.}, {Gillessen, S.}, {Habibi, M.}, {Haubois, X.}, {Henning, T.}, {Hippler, S.}, {Horrobin, M.}, {Jim\'enez-Rosales, A.}, {Jochum, L.}, {Jocou, L.}, {Kaufer, A.}, {Kervella, P.}, {Lacour, S.}, {Lapeyr\`ere, V.}, {Le Bouquin, J.-B.}, {L\'ena, P.}, {Nowak, M.}, {Ott, T.}, {Paumard, T.}, {Perraut, K.}, {Perrin, G.}, {Pfuhl, O.}, {Rodr\'{\i}guez-Coira, G.}, {Shangguan, J.}, {Scheithauer, S.}, {Stadler, J.}, {Straub, O.}, {Straubmeier, C.}, {Sturm, E.}, {Tacconi, L. J.}, {Vincent, F.}, {von Fellenberg, S.}, {Waisberg, I.}, {Widmann, F.}, {Wieprecht, E.}, {Wiezorrek, E.}, {Woillez, J.}, {Yazici, S.}, \& {Zins, G.}}]{GRAVITY+20_Schwarzschild_prec}
{GRAVITY Collaboration}, {Abuter, R.}, {Amorim, A.}, {et~al.} 2020, A\&A, 636, L5

\bibitem[{{Grould} {et~al.}(2017){Grould}, {Vincent}, {Paumard}, \& {Perrin}}]{Grould+17}
{Grould}, M., {Vincent}, F.~H., {Paumard}, T., \& {Perrin}, G. 2017, A\&A, 608, A60

\bibitem[{{Hei\ss{}el} {et~al.}(2022){Hei\ss{}el}, {Paumard}, {Perrin}, \& {Vincent}}]{Heissel+22}
{Hei\ss{}el}, G., {Paumard}, T., {Perrin}, G., \& {Vincent}, F. 2022, A\&A, 660, A13

\bibitem[{{Hei\ss{}el} {et~al.}(2025){Hei\ss{}el}, {Vincent}, {Abd El Dayem}, {Paumard}, \& {Perrin}}]{Heissel+25}
{Hei\ss{}el}, G., {Vincent}, F., {Abd El Dayem}, K., {Paumard}, T., \& {Perrin}, G. 2025

\bibitem[{Kocsis \& Tremaine(2011)}]{KocsisEtAl2011}
Kocsis, B. \& Tremaine, S. 2011

\bibitem[{Kocsis \& Tremaine(2015)}]{KocsisEtAl2015}
Kocsis, B. \& Tremaine, S. 2015

\bibitem[{{Kravchenko, K.} {et~al.}(2023){Kravchenko, K.}, {Dallilar, Y.}, {Absil, O.}, {Agudo Berbel, A.}, {Baruffolo, A.}, {Bonse, M. J.}, {Buron, A.}, {Cao, Y.}, {Cortes, A.}, {Dannert, F.}, {Davies, R.}, {De Rosa, R. J.}, {Deysenroth, M.}, {Doelman, D. S.}, {Eisenhauer, F.}, {Esposito, S.}, {Feuchtgruber, H.}, {Förster Schreiber, N.}, {Gao, X.}, {Gemperlein, H.}, {Genzel, R.}, {Gillessen, S.}, {Ginski, C.}, {Glauser, A. M.}, {Glindemann, A.}, {Grani, P.}, {Haguenauer, P.}, {Hartwig, J.}, {Hayoz, J.}, {Heida, M.}, {Kenworthy, M.}, {Kolb, J.}, {Kuntschner, H.}, {Lutz, D.}, {Liu, D.}, {MacIntosh, M.}, {Marsset, M.}, {Orban de Xivry, G.}, {Özdemir, H.}, {Puglisi, A.}, {Quanz, S. P.}, {Rau, C.}, {Riccardi, A.}, {Schuppe, D.}, {Snik, F.}, {Sturm, E.}, {Tacconi, L.}, {Taylor, W. D.}, \& {Wiezorrek, E.}}]{ERIS}
{Kravchenko, K.}, {Dallilar, Y.}, {Absil, O.}, {et~al.} 2023

\bibitem[{{Krishnendu} {et~al.}(2019){Krishnendu}, {Saleem}, {Samajdar}, {Arun}, {Del Pozzo}, \& {Chandra Kant Mishra}}]{Krishnendu2019}
{Krishnendu}, {Saleem}, {Samajdar}, {et~al.} 2019

\bibitem[{Merritt(2013)}]{Merritt2013}
Merritt, D. 2013, Dynamics and evolution of galactic nuclei, Princeton series in astrophysics (Princeton University Press)

\bibitem[{Merritt {et~al.}(2010)Merritt, Alexander, Mikkola, \& Will}]{MerrittEtAl2010}
Merritt, D., Alexander, T., Mikkola, S., \& Will, C.~M. 2010, Phys. Rev. D, 81, 062002

\bibitem[{{Poisson}(1998)}]{Poisson1998}
{Poisson}, E. 1998

\bibitem[{Poisson \& Will(2014)}]{PoissonWill2014}
Poisson, E. \& Will, C.~M. 2014, Gravity: Newtonian, Post-Newtonian, Relativistic (Cambridge University Press)

\bibitem[{Psaltis {et~al.}(2013)Psaltis, Gongjie, \& Abraham}]{PsaltisEtAl2013}
Psaltis, D., Gongjie, L., \& Abraham, L. 2013

\bibitem[{Sch{\"o}del {et~al.}(2002)Sch{\"o}del, Ott, Genzel, Hofmann, Lehnert, Eckart, Mouawad, Alexander, Reid, Lenzen, Hartung, Lacombe, Rouan, Gendron, Rousset, Lagrange, Brandner, Ageorges, Lidman, Moorwood, Spyromilio, Hubin, \& Menten}]{Schoedel+02}
Sch{\"o}del, R., Ott, T., Genzel, R., {et~al.} 2002, Nature, 419, 694

\bibitem[{{Tucker} \& {Will}(2019)}]{Tucker2019}
{Tucker}, A. \& {Will}, C. 2019

\bibitem[{Waisberg {et~al.}(2018)Waisberg, Dexter, Gillessen, Pfuhl, Eisenhauer, Plewa, Baub{\"o}ck, Jimenez-Rosales, Habibi, Ott, von Fellenberg, Gao, Widmann, \& Genzel}]{Waisberg+18}
Waisberg, I., Dexter, J., Gillessen, S., {et~al.} 2018, Monthly Notices of the Royal Astronomical Society, 476, 3600

\bibitem[{Wald(1974)}]{Wald_1974}
Wald, R. 1974, Phys. Rev. D

\bibitem[{{Will} \& {Maitra}(2016)}]{WillMaitra2016}
{Will}, C. \& {Maitra}, M. 2016

\bibitem[{{Will} {et~al.}(2023){Will}, {Naoz, S.}, {Hees, A.}, {Tucker, A.}, {Zhang, E.}, {Do, T.}, \& {Andrea, G.}}]{Will+23}
{Will}, C., {Naoz, S.}, {Hees, A.}, {et~al.} 2023

\bibitem[{Will(2008)}]{Will2008}
Will, C.~M. 2008, The Astrophysical Journal, 674, L25

\bibitem[{Will(2018)}]{Will2018}
Will, C.~M. 2018, Theory and experiment in Gravitational physics, second edition (Cambridge University Press)

\bibitem[{Yu {et~al.}(2016)Yu, Zhang, \& Lu}]{Yu+16}
Yu, Q., Zhang, F., \& Lu, Y. 2016, The Astrophysical Journal, 827, 114

\end{thebibliography}
\bibliographystyle{aa}

\begin{appendix}

\section{Orbit parametrization}\label{sec3}

Let us present the different ways to parameterize our orbits.

\subsection{Newtonian orbits}\label{sec3.1}

A Newtonian orbit is defined by 6 Keplerian parameters that correspond to the 6 Newtonian degrees of freedom (3 positions + 3 velocities). 
The typical set of Keplerian parameters is the following:
\begin{list}{$\circ$}{}  
\item $a_{\mrm{sma}}$ the semi-major axis;
\item $e$ the eccentricity of the orbit;
\item $\iota \in [0;\pi]$ is the inclination\footnote{Note that the inclination $\iota$ we use here is different from the definition of the inclination "$i$" used in \citet{GRAVITY+20_Schwarzschild_prec} which used the opposite convention for the sense of rotation of the inclination (see Appendix C of \citet{Heissel+22}).} between the fundamental and orbital planes, and encodes the rotation about $\mrm{\mbf{\mathcal{L}_\mbf{orb//sky}}}$, from $\mbf Z$ to $\mbf z_{\mathrm{orb}}$;
\item $\Omega \in [0;2\pi]$ is the position angle of the line of nodes $\mrm{\mbf{\mathcal{L}_\mbf{orb//sky}}}$, intersection between the orbital and fundamental planes; it encodes the rotation about $\mbf Z$, from $\mbf X$ to $\mrm{\mbf{\mathcal{L}_\mbf{orb//sky}}}$;     
\item $\omega \in [0;2\pi]$ is the angular position of the pericenter within the orbital plane, counted from the line of nodes; it encodes the rotation about $\mbf z_{\mathrm{orb}}$, from $\mrm{\mbf{\mathcal{L}_\mbf{orb//sky}}}$ to $\mbf x_{\mathrm{orb}}$;
\item $t_{\mathrm{peri}}$ the time of pericenter passage.
\end{list}
The first two parameters are sufficient to define (i) geometrically the ellipse of the orbit (that is, the semi-major and semi-minor axes), and (ii) the energetics of the orbit, that is, the total energy $E$ and angular momentum $L$ of the two-body system. There is a one-to-one correspondence between $(a_{\mrm{sma}}, e)$ and $(E, L)$. Note that these parameters are independent of the choice of fundamental frame. 
In addition, we defined\footnote{In this paper, all angles are defined according to the right-hand rule. Namely, they have positive values when they represent a rotation that appears counterclockwise when looking in the negative direction of the axis, and negative values when the rotation appears clockwise.} the three angles $\iota$, $\Omega$ and $\omega$ (see figure \ref{frame_orb}) that encode the orientation of the ellipse; they are the three Euler angles, allowing to rotate the orbit frame into the observer’s frame. We highlight this dependency of the 3 angular orbital parameters on the choice of fundamental plane. For instance, \citet{WillMaitra2016} use the black hole equatorial plane as their fundamental plane, while we use the plane of the sky. As a consequence, the inclination angle of our work is not the same as in theirs, and we will see that this has obviously profound consequences on the evolution equations that we will obtain later on. 
Moreover, we need the initial condition along the orbit, so we introduce $t_{\mathrm{peri}}$ the time of pericenter passage.
Finally, to characterize the position of the star along the orbit, we use the true anomaly $f$ (seefigure \ref{frame_orb}) or the azimuthal angle $\varphi=\omega+f$.

\subsection{Relativistic orbits}\label{sec3.2}

For orbits in general relativity (GR) or post-Newtonian (PN) theories, the orbital elements become time-varying. The values of the orbital parameters at a particular date can thus be seen as the orbital parameters of a Newtonian orbit, osculating the relativistic orbit at this particular date. It is customary to represent a relativistic orbit by providing the set of orbital parameters at a particular time $t_{\mathrm{osc}}$, when it is osculated by the Newtonian orbit described by this same set of orbital parameters. A relativistic orbit is thus described by the 6 usual orbital parameters, plus the osculating time $t_{\mathrm{osc}}$. The time-varying orbital parameters of a relativistic orbit can therefore be seen as the orbital parameters of a family of osculating Newtonian elliptical orbits. From this point onward, we will consider ($a_{\mrm{sma}}$, $e$, $\iota$, $\Omega$, $\omega$) as functions of time such that ($a_{\mrm{sma}}(t_{\mathrm{osc}})$, $e(t_{\mathrm{osc}})$, $\iota(t_{\mathrm{osc}})$, $\Omega(t_{\mathrm{osc}})$, $\omega(t_{\mathrm{osc}})$)=($a_{\mrm{sma,\,osc}}$, $e_{\mathrm{osc}}$, $\iota_{\mathrm{osc}}$, $\Omega_{\mathrm{osc}}$, $\omega_{\mathrm{osc}}$).

As for the osculating $t_{\mathrm{peri}}$, we take the most recent time of pericenter passage before the first observation date of each star \citep[see table D.1 of][for the first observation dates]{GRAVITY+22_mass_distribution}.

Moreover, it is worth mentioning that adding a certain time $T$ to $t_{\mathrm{osc}}$ does not mean that the orbit will be temporally translated by $T$; taking $t_{\mathrm{osc}}=t_0+T$ instead of $t_{\mathrm{osc}}=t_0$ does not mean that we will observe the same orbit if we wait for $T$. This would only be true if we also add $T$ to $t_{\mathrm{peri}}$ as well. Since it is common not to find in literature\footnote{When not mentioned, the osculation times of S stars actually correspond to the time of the most recent apocenter before the first observation date of each star, as mentioned in \citet{GRAVITY+18_redshift, GRAVITY+20_Schwarzschild_prec}.} any mention of the osculating time when giving the numerical values of the orbital parameters of S stars, one can misinterpret the true values of these parameters \citep[see][in preparation]{Heissel+25}.

Finally, it is very important to realize that the same set of initial conditions will yield different GR orbits when using different coordinate systems. This is well known in the context of relativity theory and stems from the fact that orbital elements are not covariant quantities \citep{Brumberg}. Therefore, a relativistic orbit should be described by a set of 6 orbital parameters, plus the osculating time, plus the mention of the coordinate system used in the integration. This information is necessary and sufficient for uniquely pinpointing the orbit \citep[see][in preparation]{Heissel+25}.

\section{Details on the frames of reference}\label{app_frames}

In this appendix, we gather some details about the frames of reference of section \ref{sec2}, that are important for the comprehension and reproducibility of our results.
First, concerning the orbit frame represented figure \ref{frame_orb}, we define the line of nodes as the intersection between the orbit and sky planes, and its director vector $\mrm{\mbf{\mathcal{L}_\mbf{orb//sky}}}$ goes from the black hole to $\mathcal{AN}_{\mrm{orb//sky}}$, the ascending node of the orbit with respect to the observer. The latter is the intersection between the sky plane and trajectory where the star passes from behind the sky plane ($Z<0$) to the front ($Z>0$).

\begin{figure}[htp]
    \centering
    \includegraphics[width=0.9\columnwidth]{ 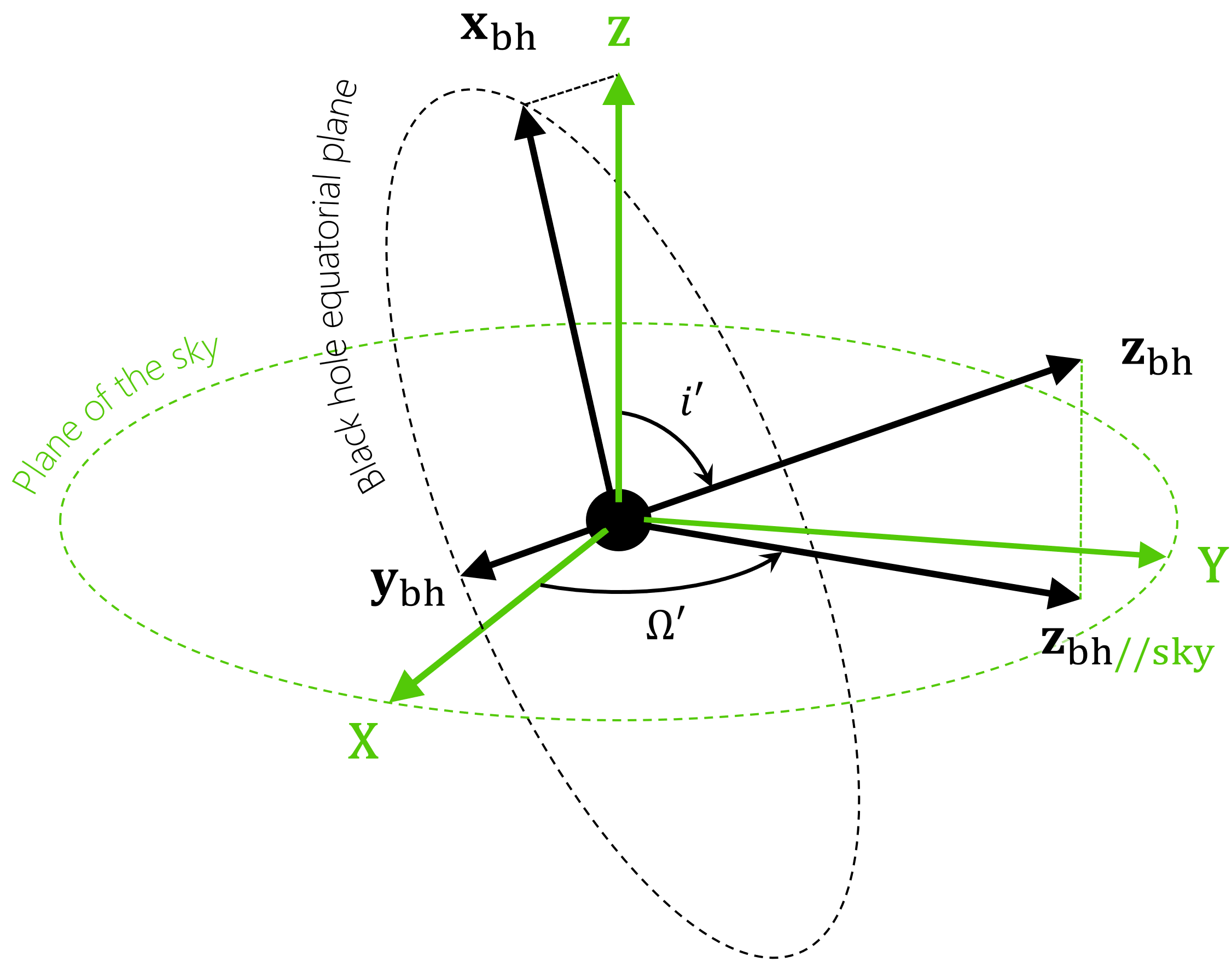}
    \caption{\label{frame_bh_sky} The angles $(i',\Omega')$ in black characterize the spin orientation of Sgr A* and give the position of the black-hole frame relative to the observer frame.}
\end{figure}

Second, concerning the black hole frame, we mentioned that the choice of the orientation of $\mbf x_{\mrm{bh}}$ and $\mbf y_{\mrm{bh}}$ inside the black hole equatorial plane has no impact on the physics of the problem. 

We have defined in figure \ref{frame_bh_orb} the orientation of $\mbf z_{\mathrm{bh}}$ in the orbit frame by introducing the angles $(\theta,\beta)$. These angles are very useful for simplifying the computations of the quantities appearing in Lagrange planetary equations. Here we introduce an alternative definition with the angles\footnote{In \citet{EHT22}, they refer to the spin parameter as "$a_*$" and the black hole inclination with respect to the line of sight as "$i$". They use the following convention $a_* \in [-1;1]$ and $i\in[0;\pi/2]$ instead of ours, which aligns with the work of \citet{Grould+17}.} $(i',\Omega')$ defined in figure \ref{frame_bh_sky}. These angles are more adapted when dealing with multi-star fit because they allow us to characterize the orientation of the black hole with respect to the observer we use the angles $(i',\Omega')$. This is particularly useful in the case of studying the spin of SgrA*; multiple stars would be used when constraining the spin and its orientation, meaning that defining the latter with respect to the orbit of one single star instead of the sky is unpractical. 
On the other hand, for the calculations of section \ref{sec4.2}, the use of $(\theta,\beta)$, which define the orientation of the spin vector relative to the star's orbit and $\mrm{\mbf{\mathcal{L}_\mbf{orb//sky}}}$, will be more practical. 
This allows us to define the black hole's angles with respect to the sky plane as follows:
\begin{list}{$\circ$}{}  
\item $i' \in [0;\pi]$ is the rotation angle from $\mbf Z$ to $\mbf z_{\mathrm{bh}}$;
\item $\Omega' \in [0;2\pi]$ is the rotation angle about $\mbf Z$, from $\mbf X$ to $\mbf z_{\mathrm{bh//sky}}$.
\end{list}

Third, we introduced in figure \ref{frame_orb} the Gaussian frame $(\mbf n_{\mrm{orb}},\mbf m_{\mrm{orb}},\mbf z_{\mrm{orb}})$ \footnote{Also denoted as $(\vec n,\vec m, \vec k)$ in Eq. (4.55) of \citet{Merritt2013}.}. This allows us to write the star's velocity vector $\vec v$ of norm $v$ as: 
\begin{align}\label{}
    \vec v=v_r \mbf n_{\mrm{orb}}+ v_t \mbf m_{\mrm{orb}},
\end{align}
with: 
\begin{align}\label{}
    v_r &=\sqrt{\frac{Gm}{p}}e\sin f ; \quad v_t =\sqrt{\frac{Gm}{p}}(1+e\cos f).  
\end{align}
This can be done, even in the presence of an "out-of-plane precession" (see section \ref{sec4}) since at any time, we can consider the osculating Keplerian orbit (see section \ref{sec4}) which contains the velocity of the star at that time. In other words, the velocity is always within the momentarily orbital plane, and thus never has a $\mbf z_{\mrm{orb}}$-component.
In addition to the true anomaly $f$ introduced in section \ref{sec2}, it is also useful to define the angle $\varphi=\omega+\varphi\in [0;2\pi]$, defined as the rotation angle about $\mbf z_{\mathrm{orb}}$, from $\mrm{\mbf{\mathcal{L}_\mbf{orb//sky}}}$ to $\mbf n_{\mathrm{orb}}$, and characterizing the position of the star along the orbit relative to the ascending note (see figure \ref{frame_orb}).
We know from Eq. (4.55) of \citet{Merritt2013} that by defining this quantity we can write: 
\begin{equation}
    \begin{aligned} 
    \mbf n_{\mrm{orb}} &= (\cos\varphi\cos\Omega-\cos \iota\sin\varphi\sin\Omega)\;\mbf X \\
    &\quad+ (\cos\varphi\sin\Omega+\cos \iota\sin\varphi\cos\Omega)\mbf Y \\
    &\quad+ \sin \iota\sin\varphi\;\mbf Z ;\\
    \mbf m_{\mrm{orb}} &= (-\sin\varphi\cos\Omega-\cos \iota\cos\varphi\sin\Omega)\;\mbf X \\
    &\quad+ (-\sin\varphi\sin\Omega+\cos \iota\cos\varphi\cos\Omega)\;\mbf Y \\
    &\quad+ \sin \iota\cos\varphi\;\mbf Z ;\\
    \mbf z_{\mrm{orb}} &= \sin \iota\sin\Omega\;\mbf X \\
    &\quad -\sin \iota\cos\Omega\;\mbf Y \\
    &\quad+ \cos \iota\;\mbf Z .\\
    \end{aligned}
\end{equation}
Similarly to the expression of $\mbf z_{\mrm{orb}}$, we can write $\mbf z_{\mrm{bh}}$ in the fundamental frame of the observer as:
\begin{equation}
    \begin{aligned} 
    \mbf z_{\mrm{bh}} &= \sin i'\cos\Omega'\;\mbf X  +\sin i'\sin\Omega'\;\mbf Y + \cos i'\;\mbf Z .\\
    \end{aligned}
\end{equation}
Therefore, we obtain: 
\begin{equation}
    \begin{aligned} 
    \mbf z_{\mrm{bh}}\cdot\mbf n_{\mrm{orb}} &= \cos i' \sin\varphi \sin\iota \\
    &\quad+ \sin i'\left( \cos\varphi \cos(\Omega-\Omega') - \sin\varphi\cos\iota\sin(\Omega-\Omega') \right)\\
    &=\sin\theta \cos(\beta-\varphi);\\ 
    \mbf z_{\mrm{bh}}\cdot\mbf m_{\mrm{orb}} &= \cos i' \cos\varphi \sin\iota \\
    &\quad -\sin i'\big( \sin\varphi \cos(\Omega-\Omega') +  \cos\varphi \cos\iota\sin(\Omega-\Omega') \big)\\
    &=\sin\theta \sin(\beta-\varphi);\\ 
    \mbf z_{\mrm{bh}}\cdot\mbf z_{\mrm{orb}} &= \cos i'\cos \iota+\sin i'\sin \iota\sin(\Omega-\Omega')\\
    &=\cos\theta .
    \end{aligned}
\end{equation}    
which will become very handy for the derivation of Eqs. \eqref{R_sch} to \eqref{W_Q}.

\section{Perturbations in the Gaussian frame}\label{RSW}

Let us express\footnote{The use of $\theta$ and $\beta-\varphi$ greatly simplifies the scalar products as detailed in the end of appendix \ref{app_frames}.} the components of the various PN accelerations in the Gaussian frame $(\mbf n_{\mrm{orb}},\mbf m_{\mrm{orb}},\mbf z_{\mrm{orb}})$:

\begin{align}    
    \begin{split}\label{R_sch}
    &\mathcal{R}_{\mrm{Sch}}^{\mrm{2PN}} = \mbf a_{\mrm{Sch}}^{\mrm{2PN}}\cdot\mbf n_{\mrm{orb}} \\
    &\quad\quad = \epsilon\small\frac{G m}{p^2} (1+e\cos f)^2 \left(3(e^2+1)+2e\cos f-4e^2\cos^2 f \right) \normalsize \\  
    & \quad\quad\quad\;\; -9\epsilon^2\frac{Gm}{p^2}(1+e\cos f)^4;
    \end{split}\\
    \label{R_chi}
    &\mathcal{R}_\mrm{LT}^{\mrm{1.5PN}} = \mbf a_\mrm{LT}^{\mrm{1.5PN}}\cdot\mbf n_{\mrm{orb}} = 2\epsilon^{3/2}\frac{Gm}{p^2} \chi(1+e\cos f)^4\cos\theta  ;\\
    \begin{split}\label{R_Q}
    &\mathcal{R}_Q^{\mrm{2PN}} =  \mbf a_Q^{\mrm{2PN}}\cdot\mbf n_{\mrm{orb}} \\
    &\quad\quad= -\frac{3}{2}\epsilon^2\frac{Gm}{p^2}\chi^2 (1+e\cos f)^4(1-3 \sin^2\theta\cos^2 (\beta-\varphi));
    \end{split}
\end{align}
\begin{align}
    \begin{split}\label{S_sch}
    &\mathcal{S}_{\mrm{Sch}}^{\mrm{2PN}} = \mbf a_{\mrm{Sch}}^{\mrm{2PN}}\cdot\mbf m_{\mrm{orb}} \\  
    & \quad\quad = 4\epsilon\frac{G m}{p^2} (1+e\cos f)^3 e\sin f -2\epsilon^2\frac{Gm}{p^2}e(1+e\cos f)^4\sin f;
    \end{split}\\
    \label{S_chi}
    &\mathcal{S}_\mrm{LT}^{\mrm{1.5PN}} = \mbf a_\mrm{LT}^{\mrm{1.5PN}}\cdot\mbf m_{\mrm{orb}} = - 2\epsilon^{3/2}\frac{Gm}{p^2}\chi e \sin f(1+e\cos f)^3\cos\theta  ; \\  
    \begin{split}\label{S_Q}
     &\mathcal{S}_Q^{\mrm{2PN}} =  \mbf a_Q^{\mrm{2PN}}\cdot\mbf m_{\mrm{orb}} \\
    &\quad\quad= -3\epsilon^2\frac{Gm}{p^2}\chi^2 (1+e\cos f)^4 \sin^2\theta\cos(\beta-\varphi)\sin(\beta-\varphi) ;
    \end{split}
\end{align}
\begin{align}
    &\mathcal{W}_{\mrm{Sch}}^{\mrm{2PN}} = \mbf a_{\mrm{Sch}}^{\mrm{2PN}}\cdot\mbf z_{\mrm{orb}} = 0 ;\label{W_sch}
\end{align}
\begin{align}
    \begin{split}
    &\mathcal{W}_\mrm{LT}^{\mrm{1.5PN}} = \mbf a_\mrm{LT}^{\mrm{1.5PN}}\cdot\mbf z_{\mrm{orb}} \\     
    &\quad\quad\quad\;= 2\epsilon^{3/2}\frac{Gm}{p^2} \chi(1+e\cos f)^3\sin\theta\\ 
    & \quad\quad\quad\;\; \times\left[ 2(1+e\cos f)\cos(\beta-\varphi) +e\sin f\sin(\beta-\varphi)  \right];
    \end{split} \label{W_chi}\\
    \begin{split}
    &\mathcal{W}_Q^{\mrm{2PN}} =  \mbf a_Q^{\mrm{2PN}}\cdot\mbf z_{\mrm{orb}} \\
    &\;\;\;\quad\quad= -3\epsilon^2\frac{Gm}{p^2}\chi^2 (1+e\cos f)^4\cos\theta \sin\theta\cos(\beta-\varphi).   \label{W_Q}
    \end{split}
\end{align}

\section{Secular evolution in the PN formalism and two-timescale analysis}\label{app2timescale}

It is possible to derive the secular evolution of the orbital parameters at an arbitrary PN order using a perturbation method called two-timescale analysis. This section tries to pedagogically
introduce the material presented in \citet{WillMaitra2016, Tucker2019}, while taking the orbital elements as the ones defined with respect to observer's frame and not the black hole's frame\footnote{Be aware that in \citet{WillMaitra2016, Tucker2019} the quantities "$\iota$", "$\omega$" and "$\Omega$" do not share the same definition as in this paper, in \citet{Will2008, Will2018}, or in \citet{Alush+22}. For example, their inclination "$\iota$" corresponds to $\theta$ in this paper and not $\iota$.}; this allows us to obtain secular shifts of quantities that are directly observable. 

\subsection{Two-timescale method}

Let us label generically all orbital elements by $X_\alpha$, where $\alpha$ is an index, running typically from 1 to 5 to label the 5 standard orbital elements $(p,e,\iota,\Omega,\omega)$.
We will use here the orbital plane azimuthal angle $\varphi=\omega+f$ as the parameter, rather than time. So we consider functions $X_\alpha(\varphi)$. Equations \eqref{dpdt} to \eqref{dwdt} give the time evolution $\mrm{d} X_\alpha / \mrm{d}t$. By using Eq. \eqref{dfdt} we can easily get:
\begin{align}\label{dvarphidt}
    \begin{split}
    \frac{\mrm{d}\varphi}{\mrm{d}t} &= \frac{\mrm{d}\omega}{\mrm{d}t} + \frac{\mrm{d}f}{\mrm{d}t} \\
    &= \left(\frac{\mrm{d}f}{\mrm{d}t}\right)_{\mrm{Kepler}}- \cos\iota \frac{\mrm{d}\Omega}{\mrm{d}t} \\
    &= \sqrt{\frac{Gm}{p^3}}(1+e\cos f)^2 - \sqrt{\frac{p}{Gm}} \cot\iota\frac{\sin\varphi}{1+e\cos f} \mathcal{W},
    \end{split}
\end{align}
and convert $\mrm{d} X_\alpha / \mrm{d} t$ to $\mrm{d}X_\alpha /\mrm{d} \varphi$.
The expressions of the various perturbing accelerations up to 2PN order are given by Eqs. \eqref{R_sch} to \eqref{W_Q}.
Using all this material, the evolution of the orbital elements will finally read
\be
\frac{\mrm{d} X_\alpha}{\mrm{d} \varphi} = \tilde{\epsilon} \, Q_\alpha^{(1)} + \tilde{\epsilon}^{3/2} \, Q_\alpha^{(3/2)} + \tilde{\epsilon}^2 \, Q_\alpha^{(2)},
\ee 

where we express the PN small parameter\footnote{Not to be confused with the varying $\epsilon$.} $\tilde{\epsilon}$ as:  
\be
\label{eq:epsPN_cst}
\tilde{\epsilon} = \frac{GM}{\tilde{p}c^2}.
\ee
Here, we use an orbit-averaged value of the semilatus rectum, $\tilde{p}$, which is \textit{constant} in the evolution. 
The functions $\tilde{\epsilon} Q_\alpha^{(1)}$, $\tilde{\epsilon}^{3/2} \, Q_\alpha^{(3/2)}$, and $\tilde{\epsilon}^2 Q_\alpha^{(2)}$, denote the 1PN, 1.5PN, and 2PN components of the evolution equation, respectively, and the $ Q_\alpha^{(i)}$ are by definition 0PN quantities.

The evolution of the orbital parameters under these PN components will take the form of a long-term (compared to the orbital period) secular evolution, on top of which there are quickly-varying (at the orbital timescale) oscillations. Such a behavior can be well captured in the framework of a two-timescale analysis. Based on this understanding of what the solution should look like, we will consider that $X_\alpha$ is no longer a function $X_\alpha(\varphi)$ of the orbital plane azimuthal angle, but rather a function of two variables, $X_\alpha(\Phi,\varphi)$, where $\Phi = \tilde{\epsilon} \varphi$ captures the long-term behavior of the element, while $\varphi$ captures the quickly-varying oscillations. We will use the trick of considering that $\varphi$ and $\Phi$ are independent variables (which is only a trick, they are actually directly related by $\Phi = \tilde{\epsilon} \varphi$), and derive evolution equations for $X_\alpha$, now considered as a function of two independent variables $\Phi$ and $\varphi$.

Let us rewrite with more details the evolution of $X_\alpha$, which now reads:
\begin{align}
\begin{split}
    \label{eq:evol_Xa}
\frac{\mrm{d} X_\alpha (\Phi,\varphi)}{\mrm{d} \varphi} &= \tilde{\epsilon} \, Q_\alpha^{(1)}\left(X_\beta (\Phi,\varphi);\varphi \right) + \tilde{\epsilon}^{3/2} \, Q_\alpha^{(3/2)}\left(X_\beta (\Phi,\varphi);\varphi \right) \\
&\quad + \tilde{\epsilon}^2 \, Q_\alpha^{(2)}(X_\beta (\Phi,\varphi);\varphi),
\end{split}
\end{align}

where we remind that $\tilde{\epsilon}$ does not depend on any variable quantity, and $Q_\alpha^{(1)}$, $Q_\alpha^{(3/2)}$ and $Q_\alpha^{(2)}$ now depend on the various orbital parameters considered as functions of two variables, $X_\beta (\Phi,\varphi)$, and explicitly also on $\varphi$. 
We will now look for a solution that reads:
\begin{align}
\begin{split}
X_\alpha (\Phi,\varphi) &= \tilde{X}_\alpha(\Phi) + \tilde{\epsilon} \, Y^{(1)}_\alpha(\tilde{X}_\beta(\Phi);\varphi) + \tilde{\epsilon}^{3/2} \, Y^{(3/2)}_\alpha(\tilde{X}_\beta(\Phi);\varphi)  \\
&\quad+ \tilde{\epsilon}^2 \, Y^{(2)}_\alpha(\tilde{X}_\beta(\Phi);\varphi),
\end{split}
\end{align}
where $\tilde{X}_\alpha(\Phi)$ captures the long-term, secular evolution of the orbital element, while $Y^{(i)}_\alpha$ capture the quickly-varying evolution due to the different perturbing accelerations, the upper index $i$ reminding the PN order of the small oscillating correction to the secular evolution. 
We will assume that these $Y^{(i)}_\alpha$ are $2\pi$-periodic in $\varphi$.  

There is not unicity of this ansatz of solution but it is natural to consider:
\be \label{ansatz}
\tilde{X}_\alpha(\Phi) = \left\langle X_\alpha (\Phi,\varphi) \right\rangle, \qquad \left\langle Y^{(i)}_\alpha(\tilde{X}_\beta(\Phi),\varphi) \right\rangle = 0,
\ee
where
\be 
\left\langle F(\Phi,\varphi) \right\rangle = \frac{1}{2\pi} \int_0^{2\pi} F(\Phi,\varphi) d \varphi, \label{phi_avg}
\ee
is the orbit average of some function $F(\Phi,\varphi)$, where we consider that $\Phi$ remains constant in the average, and that the result does not depend on the value of $\Phi$. So we consider that $\tilde{X}_\alpha$ is simply the orbit average of ${X}_\alpha$, and that the oscillating term $Y^{(i)}_\alpha$ averages to zero over one orbit. 
Now that we have this ansatz of solution, let us reexpress the evolution of the orbital parameter $X_\alpha$. We write the derivative with respect to $\Phi$ as:
\be
\frac{\mrm{d}}{\mrm{d} \varphi}= \tilde{\epsilon} \frac{\partial}{\partial \Phi} + \frac{\partial}{\partial \varphi},
\ee
and Eq. \eqref{eq:evol_Xa} becomes:
\begin{align}
\begin{split}  
\frac{\mathrm{d} X_\alpha}{\mathrm{d} \varphi} &= \frac{\mathrm{d} \tilde{X}_\alpha}{\mathrm{d} \Phi} \frac{\mathrm{d} \Phi}{\mathrm{d} \varphi} \\ 
& + \tilde{\epsilon} \sum_\beta \left(\frac{\partial Y^{(1)}_\alpha}{\partial \tilde{X}_\beta} \frac{\mathrm{d} \tilde{X}_\beta}{\mathrm{d} \Phi} \frac{\mathrm{d} \Phi}{\mathrm{d} \varphi}\right) + \tilde{\epsilon} \frac{\partial Y^{(1)}_\alpha}{\partial \varphi} \\ 
& + \tilde{\epsilon}^{3/2} \sum_\beta \left(\frac{\partial Y^{(3/2)}_\alpha}{\partial \tilde{X}_\beta} \frac{\mathrm{d} \tilde{X}_\beta}{\mathrm{d} \Phi} \frac{\mathrm{d} \Phi}{\mathrm{d} \varphi}\right) + \tilde{\epsilon}^{3/2} \frac{\partial Y^{(3/2)}_\alpha}{\partial \varphi} \\ 
& + \tilde{\epsilon}^2 \sum_\beta \left(\frac{\partial Y^{(2)}_\alpha}{\partial \tilde{X}_\beta} \frac{\mathrm{d} \tilde{X}_\beta}{\mathrm{d} \Phi} \frac{\mathrm{d} \Phi}{\mathrm{d} \varphi}\right) + \tilde{\epsilon}^2 \frac{\partial Y^{(2)}_\alpha}{\partial \varphi} \\ 
&= \tilde{\epsilon} \, Q_\alpha^{(1)} + \tilde{\epsilon}^{3/2} \, Q_\alpha^{(3/2)} + \tilde{\epsilon}^2 \, Q_\alpha^{(2)}. \\ 
\end{split}
\end{align} 

Using $\mrm{d}\Phi / \mrm{d}\varphi = \tilde{\epsilon}$ and dividing by $\tilde{\epsilon}$, we get:
\begin{align}
\begin{split}
\frac{\mathrm{d} \tilde{X}_\alpha}{\mathrm{d} \Phi} &= Q_\alpha^{(1)} + \tilde{\epsilon}^{1/2} \, Q_\alpha^{(3/2)} + \tilde{\epsilon} \, Q_\alpha^{(2)} \\ 
&- \tilde{\epsilon} \, \sum_\beta \left(\frac{\partial Y^{(1)}_\alpha}{\partial \tilde{X}_\beta} \frac{\mathrm{d} \tilde{X}_\beta}{\mathrm{d} \Phi}\right) - \tilde{\epsilon}^{3/2} \, \sum_\beta \left(\frac{\partial Y^{(3/2)}_\alpha}{\partial \tilde{X}_\beta} \frac{\mathrm{d} \tilde{X}_\beta}{\mathrm{d} \Phi}\right) \\ 
&- \tilde{\epsilon}^2 \,\sum_\beta \left( \frac{\partial Y^{(2)}_\alpha}{\partial \tilde{X}_\beta} \frac{\mathrm{d} \tilde{X}_\beta}{\mathrm{d} \Phi}\right)  - \frac{\partial Y^{(1)}_\alpha}{\partial \varphi} - \tilde{\epsilon}^{1/2} \frac{\partial Y^{(3/2)}_\alpha}{\partial \varphi} - \tilde{\epsilon} \frac{\partial Y^{(2)}_\alpha}{\partial \varphi}.  
\end{split}
\end{align}
Keep in mind that we have divided by $\tilde{\epsilon}$, so a term of order $k$PN in $\mrm{d}X_\alpha / \mrm{d}\varphi$ will become of order $(k-1)$PN in $\mrm{d}\tilde{X}_\alpha / \mrm{d}\Phi$. 

We know that:
\be
\left\langle \frac{\partial Y^{(i)}_\alpha}{\partial \varphi} \right\rangle = \frac{1}{2\pi} \int_0^{2\pi} \frac{\partial Y^{(i)}_\alpha}{\partial \varphi}  d \varphi  = \frac{1}{2\pi} \left[ Y^{(i)}_\alpha \right]_0^{2\pi} = 0,
\ee
considering that $Y^{(i)}_\alpha$ is $2\pi$-periodic. Also, if we assume that $\Phi$ and $\varphi$ are independent, we can write:
\be
\label{eq:Yalpha_der}
\left\langle \frac{\partial Y^{(i)}_\alpha}{\partial \tilde{X}_\beta} \right\rangle \equiv \left\langle Y^{(i)}_{\alpha,\beta} \right\rangle = 0,
\ee
where we use the standard coma notation for partial derivatives. Then:
\be
\left\langle \frac{\partial Y^{(i)}_\alpha}{\partial \tilde{X}_\beta} \frac{\mrm{d} \tilde{X}_\beta}{\mrm{d} \Phi} \right\rangle = \left\langle \frac{\partial Y^{(i)}_\alpha}{\partial \tilde{X}_\beta} \right\rangle \frac{\mrm{d} \tilde{X}_\beta}{\mrm{d} \Phi} = 0.
\ee
and by taking the orbit average, we get:
\begin{align}
\begin{split}
    \label{eq:evol_Xtilde_a}
\frac{\mrm{d} \tilde{X}_\alpha}{\mrm{d} \Phi} &= \left\langle Q_\alpha^{(1)}(X_\beta(\Phi,\varphi),\varphi) \right\rangle + \tilde{\epsilon}^{1/2} \, \left\langle Q_\alpha^{(3/2)}\left(X_\beta (\Phi,\varphi);\varphi \right)\right\rangle \\
&\quad + \tilde{\epsilon} \, \left\langle Q_\alpha^{(2)} (X_\beta(\Phi,\varphi),\varphi) \right\rangle.
\end{split}
\end{align}



Let us now consider the Taylor expansions:
\begin{align}
\begin{split}
\label{eq:QTaylor}
Q_\alpha^{(i)}(X_\beta;\varphi) &= Q_\alpha^{(i)}(\tilde{X}_\beta + \tilde{\epsilon} \, Y^{(1)}_\beta + \tilde{\epsilon}^{3/2} \, Y^{(3/2)}_\beta + \tilde{\epsilon}^2 \, Y^{(2)}_\beta;\varphi) \\ 
&= Q_\alpha^{(i)}(\tilde{X}_\beta;\varphi)  \\
&\quad+ \sum_\gamma \left( \tilde{\epsilon} \, Y^{(1)}_\gamma + \tilde{\epsilon}^{3/2} \, Y^{(3/2)}_\gamma + \tilde{\epsilon}^2 \, Y^{(2)}_\gamma\right)Q_{\alpha,\gamma}^{(i)}+... \\ 
&= Q_\alpha^{(i)}(\tilde{X}_\beta;\varphi) + \tilde{\epsilon} \, \sum_\gamma \left(Y^{(1)}_\gamma \, Q_{\alpha,\gamma}^{(i)} (\tilde{X}_\beta;\varphi)\right) \\
& \quad+ \, \tilde{\epsilon}^{3/2} \, \sum_\gamma \left(Y^{(3/2)}_\gamma \, Q_{\alpha,\gamma}^{(i)} (\tilde{X}_\beta;\varphi)\right) + O(\tilde{\epsilon}^2), \\
\end{split}
\end{align}
where the same coma notation for partial derivative is used. The summation is done on all parameters $\tilde{X}_\gamma$ that appear in $Q_\alpha^{(i)}$. Note that $Q_\alpha^{(i)} (\tilde{X}_\beta,\varphi)$ in this expression means that we consider the orbital parameters to be only secularly varying, and constant at the orbital timescale considered in the average process.
So considering all terms up to 1PN order, Eq. \eqref{eq:evol_Xtilde_a} becomes:
\begin{align}
\begin{split}
\label{eq:dXdth_0}
\frac{\mathrm{d} \tilde{X}_\alpha(\Phi)}{\mathrm{d} \Phi} &= \langle Q_\alpha^{(1)}(\tilde{X}_\beta;\varphi) \rangle \\ 
&\quad+ \tilde{\epsilon}^{1/2}\langle Q_\alpha^{(3/2)}(\tilde{X}_\beta;\varphi) \rangle\\ 
&\quad+ \tilde{\epsilon} \left[\sum_\gamma \left(\langle Y^{(1)}_\gamma \, Q_{\alpha,\gamma}^{(1)} (\tilde{X}_\beta;\varphi) \rangle\right) + \langle Q_\alpha^{(2)}(\tilde{X}_\beta;\varphi) \rangle \right]. \\ 
\end{split}
\end{align}
This is the final expression of the secular evolution of the orbital parameters. This expression is in agreement with the formulas of \citet{WillMaitra2016,Tucker2019} provided that only $Q_\alpha^{(1)}$ is considered\footnote{Note that \citet{WillMaitra2016,Tucker2019} call $Q_\alpha^{(0)}$ the quantity labeled $Q_\alpha^{(1)}$ in these notes. This is because they consider it as a 0PN quantity in $d \tilde{X} / d \Phi$, while we consider it as a 1PN quantity in $d X / d \varphi$. The two points of view are exactly equivalent.}.
\citet{WillMaitra2016,Tucker2019} derive the following expression for $Y^{(1)}_\alpha$:
\be
\label{eq:Y1alpha}
Y^{(1)}_\alpha = \int_0^\varphi Q^{(1)}_\alpha \, d \varphi' - \left( \varphi + \pi \right) \left\langle Q^{(1)}_\alpha \right\rangle + \left\langle \varphi Q^{(1)}_\alpha \right\rangle.
\ee

The first term in the square brackets of Eq. \eqref{eq:dXdth_0} then becomes:
\be
\label{eq:Yterms}
\begin{split}
\left\langle Y^{(1)}_\gamma \, Q_{\alpha,\gamma}^{(1)} (\tilde{X}_\beta,\varphi) \right\rangle  &= \left\langle Q_{\alpha,\gamma}^{(1)} \int_0^\varphi Q_\gamma^{(1)} \right\rangle + \left\langle \varphi Q_\gamma^{(1)} \right\rangle \left\langle Q_{\alpha,\gamma}^{(1)} \right\rangle \\
&- \pi \left\langle Q_\gamma^{(1)} \right\rangle \left\langle Q_{\alpha,\gamma}^{(1)} \right\rangle - \left\langle Q_\gamma^{(1)} \right\rangle \left\langle \varphi Q_{\alpha,\gamma}^{(1)} \right\rangle,
\end{split}
\ee
so that provided that the $Q_\alpha^{(i)}$ are known for all elements, the secular evolution can be derived. Finally we can come back to the evolution equation as a function of $\varphi$ by remembering that $\Phi$ and $\varphi$ are actually the same quantity up to a factor $\tilde{\epsilon}$. So we get:
\begin{align}
\begin{split}
\label{eq:dXdphi_final}
\frac{\mathrm{d} \tilde{X}_\alpha(\varphi)}{\mathrm{d} \varphi} &= \tilde{\epsilon} \, \langle Q_\alpha^{(1)}(\tilde{X}_\beta;\varphi) \rangle \\ 
&\quad+ \tilde{\epsilon}^{3/2} \, \langle Q_\alpha^{(3/2)}(\tilde{X}_\beta;\varphi) \rangle \\
&\quad+ \tilde{\epsilon}^2 \left[\sum_\gamma \left(\langle Y^{(1)}_\gamma \, Q_{\alpha,\gamma}^{(1)} (\tilde{X}_\beta;\varphi) \rangle\right) + \langle Q_\alpha^{(2)}(\tilde{X}_\beta;\varphi) \rangle \right], \\ 
\end{split}
\end{align}
which is the second order secular evolution of the orbital element, here considered as a function of $\varphi$ only (the variable $\Phi$ was just a mathematical trick to derive the equation).

\subsection{Application to the 2PN Schwarzschild shift}\label{app2timescale_sch}

In this section we neglect the spin effects and consider a 2PN Schwarzschild spacetime. So in particular the $Q_\alpha^{(3/2)}$ is zero.

Let us start from the expression of $d \omega / d t$ given in Eq. \eqref{dwdt}.
Eq. \eqref{dvarphidt} leads to:
\be
\label{dvarphidt_sch}
\frac{\mrm{d} \varphi}{\mrm{d} t} = \sqrt{\frac{GM}{p^3}} \left( 1 + e \cos f\right)^2,
\ee
as $\Omega$ is constant in Schwarzschild (the evolution of $\Omega$ depends
on $\mathcal{W}$, which is linked to the spin). 
The evolution of $\omega$ as a function of $\varphi$ then reads:
\be
\begin{split}
\frac{\mrm{d} \omega}{\mrm{d} \varphi} &= \frac{1}{e} \sqrt{\frac{p}{GM}} \left[ - \cos f \mathcal{R} + \frac{2 + e \cos f}{1 + e \cos f} \sin f \mathcal{S} \right] \\
&\times \sqrt{\frac{p^3}{GM}} \left( 1 + e \cos f\right)^{-2},
\end{split}
\ee
by combining Eqs. \eqref{dpdt} to \eqref{dwdt} and \eqref{dvarphidt_sch}.
Using the 2PN expressions of $\mathcal{R}$ and $\mathcal{S}$ given in Eqs. \eqref{R_sch} and \eqref{S_sch}, we get:
\be
\begin{split}
\frac{\mrm{d} \omega}{\mrm{d} \varphi} &= \epsilon \, \frac{1}{e} \left[ -\cos f \left( 3 + 2 e \cos f + e^2 \left(4 \sin^2 f - 1\right) \right) \right.\\
& \qquad \;\left.+ 4\, \frac{2 + e \cos f}{1 + e \cos f}  \sin f \left( 1 + e \cos f \right) e \sin f \right] \\
&+ \epsilon^2 \, \frac{1}{e} \left[ -\cos f \left(- 9 \left( 1 + e \cos f \right)^2 \right) \right.\\
& \qquad \;\left.-  2 \, \frac{2 + e \cos f}{1 + e \cos f}  \sin f\left( 1 + e \cos f \right)^2 \, e \sin f \right].
\end{split}
\ee
We want to replace $\epsilon$ by $\tilde{\epsilon} = GM/\tilde{p}c^2$, for a constant value of the semilatus rectum $\tilde{p}$ to obtain:
\be
\begin{split}
\frac{\mrm{d} \omega}{\mrm{d} \varphi} &= \tilde{\epsilon} \, \frac{1}{e} \, \frac{\tilde{p}}{p}\, \left[ -\cos f \left( 3 + 2 e \cos f + e^2 \left(4 \sin^2 f - 1\right) \right) \right.\\
& \qquad \;\left.+ 4\, \frac{2 + e \cos f}{1 + e \cos f}  \sin f \left( 1 + e \cos f \right) e \sin f \right] \\ 
&+ \tilde{\epsilon}^2 \, \frac{1}{e} \left( \frac{\tilde{p}}{p}\right)^2 \left[ -\cos f \left(- 9 \left( 1 + e \cos f \right)^2 \right) \right.\\
& \qquad \;\left.-  2 \, \frac{2 + e \cos f}{1 + e \cos f}  \sin f\left( 1 + e \cos f \right)^2 \, e \sin f \right],
\end{split}
\ee
which takes the exact same form as the general expression of Eq. \eqref{eq:evol_Xa}. The quantity in factor of $\tilde{\epsilon}$ in the first line corresponds to $Q_\omega^{(1)}$, while that in factor of $\tilde{\epsilon}^2$ in the second line corresponds to $Q_\omega^{(2)}$. They both depend on three orbital parameters, $\omega$ itself (through $f = \varphi - \omega$), $p$ and $e$, as well as explicitly on $\varphi$ (again through $f$). They are $2\pi$-periodic in $\varphi$. Let us simplify a bit to get:
\be
\begin{split}
Q_\omega^{(1)} &= \frac{\tilde{p}}{p} \left(8 + \left(e - \frac{3}{e} \right) \cos f - 10 \cos^2 f \right); \\
Q_\omega^{(2)} &= \left( \frac{\tilde{p}}{p}\right)^2 \left(-4 + 3 \left( \frac{3}{e} - 2e \right) \cos f + 2 \left(11 - e^2\right) \cos^2 f \right.\\
& \qquad \;\left.+ 15 e \cos^3 f + 2 e^2 \cos^4 f\right).
\end{split}
\ee

We now want to compute Eq. \eqref{eq:dXdphi_final} for $\omega$, which reads:
\be
\begin{split}
\label{eq:dom_dphi}
\frac{\mrm{d} \tilde{\omega}}{\mrm{d} \varphi} &= \tilde{\epsilon} \, \left\langle Q_\omega^{(1)}(p, e,\omega,\varphi) \right\rangle \\ 
& + \tilde{\epsilon}^2 \left[\left\langle Y^{(1)}_p \, Q_{\omega,p}^{(1)} (p, e,\omega,\varphi) \right\rangle + \left\langle Y^{(1)}_e \, Q_{\omega,e}^{(1)} (p, e,\omega,\varphi) \right\rangle \right.\\
& \qquad \;\left.+ \left\langle Y^{(1)}_\omega \, Q_{\omega,\omega}^{(1)} (p, e,\omega,\varphi) \right\rangle  + \,\left\langle Q_\omega^{(2)}(p, e,\omega,\varphi) \right\rangle \right],
\end{split}
\ee
where we remind that the orbital elements appearing in the various functions $Q$ and $Y$ are considered to vary only secularly (see the comment following Eq. \eqref{eq:QTaylor}): they are functions of $\Phi$ alone, they do not depend on $\varphi$, so they are not affected by the orbit averaging. For instance, we have $\left\langle e \cos f \right\rangle = e \left\langle \cos f \right\rangle $, which would not be true if the full dependence of $e$ on both $\varphi$ and $\Phi$ was considered.

The first and last terms of Eq. \eqref{eq:dom_dphi} are trivial to compute:
\be
\begin{split}
\left\langle Q_\omega^{(1)} \right\rangle &= 3, \\
\left\langle Q_\omega^{(2)} \right\rangle &= 7 - \frac{e^2}{4},
\end{split}
\ee
where we use $p=\tilde{p}$, which is true here given that the orbital-timescale variability of $p$ is discarded, only its secular variation is considered. 

The value of $\left\langle Q_\omega^{(1)} \right\rangle$ leads to the well-known textbook result: $\mrm{d} \tilde{\omega} / \mrm{d}\varphi = 3 \tilde{\epsilon}$ at 1PN order. However here, we need to go to 2PN, so a we also need to consider the three $\left\langle YQ \right\rangle$ terms in the bracket of Eq. \eqref{eq:dom_dphi}. They read:
\be
\begin{split}
\label{eq:Yterms_om}
&\left\langle Y^{(1)}_p \, Q_{\omega,p}^{(1)} \right\rangle + \left\langle Y^{(1)}_e \, Q_{\omega,e}^{(1)} \right\rangle + \left\langle Y^{(1)}_\omega \, Q_{\omega,\omega}^{(1)} \right\rangle = \\ 
&\qquad \;\qquad \; \left\langle Q_{\omega,p}^{(1)} \int_0^\varphi Q_p^{(1)} \right\rangle + \left\langle Q_{\omega,e}^{(1)} \int_0^\varphi Q_e^{(1)} \right\rangle + \left\langle Q_{\omega,\omega}^{(1)} \int_0^\varphi Q_\omega^{(1)} \right\rangle \\ 
&\qquad \;\qquad \; + \, \left\langle \varphi Q_p^{(1)} \right\rangle \left\langle Q_{\omega,p}^{(1)} \right\rangle + \left\langle \varphi Q_e^{(1)} \right\rangle \left\langle Q_{\omega,e}^{(1)} \right\rangle + \left\langle \varphi Q_\omega^{(1)} \right\rangle \left\langle Q_{\omega,\omega}^{(1)} \right\rangle \\ 
&\qquad \;\qquad \; - \, \pi \left\langle Q_p^{(1)} \right\rangle \left\langle Q_{\omega,p}^{(1)} \right\rangle - \pi \left\langle Q_e^{(1)} \right\rangle \left\langle Q_{\omega,e}^{(1)} \right\rangle - \pi \left\langle Q_\omega^{(1)} \right\rangle \left\langle Q_{\omega,\omega}^{(1)} \right\rangle \\
&\qquad \;\qquad \; - \, \left\langle Q_p^{(1)} \right\rangle \left\langle \varphi Q_{\omega,p}^{(1)} \right\rangle -  \left\langle Q_e^{(1)} \right\rangle \left\langle \varphi Q_{\omega,e}^{(1)} \right\rangle - \left\langle Q_\omega^{(1)} \right\rangle \left\langle \varphi Q_{\omega,\omega}^{(1)} \right\rangle,
\end{split}
\ee
where we use Eq. \eqref{eq:Yterms} to express the $Y^{(1)}_\alpha$. We thus need to compute quite a few more terms.
Let us start by:
\be
\begin{split}
Q_{\omega,p}^{(1)} &= -\frac{1}{p} Q_{\omega}^{(1)}, \\ 
Q_{\omega,e}^{(1)} &= \cos f \left( 1 + \frac{3}{e^2}\right), \\
Q_{\omega,\omega}^{(1)} &= \sin f \left( e - \frac{3}{e} - 20 \cos f\right).
\end{split}
\ee
The last two have zero average, because $\left\langle \cos f \right\rangle = \left\langle \sin f \right\rangle = \left\langle \cos f \sin f \right\rangle = 0$. So, this cancels a few terms in the second and third lines of the RHS of Eq. \eqref{eq:Yterms_om}. Following the exact same procedure as for $\omega$, we can write $\mrm{d} p / \mrm{d}\varphi$ and $\mrm{d}e / \mrm{d}\varphi$, and extract their 1PN component to get:
\be
\begin{split}
Q_p^{(1)} &= 8\, \tilde{p} \,e \,\sin f, \\ 
Q_e^{(1)} &= \left(3 + 7 e^2\right) \sin f + 10 e \cos f \sin f, 
\end{split}
\ee
both of which have zero average (no secular evolution of $p$ and $e$ at 1PN order). This cancels the first two terms of the last two lines in Eq. \eqref{eq:Yterms_om}. So we are left with:
\be
\begin{split}
\label{eq:Yterms_om_}
&\left\langle Y^{(1)}_p \, Q_{\omega,p}^{(1)} \right\rangle + \left\langle Y^{(1)}_e \, Q_{\omega,e}^{(1)} \right\rangle + \left\langle Y^{(1)}_\omega \, Q_{\omega,\omega}^{(1)} \right\rangle = \\
&\qquad \;\qquad \; \left\langle Q_{\omega,p}^{(1)} \int_0^\varphi Q_p^{(1)} \right\rangle + \left\langle Q_{\omega,e}^{(1)} \int_0^\varphi Q_e^{(1)} \right\rangle + \left\langle Q_{\omega,\omega}^{(1)} \int_0^\varphi Q_\omega^{(1)} \right\rangle \\ 
&\qquad \;\qquad \; + \, \left\langle \varphi Q_p^{(1)} \right\rangle \left\langle Q_{\omega,p}^{(1)} \right\rangle  -  3 \left\langle \varphi Q_{\omega,\omega}^{(1)} \right\rangle .
\end{split}
\ee
It is easy to get:
\be
\begin{split}
\left\langle \varphi Q_p^{(1)} \right\rangle &= -8 \, \tilde{p} \,e \,\cos \omega, \\ 
\left\langle \varphi Q_{\omega,\omega}^{(1)} \right\rangle &= - \cos \omega \left(e - \frac{3}{e} \right) + 5 \, \cos 2\omega,
\end{split}
\ee
using:
\be
\begin{split}
\left\langle \varphi \sin f \right\rangle = - \cos \omega, \quad \left\langle \varphi \cos f \sin f \right\rangle = - \frac{1}{4} \cos 2\omega.
\end{split}
\ee
The terms with integrals of $Q_\beta^{(1)}$ are a bit more cumbersome, but it is still straightforward to obtain:
\be
\begin{split}
\left\langle Q_{\omega,p}^{(1)} \int_0^\varphi Q_p^{(1)} \right\rangle &= -24 e \cos \omega + 4 e \left( e - \frac{3}{e} \right), \\ 
\left\langle Q_{\omega,e}^{(1)} \int_0^\varphi Q_e^{(1)} \right\rangle &=  -12 - \frac{7}{2}e^2 - \frac{9}{2 e^2}, \\
\left\langle Q_{\omega,\omega}^{(1)} \int_0^\varphi Q_\omega^{(1)} \right\rangle &= \frac{25}{2}+ \left( e - \frac{3}{e} \right) \left(-3 \cos \omega + \frac{1}{2} \left( e - \frac{3}{e}\right) \right) \\
&\;\;\; + 15 \cos 2\omega.
\end{split}
\ee

Putting everything together, we finally arrive at:
\be
\label{eq:domdphi_fin}
\frac{\mrm{d} \tilde{\omega}}{ \mrm{d} \varphi} = 3\,\tilde{\epsilon} -\frac{3}{4}\tilde{\epsilon}^2 \, \left( 10 -  \tilde{e}^2\right).
\ee
Eq. \eqref{eq:domdphi_fin} agrees with the 2PN result of \citet{WillMaitra2016}, see their "Eq. (3.23a)". Finally, we can multiply Eq. \eqref{eq:domdphi_fin} by $2\pi$ to have the 2PN expression of the secular shift of the pericenter/apocenter in Schwarzschild:
\be
\label{eq:Deltaom_sch}
\Delta\omega_{\mrm{Sch}}  = 6\pi\,\tilde{\epsilon} -\frac{3\pi}{2}\tilde{\epsilon}^2 \, \left( 10 -  \tilde{e}^2\right).
\ee

If in the 2PN term of Eq. \eqref{eq:Deltaom_sch}, if we replace $\tilde{\epsilon}$ and $\tilde{e}$ by $\epsilon$ and $e$, the contribution of this replacement will only affect the expression at the 3PN order (the 2PN term will still be reduced to the same expression). Similarly, in the 1PN term of Eq. \eqref{eq:Deltaom_sch}, if we replace $\tilde{\epsilon}$ by $\epsilon$, the contribution of this replacement will impact the expression at the 2PN order. This is why it is crucial to use $\tilde{\epsilon}$ in the 1PN term of Eq. \eqref{eq:Deltaom_sch} when expressing the Schwarzschild precession at the 2PN order using Eq. \eqref{eq:Deltaom_sch}. If one wants to use the osculating elements, then the choice of Eq. \eqref{ansatz} would be different as well as the 2PN term of the final expression. Conversely, when using the Lense-Thirring and quadrupole moment 2PN secular shifts we can use any osculating or average value of $p$, $\theta$ and $\psi$ because these expressions contain only the leading contribution.

\section{Mid-inclined orbits}\label{sec_outofplane_mid}

\begin{figure}[htp]
    \centering
    \subfloat[Schematic illustration of the osculating Keplerian orbit]
    {\includegraphics[clip,width=0.7\columnwidth]{ 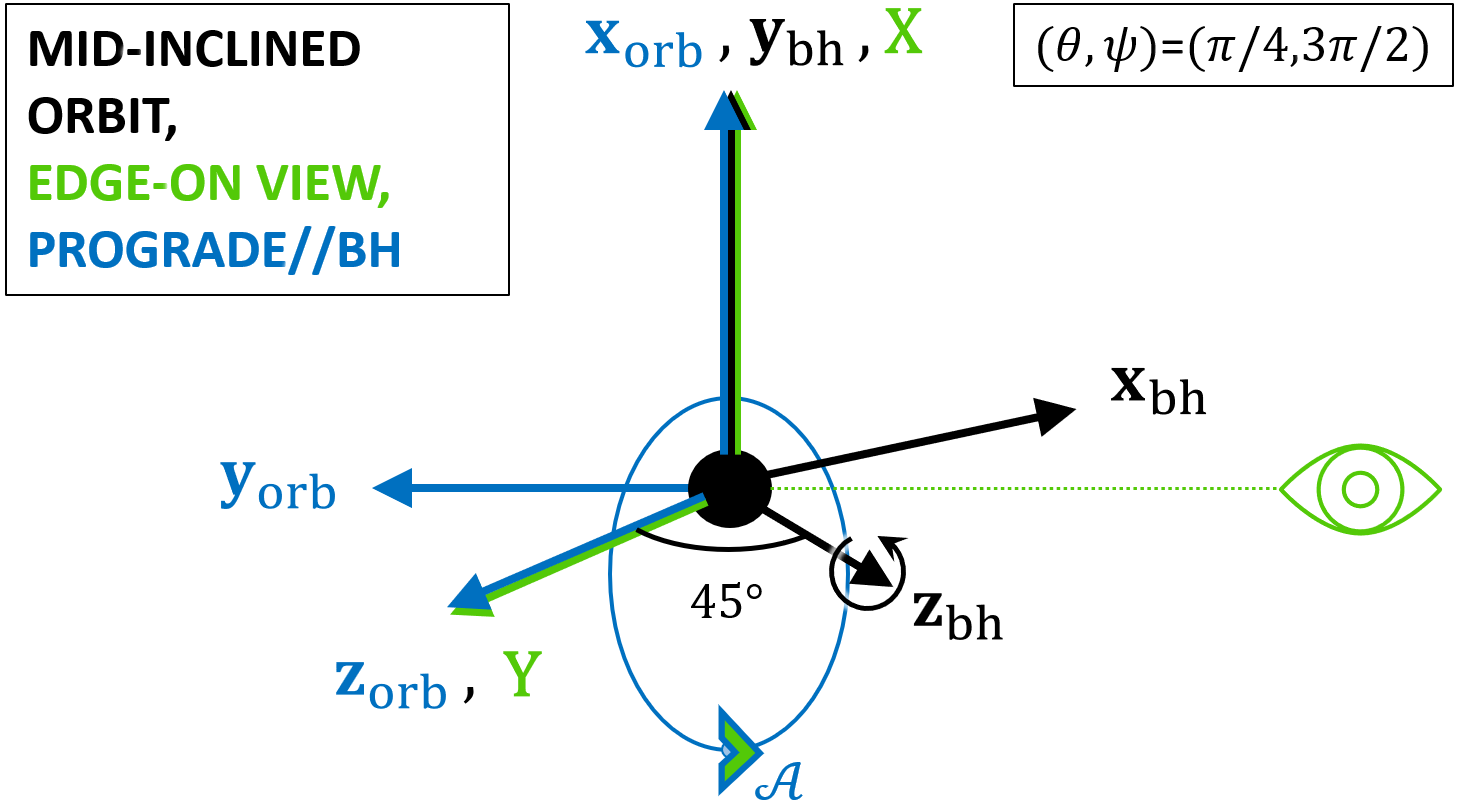}\label{illustr_mid_pro}}\\
    \vspace{3mm}
    \subfloat[Simulation using the 2PN code]
    {\includegraphics[clip,width=0.87\columnwidth]{ 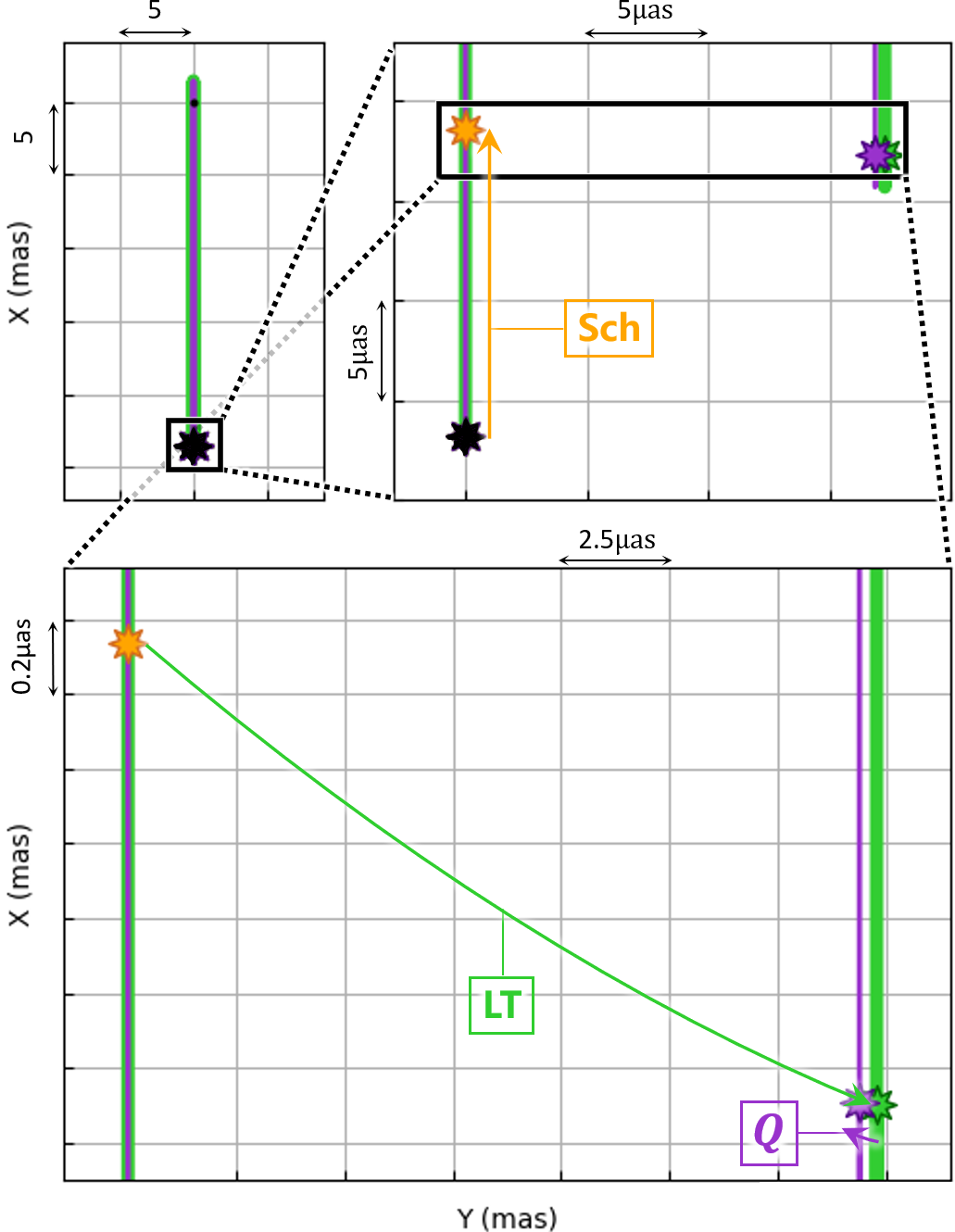}\label{simu_mid_pro}}\\
    \caption{\label{mid_pro} Same as figure \ref{eq_pro} but with mid-inclined orbits relative to the equatorial plane of the black hole. We see that that the apocenter experiences both an in-plane and out-of-plane precession due to the Lense-Thirring effect. The in-plane precession is clockwise as seen on the face-on projection of figure \ref{illustr_eq_pro}, and the out-of-plane precession is mainly for the major axis but also acts on the minor axis. We also see that the quadrupole moment appears to have the opposite behavior to the Lense-Thirring effect for both the in-plane and out-of-plane shifts.}
\end{figure}

\begin{figure}[htp]
    \centering
    \subfloat[Schematic illustration of the osculating Keplerian orbit]
    {\includegraphics[clip,width=0.7\columnwidth]{ 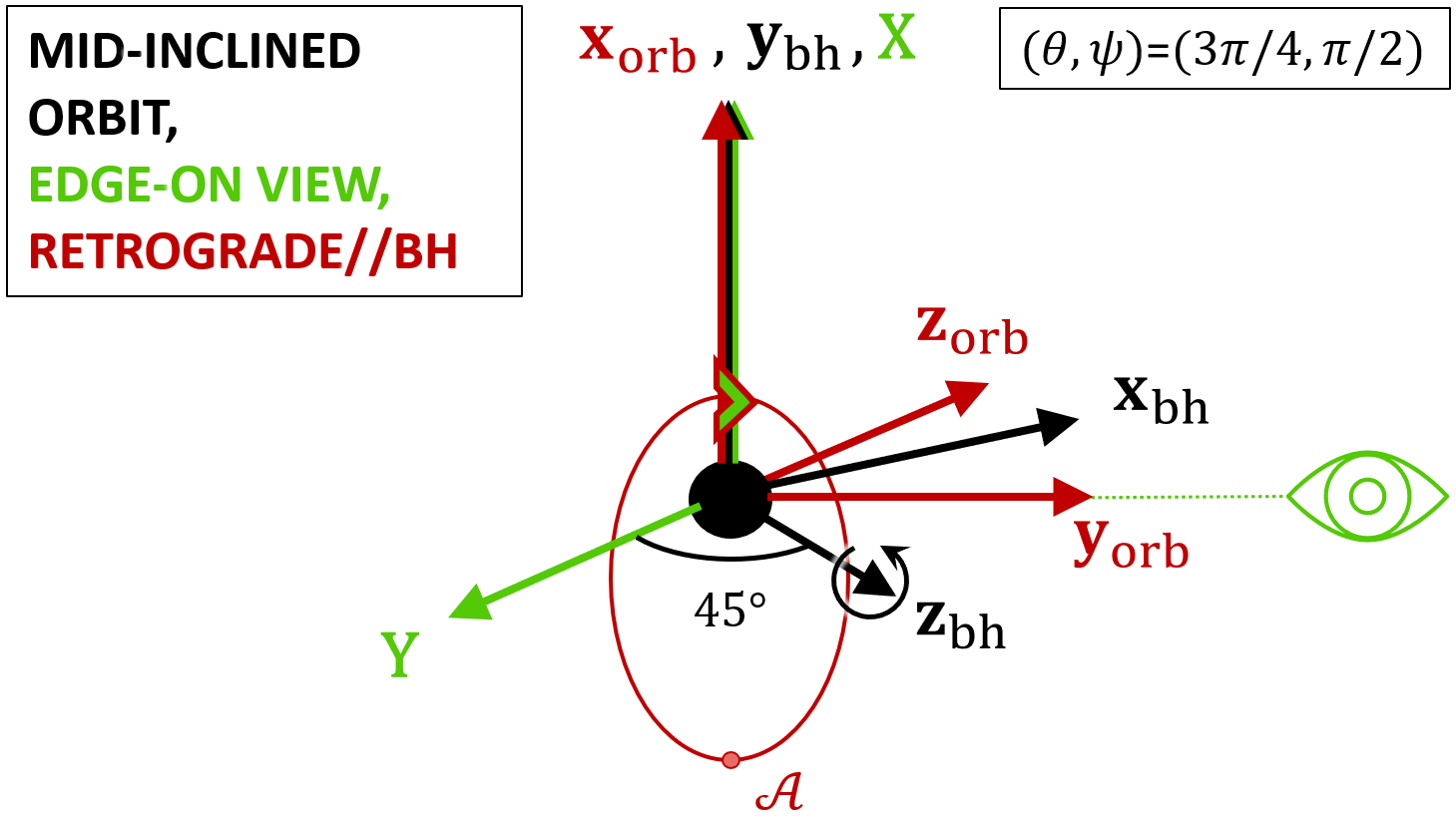}\label{illustr_mid_ret}}\\
    \vspace{3mm}
    \subfloat[Simulation using the 2PN code]
    {\includegraphics[clip,width=0.9\columnwidth]{ 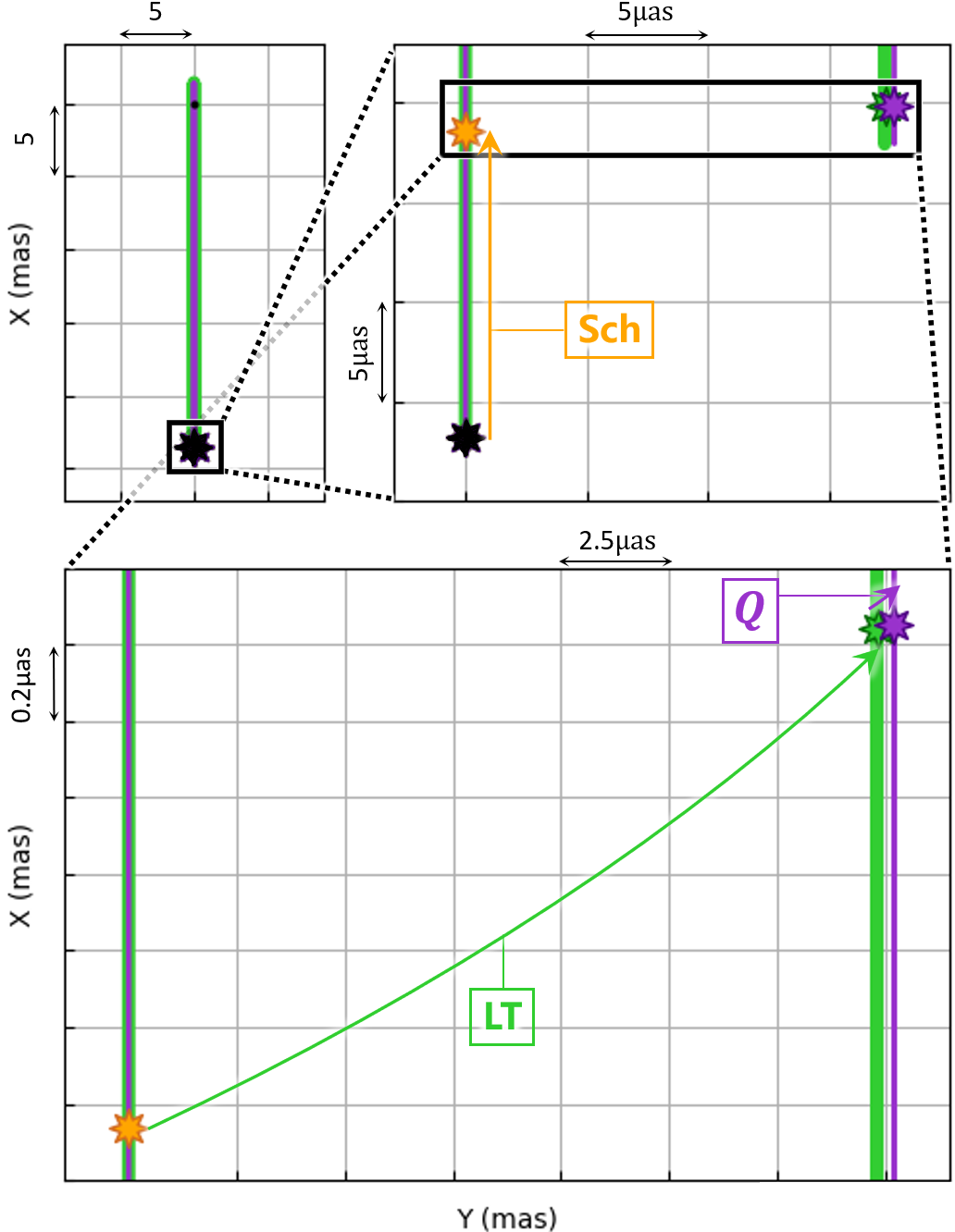}\label{simu_mid_ret}}\\
    \caption{\label{mid_ret} Same as figure \ref{mid_pro} but for retrograde orbits, relative to the black hole. We see that the apocenter experiences both an in-plane and out-of-plane precession due to the Lense-Thirring effect. The in-plane precession is clockwise as seen on the face-on projection of figure \ref{illustr_eq_ret}, and the out-of-plane precession is mainly for the major axis but also acts on the minor axis. We also see that the quadrupole moment appears to have the same behavior as the Lense-Thirring effect for the both types of precession.}
\end{figure}

\begin{figure}[htp]
    \centering
    \includegraphics[trim={0 0 0 0},clip,width=0.75\columnwidth]{ 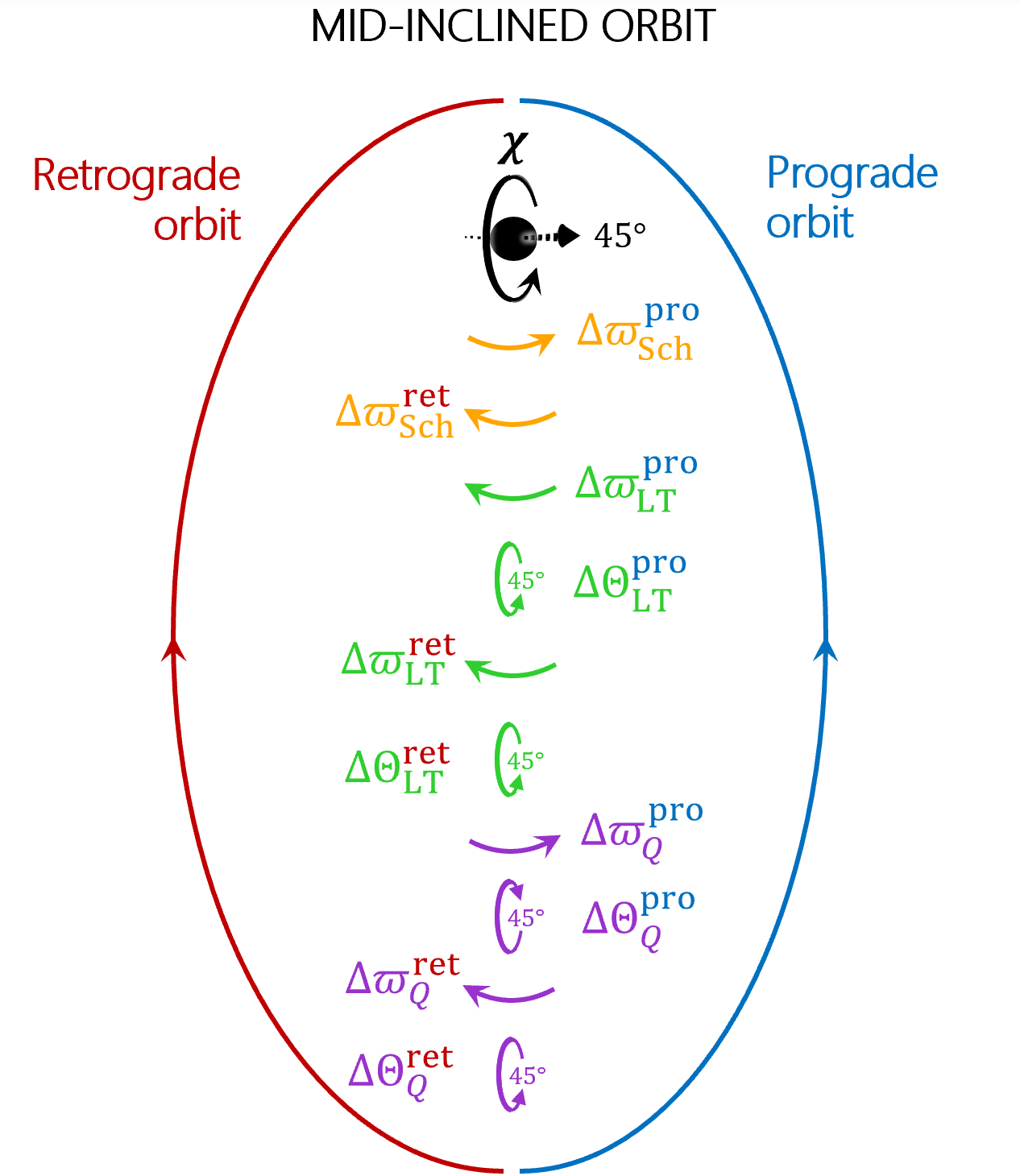}
    \caption{\label{mid_theory} Schematic illustration of the secular precessions for a mid inclined orbit relative to the equatorial plane of the black hole. When we have $\mbf z_{\mathrm{bh}}=-\mbf y^\mrm{pro}_{\mathrm{orb}}+\mbf z^\mrm{pro}_{\mathrm{orb}}=\mbf y^\mrm{ret}_{\mathrm{orb}}-\mbf z^\mrm{ret}_{\mathrm{orb}}$ for example, the in-plane precessions act in the same way as for equatorial orbits (see \ref{eq_plane_theory}). As for the out-of-plane precessions, the Lense-Thirring precession pushes the apocenter in the same direction as the spin, and the quadrupole moment effect as well for the retrograde orbit but in the direction opposite to the the spin for the prograde orbit.}
\end{figure}

Let us investigate the case of a mid-inclined orbit, for example, the case where $\mbf z_{\mathrm{bh}}=-\mbf y^\mrm{pro}_{\mathrm{orb}}+\mbf z^\mrm{pro}_{\mathrm{orb}}=\mbf y^\mrm{ret}_{\mathrm{orb}}-\mbf z^\mrm{ret}_{\mathrm{orb}}$, i.e $(\theta,\psi)=(\pi/4,3\pi/2)$ or $(\theta,\psi)=(3\pi/4,\pi/2)$. we see from Eqs. \eqref{Dvarpi_chi}, \eqref{DTh_chi}, and \eqref{DXi_chi}, that with these values of $\theta$ and $\psi$ we will observe both in-plane and out-of-plane precessions due to the Lense-Thirring effect, at moderate values. However, when looking at Eqs. \eqref{DTh_q} and \eqref{DXi_q} we see that the orbits will experience the most out-of-plane shift due to the quadrupole moment in this configuration.
Indeed, when simulating a prograde and retrograde orbit relative to the black hole, we see in figures \ref{mid_pro} and \ref{mid_ret} that the Lense-Thirring and quadrupole moment effects each make the apocenters experience both an in-plane and out-of-plane precession. The in-plane precessions are similar the the equatorial orbit case. As for the out-of-plane precessions, they mainly act on the major axis but also on the minor axis on much longer timescales (see Eq. \eqref{P_cone} below). We also see that the quadrupole moment appears to oppose the Lense-Thirring effect for the out-of-plane shift when the orbit is prograde, but not for the retrograde orbit.


In practice, unless we have an equatorial (no out-of-plane precession) or polar (surface of the out-of-plane precession cone represented in figure \ref{split_out} being a plane for $\theta=\pi/2[\pi]$) orbit with the major axis being equatorial or polar, both the major and minor axis will precess. However, the out-of-plane shift remains mainly along the tangent to the surface of the out-of-plane precession cone as long as the period of the orbit is much less than the precession period $P^\mrm{\,cone}$ of the angular momentum of the orbit around the out-of-plane precession cone. To check that, we derive $P^\mrm{\,cone}/P$ using Eqs. \eqref{rate_LT} and \eqref{rate_Q} at the 2PN order: 
 \begin{equation}\label{P_cone}
     \frac{P^\mrm{\,cone}_\mrm{LT}}{P} = \frac{2\pi}{\upomega_\mrm{LT}P} \quad ; \quad \frac{P^\mrm{\,cone}_Q}{P} = \frac{2\pi}{\upomega_Q P}
 \end{equation}
If we apply this estimation for "S2/10" for example, we get $\frac{P^\mrm{\,cone}_\mrm{LT}}{P} (\chi=0.99)\sim 10^3$, $\frac{P^\mrm{\,cone}_\mrm{LT}}{P} (\chi=0.1)\sim 10^4$, $\frac{P^\mrm{\,cone}_Q}{P} (\chi=0.99)\sim 10^5$ and $\frac{P^\mrm{\,cone}_Q}{P} (\chi=0.1)\sim 10^7$.

We summarize in a schematic way the results of figures \ref{mid_pro} and \ref{mid_ret} in figure \ref{mid_theory}, and how the quadrupole moment out-of-plane precession of the major axis, is studied through a mid-inclined orbits with $\mbf z_\mrm{bh//orb}//\mbf y_\mrm{orb}$.  Alternatively, in order to maximize the quadrupole moment out-of-plane precession of the minor axis instead of the major axis, one would need to study a mid-inclined orbits with $\mbf z_\mrm{bh//orb}//\mbf x_\mrm{orb}$.

In conclusion, we now have insight into the different types of precession that can manifest around a rotating black hole and comprehend how a random relative orientation between the orbit and black hole can be decomposed into these precession categories. With this new understanding, one can have a better intuition over the possible black hole orientations that can generate the secular precessions that we observe on the orbits of currently, and potential closer-in stars.

\end{appendix}

\end{document}